\documentclass[letter,12pt]{article}
\pdfoutput=1 

\usepackage{jheppub} 

\usepackage[utf8]{inputenc}

\usepackage{epsfig}
\usepackage{amssymb}
\usepackage{amsfonts}
\usepackage{amsmath}
\usepackage{titlesec}
\usepackage{yfonts}

\usepackage{adjustbox}

\usepackage{arydshln}

\usepackage{ dsfont }

\usepackage[russian,english]{babel}
\usepackage[T1,T2A]{fontenc} 

\usepackage[utf8]{inputenc}
\usepackage{amsmath}
\usepackage{calc}

\usepackage{accents}
\newcommand{\dbtilde}[1]{\accentset{\approx}{#1}}

\def\g{{\rm g}}

\def\H{{\mathcal H}}

\def\C{{\mathcal C}}

\def\CC{{\mathbb C}}

\def\Im{ \hbox{\rm Im}}
\def\Ker{ \hbox{\rm Ker}}

\newcommand{\Z}{\ensuremath{\mathbb Z}}
\newcommand{\R}{\ensuremath{\mathbb R}}

\usepackage{amsmath}

\usepackage{ upgreek }
\usepackage{physics}

\def\bea{\begin{eqnarray}}
\def\eea{\end{eqnarray}}

\def\w{{\rm w}}

\def\n{{\rm n}}

\mathchardef\mhyphen="2D

\title{Quantum stabilizer codes, lattices, and CFTs}

\author[a,b]{Anatoly Dymarsky,} 
\author[a]{and Alfred Shapere} 
 
\affiliation[a]{Department of Physics and Astronomy, \\ University of Kentucky,\\ 505 Rose Street, Lexington, KY,  40506, USA\\}
\affiliation[b]{Skolkovo Institute of Science and Technology,\\Skolkovo Innovation Center, Moscow, Russia\\}
\emailAdd{a.dymarsky@uky.edu}
\emailAdd{shapere@g.uky.edu}

\abstract{
There is a rich connection between classical error-correcting codes, Euclidean lattices, and chiral conformal field theories. Here we show that quantum error-correcting codes, those of the stabilizer type, are related to Lorentzian lattices and non-chiral CFTs. More specifically, real self-dual stabilizer codes can be associated with even self-dual Lorentzian lattices, and thus define Narain CFTs.  We dub the resulting theories code CFTs and study their properties. T-duality transformations of a code CFT, at the level of the underlying code, reduce to code equivalences.  By means of such equivalences, any stabilizer code can be reduced to a graph code. We can therefore represent code CFTs by graphs. We study code CFTs with small central charge $c=n\leq 12$, and find many interesting examples. Among them is a non-chiral $E_8$ theory, which is based on the root lattice of $E_8$ understood as an even self-dual Lorentzian lattice. 
By analyzing all graphs with $n\leq 8$ nodes we find many pairs and triples of physically distinct isospectral theories. We also construct numerous modular invariant functions satisfying all the basic properties expected of the CFT partition function, yet which are not partition functions of any known CFTs. 
We consider the ensemble average over all code theories, calculate the corresponding partition function, and discuss its possible holographic interpretation. The paper is written in a self-contained manner, and includes an extensive pedagogical introduction and many explicit examples. \\
}

\dedicated{This paper is dedicated to the memory of John Horton Conway}

\begin{document} 
\maketitle
\flushbottom

\section{Introduction}
\label{sec:intro}

It has been recognized for many years that  codes, lattices, and conformal field theories (CFTs) are deeply intertwined.  
Perhaps the best known example of this relation  is the construction of the Leech lattice from the extended Golay code.  The Leech lattice subsequently played a central role in the discovery of the monster group, which appears naturally as the symmetry of the Monster CFT -- a particular orbifold of the chiral CFT associated with the Leech lattice  \cite{frenkel1984natural,frenkel1989vertex}.  
More generally, classical self-dual binary linear codes are naturally associated with Euclidean even self-dual lattices, which in turn give rise to chiral bosonic CFTs \cite{dolan1996conformal}. This relation is not exclusive -- there are other known ways in which classical codes are related to chiral theories
\cite{dong1998framed,gaiotto2018holomorphic}.  

In view of the fruitful connections between classical codes, Euclidean lattices, and chiral CFTs, one may wonder if there is a corresponding hierarchy based on quantum codes.  After all, conformal field theories are fundamentally quantum in nature, and so it is natural to expect that their relation to codes extends to include quantum codes.  
In this paper we will develop this idea, and show that there is indeed a natural and compelling correspondence between an important class of quantum codes, real self-dual binary stabilizer codes, and even self-dual Lorentzian lattices.  These lattices define a class of nonchiral CFTs  that arise from toroidal compactifications of strings with quantized B-flux, a subset of the family of Narain CFTs.  In other words, real self-dual stabilizer codes are in one-to-one correspondence with a family of CFTs of a particular kind, which we call code CFTs.

The connection  between CFTs and quantum codes becomes most explicit at the level of the partition function, or, in the case of the underlying code, at the level of the code's enumerator polynomial.  
Analogously to the classical case, the (refined) enumerator polynomial of a real self-dual quantum code lifts to the Siegel theta function of the associated Lorentzian lattice, which becomes the CFT partition function upon multiplication by the appropriate power of $|\eta(\tau)|^2$ required for modular invariance.  The constraints of modular invariance of the partition function reduce to a set of simple algebraic relations that must be satisfied by the enumerator polynomial, making it possible to implement a ``baby'' analogue of the CFT modular bootstrap for quantum codes. Thus, the maximization of the spectral gap over CFTs of given $c$ becomes with some nuances a modification of the problem of maximizing error-correcting capacity,  as measured by the number of qubits whose decoherence a real self-dual quantum code of given length can detect. The number  ``controlling'' the spectral gap $d_{\rm b}$ can be bounded from above as a linear programming optimization problem, which we solve numerically for codes of length $n\le 32$; we verify explicitly that the bound is tight for $n\le 8$.  It is worth noting that the problem of finding codes with maximum error-correcting capacity
(which maximize the Hamming distance for a given code length) is closely related to  the problem of finding the maximum possible density of a lattice sphere packing in a given number of dimensions \cite{elkies2000lattices,elkies2000lattices2}; essentially, it is a version of the sphere-packing problem with respect to the distance measure appropriate for quantum codes rather than the Euclidean metric.  The sphere-packing problem has recently been recast in terms the CFT modular bootstrap \cite{hartman2019sphere,afkhami2020high,afkhami2020free}, and has been analyzed numerically, leading to improved bounds at finite $n$.  
Our work complements these studies of sphere packing by introducing a new relation between the modular bootstrap and codes.

Another question on which the connection to codes sheds new light is that of
the space of solutions of the modular bootstrap constraints, namely modular invariant functions $Z(\tau,\bar\tau)$ which are  sums of characters with positive integer coefficients, but which are not partition functions of any known CFTs. A family of chiral $Z(\tau)$ of this sort with central charge $c\geq 24$ has been previously discussed in \cite{witten2007three}. 
Furthermore, there are simple examples of $Z(\tau)$ discussed later in the text which do not correspond to any CFT at all. 
The connection to codes leads to many such examples, both chiral and non-chiral, with small central charge in the latter case, for which the CFT is not known or may not exist. At the level of codes, the question of finding such $Z(\tau,\bar\tau)$ reduces to constructing multivariate polynomials obeying all symmetry and positivity constraints that must be satisfied by enumerator polynomials, yet which are not enumerators of any  code.  Solving a simple linear programming problem yields many thousands of examples of ``fake'' enumerators, already for small central charge $c={\bar c}\leq 8$. 

Code CFTs form a discrete subset of the continuous moduli space of Narain CFTs.  We show that this subset, and hence the space of codes itself, can be described as a coset of discrete groups.  Acting on this coset are symmetries relating equivalent codes; code CFTs that are T-dual to each other correspond to 
 equivalent codes.  By making use of code equivalences, we are able to reduce a general code to an equivalent code of canonical form.  Each canonical representative is associated with an undirected graph, and equivalent codes ({\it i.e.} T-dual code CFTs) map to graphs related by a particular graph transformation, known as edge local complementation.  The representation by graphs provides a convenient way to classify the equivalence classes of codes of a given length.  We do this for $n\le 8$, and in the process find many interesting examples. 

One striking finding is the multitude of inequivalent codes sharing the same enumerator polynomial, which implies the existence of many examples of isospectral lattices and inequivalent isospectral code CFTs. The first such example appears for $n=7$; it corresponds to a pair of isospectral even self-dual Lorentzian lattices in $\R^{7,7}$. This is the lowest-dimensional example among lattices associated with the stabilizer codes, and in many ways is analogous to Milnor's example of the isospectral pair of even self-dual lattices in $\R^{16}$. But unlike the Euclidean case, where the next example occurs in $24$ dimensions, in the Lorentzian case we find many dozens of  pairs and even triplets of isospectral lattices in $\R^{8,8}$, and correspondingly many isospectral CFTs with $c={\bar c}=n=8$.

%

One of our original motivations for the present work came from quantum gravity.  In the context of the AdS/CFT correspondence,  information is understood to be stored at the boundary of spacetime, in a highly nonlocal and redundant form strongly reminiscent of  error-correcting codes \cite{almheiri2015bulk, pastawski2015holographic}.  This observation begs the question of exactly how information is stored in the dual CFT, and in particular, how the form of error correction seen in the bulk gravitational theory is implemented in the CFT.  While we do not claim to have a complete answer to this question at present, we have at least identified error-correcting codes within an important class of CFTs.  This same class of CFTs has recently been studied as a toy model of holography.  In papers by two sets of authors \cite{afkhami2020free,maloney2020averaging}, it has been shown that the {\it average} over moduli space of Narian CFTs can be reinterpreted as a sum over three-dimensional topologies.  The authors conjecture that the moduli-averaged CFT is dual to a three-dimensional gravitational theory with $U(1)^n \times U(1)^n$ gauge symmetry.  This suggests that if we are looking for a holographic version of our code construction, we should consider what happens when we average over codes.  We perform the average over  
a class of codes to obtain the averaged partition function of the corresponding CFTs, and  note the possibility of reinterpreting the partition function as a sum over handlebodies.  

Quantum code CFTs are a small subset of the space of all Narain CFTs, but our results indicate that they might be representative in a certain sense.  As evidence for this claim, we cite the fact that our numerical bound  on $d_{\rm b}/n$ is comparable to numerical bounds on  spectral gap $\Delta$ \cite{afkhami2020free}, and also the observation that the average over a class of code CFTs appears to have a holographic interpretation.  Future explorations may uncover further indications that code CFTs can provide a useful, stripped-down setting for studying holographic phenomena.

This paper is organized as follows. It includes an extensive pedagogical introduction. Section \ref{sec:cecc} discusses classical codes,  both binary codes (subsection \ref{sec:binary}) and codes over ${\rm GF}(4)$ (subsection \ref{sec:GF4}) as well as their relation to lattices, MacWilliams identities, Hamming and Gilbert-Varshamov bounds, and other related  questions. 
With some exceptions, most of the material presented in Section \ref{sec:cecc} is not new, and can be skipped by a reader with sufficient background.  Section \ref{sec:3} contains both pedagogical and original material. Subsection \ref{sec:3.1} 
introduces quantum error-correcting codes of the stabilizer type, their relation to self-orthogonal classical codes over ${\rm GF}(4)$, and the quantum version of the MacWilliams identities. Most of it can be skipped by the knowledgeable reader. Subsection \ref{sec:newA} introduces a crucial new ingredient -- the relation between  quantum codes and Lorentzian lattices. A reader with background in both classical and quantum codes can start reading the paper from this subsection. Section \ref{sec:4} is similarly mixed. Subsection \ref{NarainCFTs} introduces Narain CFTs, and is intended for readers with a background in classical or quantum codes, but no prior exposure to String Theory. It can be skipped by anyone familiar with toroidal compactifications. Subsection \ref{sec:codeCFTs} is again crucial -- it introduces the basic elements of  our construction relating quantum codes to CFTs.  All subsequent sections contain original material.

\section{Classical error-correcting codes}
\label{sec:cecc}
We start by reviewing classical error-correcting codes, focusing on aspects important for understanding quantum codes and their relation to CFTs. For a more in-depth treatment we recommend Elkies's comprehensive yet concise review \cite{elkies2000lattices,elkies2000lattices2}.

\subsection{Binary codes}
\label{sec:binary}
A binary code $\mathcal C$ is a collection of binary ``codewords,''  vectors of length $n$ consisting of zeros and ones, ${\mathcal C} \subset \Z_2^n$. Components of codewords $c\in {\mathcal C}$ are called bits. Each codeword encodes a particular message. When sent over a noisy channel, a codeword may be corrupted, {\it i.e.}, certain bits may be changed to their opposite values. The encoding procedure is designed to make it possible to restore the original form of the codeword and thus recover the message. This is done by replacing the corrupted codeword $c'\notin {\mathcal C}$ with the closest proper codeword, defined with some appropriate norm.  The most widely used norm is known as the Hamming distance.  Given two vectors $c_1,c_2\in \Z_2^n$, the Hamming distance between them $d(c_1,c_2)$ is the number of corresponding bits in $c_1$ and $c_2$ that are different. The Hamming distance of a code $d$ is the smallest Hamming distance between any two codewords  
\bea
d({\mathcal C})=\min_{c_1,c_2\in\C} d(c_1,c_2).
\eea
A code containing $K$ codewords  with Hamming distance $d$ is said to be of type $[n,K,d]$. Such a code can  correct an error corrupting up to $t = [(d-1)/2]$ bits, where $[x]$ denotes the greatest integer $\le x$.

Colloquially, an optimal code for given $n$ and $K$ is one with the maximum possible $d$, {\it i.e.}~one which can correct errors involving the maximum possible number of bits.  When $n$ goes to infinity, with $\log_2(K/2^n)$ approaching a finite limit, the maximum possible ratio $d/n$ controls the amount of information which can be sent over a noisy channel. There are numerous bounds on $d/n$, but the exact limiting value is not known. 

Codewords can be visualized as vertices of a unit cube in $n$ dimensions. To design a good code, one should place as many points at the cube's vertices as possible, making sure they are located  far away from each other. The distance $d$ between two vertices is calculated either with the Manhattan norm (the minimum distance an ant would need to travel along the edges to get from one vertex to the other), or equivalently, the Euclidean distance squared $\ell^2$. 

A code is linear if the sum of any two codewords $c_1,c_2\in\C$, obtained by adding the components modulo 2, is also a codeword $c_1+c_2\in \C$. In other words, a
classical linear code $\mathcal C$ is a vector space over the field $F=
=\Z_2$ consisting of two elements $\{0,1\}$. There are necessarily $K=2^k$ distinct codewords for some nonnegative integer $k$, which counts the number of ``logical'' bits.  All codewords are specified by a binary $n\times k$ generator matrix $G$,
\bea
c(x)=G\, x\in  F^n,\qquad x\in F^k , \label{linear}
\eea
where matrix multiplication is performed over the field $F$.
We use the notation $[n,k,d]$ to describe linear codes with Hamming distance $d$ that encode $k$ logical bits into $n$ physical bits. 

A linear code can equivalently be specified by a ``parity check'' matrix $H$ defined such that $\Ker(H)=\Im(G)$, so that $HG=0$, and $H c=0$ if and only if $c$ is a proper codeword (all algebra is mod 2). The parity check matrix is an $(n-k)\times n$ binary matrix of maximal rank. 

Linear codes always include the zero vector, i.e. the vector consisting of $n$ zeros. 
We introduce the Hamming weight ${\rm w}(c)$ as the sum of all elements of a code vector  (with the sum taken using conventional algebra, not mod 2). Then the Hamming distance is the minimal Hamming weight of all non-trivial codewords
\bea
d(\C)=\min_{c\in \C,c\neq 0} {\rm w}(c).
\eea

If a codeword $c$ has been corrupted by some error $e$, $c\rightarrow c'=c+e$, the error can be detected by applying the parity check matrix 
\bea
y(c')=H c'=H e. \label{y}
\eea
If $y\neq 0$, an error has occurred. However, the converse is not necessarily true. A vanishing result $y=0$ could mean that an undetectable error has occurred, one for which the error vector $e$ is a proper nonzero codeword, $e\in \C$. Clearly, these undetectable errors must simultaneously corrupt at least $d(\C)$ bits.  Therefore, increasing  $d$ improves the quality of the code by making it less likely for undetectable errors to occur.  

The name of the parity check matrix comes from its role of detecting errors.  Typical architectures for semiconductor computer memory supplement each byte (8 bits) of memory by an additional physical bit that has no effect on logical operations, which automatically takes the value that makes the sum of all nine bits even \cite{abel1995ibm}. A violation of that condition is an indication that a hardware error has occurred.

\subsubsection*{Example: repetition code}
The repetition code is, perhaps, the simplest example of a code. It encodes $k=1$ logical bit by repeating it $n$ times: 
\bea
G^T=\vec{1}\equiv (\underbrace{1,\dots,1}_n). \label{repeat}
\eea
This code, denoted as $i_n$, has two codewords and Hamming distance $d=n$.  Its  parity check matrix is the $(n-1)\times n$ matrix
\bea
\label{prc}
H=\left(\begin{array}{ccccc}
1 & 1 & 0 & \dots & 0 \\
0 &1 & 1 &  \dots & 0 \\
\dots & \dots & \dots & \dots & \dots\\
0 &\dots & 0 & 1 & 1
\end{array}\right).
\eea
If an error occurs that corrupts $[(n-1)/2]$ bits or fewer, one can restore the original message by rounding ${\rm w}(c')/n$ to the closest integer. This code has a small ratio of logical bits to physical bits, $k/n = 1/n$, and  is therefore not very efficient. 

\subsubsection*{Example: Hamming $[7,4,3]$ code}
A  more interesting example is the Hamming $[7,4,3]$ code, defined by the following parity check matrix 
\bea
\label{H7}
H=\left(
\begin{array}{ccccccc}
 1 & 0 & 1 & 0 & 1 & 0 & 1 \\
 0 & 1 & 1 & 0 & 0 & 1 & 1 \\
 0 & 0 & 0 & 1 & 1 & 1 & 1 \\
\end{array}
\right).
\eea
It has $d=3$ and therefore can detect and correct any one-bit error, $t=[(d-1)/2]=1$. Indeed, let $e_i$ for $1\leq i\leq 7$ be a one-bit error, a vector consisting of $6$ zeros and a one in the $i$-th position. Such an error can be detected and uniquely identified via \eqref{y}. For $c'=c+e_i$, 
\bea
y(c')=H c'=H e_i
\eea
is simply a binary 3-vector whose components are equal to the digits of the number $i$ written in base 2. Thus the value of $y(c')$ unambiguously indicates which bit should be flipped to restore the original message. 
All of the algebra above (except where explicitly noted) is to be understood mod 2. \\

Let us assume that a code can correct any error affecting $t=[(d-1)/2]$ bits  or fewer. There are $C_n^l=n!/l!(n-l)!$  errors affecting exactly $l$ bits and therefore $\sum_{l=1}^t C_n^l$
such errors overall. This number should not exceed the total number $2^{n-k}-1$ of all possible non-trivial values of $y$. Otherwise different errors would yield the same $y$  making them indistinguishable (and their sum, which would affect $2t<d$ bits, would be annihilated by $H$, leading to a contradiction). We therefore find the following bound on $t=[(d-1)/2]$,
\bea
V(t,n):=\sum_{l=0}^t {n!\over l! (n-l)!}\leq 2^{n-k}. \label{Hammingbound}
\eea
This is known as the Hamming bound. It constrains $d$ in terms of $n$ and $k$. A code saturating the Hamming bound is called perfect. The Hamming $[7,4,3]$ code is a perfect code; the repetition code is not. The Hamming bound has a simple geometric interpretation. We can define a ball in the space of codewords with radius $t$ centered at the codeword $c$ to be the set of all codewords $c'$ with $d(c,c')\leq t$. Then  $V(t,n)$ is the volume of this ball, {\it i.e.}~the total number of codewords it contains. The bound \eqref{Hammingbound} simply states that since the balls of radius $t=[(d-1)/2]$ centered at each of the codewords of a given code should not overlap, the total volume of all $2^k$ balls  can not exceed the total volume of the space of codewords $2^n$.

It is useful to think of the elements of $\Z_2= \{0,1\}$ as the equivalence classes of even and odd integers. We can further view the set of all integers $\Z$ as a lattice in $\R^1$, with the lattice $2\Z$ of even integers being a sublattice. Then $\Z_2$ is the lattice quotient $\Z/(2\Z)$, {\it i.e.}~equivalence classes of lattice vectors in $\Z$ modulo shifts by elements of the sublattice $2\Z$. For the sake of mathematical elegance (and for reasons explained below) we will rescale both of these lattices by $1/\sqrt{2}$. Then if $\Gamma=\Z/\sqrt{2}$, $\Gamma^*=\sqrt{2}\,\Z$ is its lattice dual, and  $\Z_2$ can be thought of as the quotient $\Gamma/\Gamma^*$.   

This identification is the basis for a construction of Leech and Sloane \cite{leech1971sphere,conway2013sphere}, known as Construction A, which associates a lattice to any binary linear code.  A codeword is a vector in $c\in (\Z_2)^n$ and therefore can be thought of as an equivalence class of lattice points in $\Gamma=(\Z/\sqrt{2})^n$ modulo shifts by vectors in $\Gamma^*=(\sqrt{2}\,\Z)^n$. All codewords of a given code give rise to the following set of points in $\Gamma$:
\bea
\Lambda(\C)=\{v/\sqrt{2}\, |\, v\in \Z^n , \, \,  v\equiv c \, \, \, ({\rm mod}\, \, 2), \, c\in \C\},\qquad \Gamma^*  \subset \Lambda(\C)  \subset \Gamma.
\eea
Provided $\C$ is a linear code, $\Lambda(\C)$ is a lattice. It is easy to see that $\Lambda(\C)$ uniquely characterizes $\C$. In other words, for given $n$, linear binary codes are in one to one correspondence with lattices $\Lambda$ satisfying $\Gamma^*  \subset \Lambda  \subset \Gamma$.

For a given  linear code of type $[n,k,d]$, one can define its dual, which is an $[n,n-k,d']$ code consisting of all codewords orthogonal to $\C$ mod 2, 
\bea
\C^\perp=\{\tilde{c}\, |\, \tilde{c}\in \Z_2^n\, , \, \,  \tilde{c}\cdot c \, \, \equiv \,  0\,\,({\rm mod}\, \, 2)\,\, \forall\, c\in \C\}. 
\eea
The generator matrix of the dual code $\C^\perp$ is the parity matrix of $\C$ and vice versa.
The code dual to $\C^\perp$ is the original code \,$\C$. Rescaling by the factor $1/\sqrt{2}$ introduced above is necessary for the following fundamental property: the lattice of the dual code $\Lambda(\C^\perp)$ is dual (in the lattice sense) to $\Lambda(\C)$,
\bea
\Lambda(\C^\perp)=\Lambda(\C)^*.
\eea

A linear code is called self-orthogonal if, as a linear space, it is a subcode of its dual, $\C \subset \C^\perp$. At the level of lattices, $\Lambda(\C)$ of a self-orthogonal code  is an integral lattice.  A code is called self-dual if it is equal to its dual; its corresponding lattice is then self-dual (unimodular).  Self-orthogonality  requires $k\leq n/2$, and self-duality implies $k=n/2$. Therefore self-dual codes can exist only for even values of $n$. 

A binary code is called even if the Hamming weight ${\rm w}(c)$ of all of its $2^k$ codewords is even. Since all codewords of a self-dual code are self-orthogonal (mod 2), self-dual codes are necessarily even.   
At the level of lattices, when the code is even, the norm-squared of any lattice vector is integer.

A binary code is called doubly-even if the Hamming weights of all codewords are divisible by four. The corresponding lattice is then even. It is then an elementary consequence, both for codes and lattices, that any doubly-even code (any even lattice) is self-orthogonal (lattice is integral), and vice versa.
We therefore arrive at the following  conclusion: doubly-even self-dual codes are in one to one correspondence with even self-dual lattices, which are sublattices of $\Gamma\subset \R^n$. 

Binary doubly-even self-dual codes, which correspond to even self-dual lattices, are said to be of type II; the class of type II codes is denoted $2_{\rm II}$. Even but not doubly-even self-dual codes, which correspond to odd lattices, are of type I and are in the class $2_{\rm I}$. In some treatments, the class $2_{\rm I}$ is defined to include doubly-even codes as well, in which case it corresponds to the set of all integral unimodular lattices. 

The vector $\vec{1}$ has the following special property. For any $c\in \Z_n^2$, its scalar product with $\vec{1}$ (taken using conventional algebra) is equal to the Hamming weight of $c$, $\vec{1}\cdot c=\w(c)$. For any even code, $\vec{1}$ is orthogonal to all codewords (with algebra mod 2) and therefore $\vec{1}$ belongs to the dual code. If a code is doubly-even and self-dual,  $\vec{1}$ belongs to the code, and hence $n$ must be divisible by four. In fact, doubly-even self-dual codes can exist only for $n$ divisible by eight. 

Two codes $[n_1,k_1,d_1]$ and $[n_2,k_2,d_2]$ can be combined together into a new $[n_1+n_2,k_1+k_2,\min(d_1,d_2)]$ code.  A code which is not a composition  is called indecomposable. 
The Construction A lattice  of a decomposable code is a direct sum of two lattices.

\subsubsection*{Example: repetition code and checkerboard lattice}
We apply Construction A to the repetition  code $i_n$ \eqref{repeat}.  The corresponding lattice $\Lambda$ includes the vector $\vec{1}/\sqrt{2}$ and $n$ vectors $2e_i/\sqrt{2}$, where $e_i$ is a basis vector in $\R^n$. One of these vectors, say $2e_n/\sqrt{2}$, is linearly dependent and can be dropped. Thus $\Lambda$ is a linear span of the following $n$ vectors, $\vec{1}/\sqrt{2}$ and $2e_i/\sqrt{2}$, $1\leq i \leq n-1$. This is the checkerboard lattice, isomorphic to the root lattice of the $B_n$ series rescaled by $1/\sqrt{2}$. 

The lattice of the dual code includes vectors of the form $(e_i+e_{i+1})/\sqrt{2}$, $1\leq i\leq n-1$, coming from the rows of \eqref{prc}, and $2e_n/\sqrt{2}$ (all other vectors $2e_i/\sqrt{2}$ are linearly dependent). This is the root lattice of the $C_n$ series rescaled by $1/\sqrt{2}$. In the special case $n=2$, the lattices $B_2/\sqrt{2}$ and $C_2/\sqrt{2}$ coincide, reflecting that the repetition code $i_2$ is self-dual.

\subsubsection*{Example: Hamming $[7,3,4]$ code and $E_7$ lattice}
The code dual to the Hamming $[7,4,3]$ code is known as the Hamming $[7,3,4]$ code. Its generator  matrix is given by the transpose of \eqref{H7}. Its parity check matrix, besides the rows of \eqref{H7}, includes an additional row (which can be chosen in more than one way),
\bea
\left(
\begin{array}{ccccccc}
 1 & 1 & 1 & 0 & 0 & 0 & 0 \\
\end{array}
\right). \label{additionalrowH7}
\eea
Construction A applied to the Hamming $[7,3,4]$ code yields the root lattice of Lie algebra $E_7$. The conventional basis of the $E_7$ root lattice includes integer and half-integer coordinates, which does not match the factors $1/\sqrt{2}$ appearing via Construction A. These  lattices are isomorphic, meaning that they  are equivalent up to a rotation. We establish this isomorphism explicitly in Appendix \ref{sec:aE7}.\\

The generator matrix of a code is not unique. It can be multiplied from the right by any non-degenerate $k\times k$ binary matrix (all algebra mod 2) without changing the code. Usually the particular order of the bits -- the components of the codewords -- does not matter. Therefore two codes $\C$ and $\C'$ are called equivalent if they are related by a permutation of bits. At the level of generator matrices 
\bea
G'\sim O G Q, \qquad O\in {\rm O}(n,\Z),\quad {\rm det}(Q)\neq 0, \label{equivalence}
\eea
where the permutation matrix $O$ is a binary $n\times n$ matrix with only one non-zero element in each row, which is  equal to $1$, and $O$ is non-degenerate. The matrix $Q$ is an arbitrary nondegenerate binary matrix. 
The full equivalence group has $n!$ elements, but some of them may act trivially. The automorphism group of a particular code ${\rm Aut}(\C)$ is the subgroup of permutation group which leaves the given code $\C$ invariant. 

By using an equivalence transformation of the form \eqref{equivalence}, the generator matrix of  
any code  can be brought to the canonical form 
\bea
G^T=\left(\, {\rm I}\, |\, {\rm B}\right), \label{canonical}
\eea
where $\rm I$ is the $k \times k$ identity matrix and $\rm B$ is some $ k\times (n-k)$ binary matrix. 
The representation \eqref{canonical} is not unique; one can still simultaneously permute the rows and columns of $\rm B$. The equivalence transformations (\ref{equivalence}) which permute the first $k$ and last $n-k$ bits act in a more complicated way, see Appendix~\ref{appdx:g24}.

Given a code with generator matrix of the canonical form (\ref{canonical}), the binary matrix $\rm B$ can be used to define an unoriented 
bipartite graph. 
At the level of graphs, code equivalences (\ref{equivalence}) are mapped to equivalences of graphs under the operation of edge local complementation  \cite{danielsen2008edge}. The relation between codes and graphs provides useful way to analyze and design new codes \cite{ostergaard2002classifying,tonchev2002error}, and has been used to classify all inequivalent codes for $n\leq 24$. A generalization of the relation between codes and graphs to the quantum case is an important part of our discussion in Section \ref{sec:codeCFTs}.

When $n=2k$, the matrix $\rm B$ is square. In this case the parity check matrix is given by 
\bea
H=\left({\rm B}^T|\, {\rm I}\, \right).
\eea
When the code is self-dual $G$ and $H$ must generate the same code, and therefore ${\rm B}\,{\rm B}^T={\rm I}$, understood mod 2. The same conclusion follows from the explicit form  of the generator matrix of the Construction A lattice, 
\bea
\Uplambda=\left(\begin{array}{c|c}
2\,{\rm I}\, \,  &\,\,  {\rm B}^T \\ \hline
\, 0\, \, &\,  {\rm I}\, \end{array}\right)/\sqrt{2}. \label{Lambdabinary}
\eea

\subsubsection*{Example: Extended Hamming $[8,4,4]$ code and $E_8$ lattice}
The extended Hamming $[8,4,4]$ code is obtained from the Hamming $[7,4,3]$ code by extending all rows of its generator matrix $G^T$ by one bit, assigning it the value such that the Hamming weight of each row is even. Starting from \eqref{H7} and \eqref{additionalrowH7} we obtain 
\bea
\label{H8}
G^T=\left(
\begin{array}{cccccccc}
 1 & 0 & 1 & 0 & 1 & 0 & 1 & 0 \\
 0 & 1 & 1 & 0 & 0 & 1 & 1 & 0 \\
 0 & 0 & 0 & 1 & 1 & 1 & 1 & 0 \\
 1 & 1 & 1 & 0 & 0 & 0 & 0 & 1 \\
\end{array}
\right) \sim \left({\rm  I}\, |\, {\rm B}\right),\quad B=\left(
\begin{array}{c c c c}
 0\, & 1\, & 1\, & 1 \\
 1\, & 0\, & 1\, & 1 \\
 1\, & 1\, & 0\, & 1 \\
 1\, & 1\, & 1\,& 0 \\
\end{array}
\right),
\eea
where we used the equivalence condition \eqref{equivalence} to bring $G^T$  to the canonical form \eqref{canonical}.
It can be easily checked that ${\rm B}\, {\rm B}^T={\rm I}$.

The extended Hamming $[8,4,4]$ code is denoted $e_8$. It is the unique  doubly-even self-dual code with $n=8$. Via Construction A, it gives rise to the unique even self-dual lattice in eight dimensions -- the root lattice of the Lie algebra $E_8$. 

A linear $[n,k,d]$ code has $2^k$ codewords. To summarize information about its spectrum of Hamming weights it is convenient to define the enumerator polynomial
\bea
\label{enumerator}
W_\C(x,y)=\sum_{c\in \C} x^{n-\w(c)} y^{\w(c)}.
\eea
$W_\C$ is a homogeneous polynomial of degree $n$, with positive integer coefficients and 
$W_\C(1,0)=1$,  $W_\C(1,1)=2^k$.  Under the operation of code duality, the enumerator polynomial transforms according to the MacWilliams identity
\bea
W_{\C^\perp}(x,y)=2^{n/2-k}\, W_\C\left({x+y\over \sqrt{2}},{x-y\over \sqrt{2}}\right).
\eea
In other words enumerator polynomial of a self-dual code must be invariant under 
\bea
\label{binaryMacWilliams}
x\rightarrow {x+y\over \sqrt{2}},\quad y\rightarrow {x-y\over \sqrt{2}}.
\eea
When the dual code $\C^\perp$ is not equal but is equivalent in the sense of \eqref{equivalence} to the original code, its enumerator polynomial must also be invariant under \eqref{binaryMacWilliams}. Such a code is said to be isodual. At the level of lattices, $\Lambda(\C)$ of an isodual code is isomorphic to its dual,  {\it i.e}~related to its dual by a rotation. Such lattices are called isodual, in contrast to self-dual lattices.
 Finally, there are codes $\C$ which are  not  equivalent to $\C^\perp$ in any conventional way, yet $W_\C$ is  invariant under \eqref{binaryMacWilliams}. Such codes are called formally self-dual.  All formally self-dual codes are $[n,k=n/2,d]$, {\it i.e.}~they exist only when $n$ is even. Schematically, 
\bea\nonumber
{\rm self\mhyphen dual}\, \, \subset \, \, {\rm iso\mhyphen dual} \, \, \subset \, \, {\rm formally\, \, self\mhyphen dual} .
\eea

\subsubsection*{Example: isodual $[2,1,1]$ code}
The simplest example of an isodual but not self-dual code is the $[2,1,1]$ code, which includes one trivial and one non-trivial codeword $c=(1,0)$. This code is not even and therefore not self-dual. The corresponding lattice is a non-integral lattice with a rectangular unit cell, with sides of length $\sqrt{2}$ and $1/\sqrt{2}$. Its dual lattice coincides with the original lattice after rotation by $\pi/2$.  The enumerator polynomial of this code, $W=x^2+xy$, is invariant under \eqref{binaryMacWilliams}.\\

Any enumerator polynomial of a self-dual code must be invariant  under \eqref{binaryMacWilliams} and also under $y\rightarrow -y$ since the code is even. All polynomials invariant under these symmetries are in the polynomial ring 
${\mathcal P}(W_{i_2},W_{e_8})$ generated by 
\bea
\qquad W_{i_2}=x^2+y^2 \ \ {\rm and}\  \ W_{e_8}=x^8 + 14 x^4 y^4 + y^8. \label{Wi2-binary}
\eea
This is known as Gleason's theorem. 
The generator polynomials $W_{i_2}$ and $W_{e_8}$ are themselves the invariant enumerator polynomials of the self-dual repetition code $i_2$ and the extended Hamming $[8,4,4]$ code, respectively. 

Any doubly-even formally self-dual code, provided it is a linear code, is automatically self-dual. This follows immediately from the fact that an even lattice is necessarily integral, and hence is included in its dual $\Lambda(\C)\subset \Lambda(\C^\perp)$. The same argument applies to the dual code, and its dual lattice, yielding $\Lambda(\C)=\Lambda(\C^\perp)$,  $\C=\C^\perp$.
The enumerator polynomial of a doubly-even code is invariant under  \eqref{binaryMacWilliams} and $y\rightarrow i y$. All such polynomials lie in the polynomial ring ${\mathcal P}(W_{e_8},W_{g_{24}})$ generated by $W_{e_8}$ and
\bea
W_{g_{24}}&=&x^{24} + 759 x^{16} y^8 + 2576 x^{12} y^{12} + 759 x^8 y^{16} + y^{24}.
\label{Wg24}
\eea
Here $W_{g_{24}}$ is enumerator polynomial of the extended $[24,12,8]$ Golay code, introduced below.  Instead of $W_{e_8}$ and $W_{g_{24}}$ it is sometimes convenient to use $W_{e_8}$ and $(xy(x^4-y^4))^4$.

Not all polynomials invariant under appropriate symmetries are enumerator polynomials of self-dual codes. The coefficients of bona fide enumerator polynomials are positive integers and they  additionally must satisfy $W_\C(1,0)=1$. (The condition $W_\C(1,1)=2^{n/2}$  follows from $W_\C(1,0)=1$ when $W_\C(x,y)$  is a polynomial in ${\mathcal P}(W_{i_2},W_{e_8})$ or ${\mathcal P}(W_{e_8},W_{g_{24}})$.) In what follows we  refer to polynomials that satisfy these additional conditions  as invariant polynomials.


An arbitrary lattice $\Lambda$ is characterized by its theta-function,
\bea
\label{Thetadef}
\Theta_\Lambda(\tau)=\sum_{v\in \Lambda} q^{|v|^2/2},\qquad q=e^{2\pi i \tau},
\eea
which is a holomorphic function of $q$. Using the Poisson resummation formula, and for simplicity assuming the lattice is unimodular, one can express the theta function of the dual lattice in terms of $\Theta_\Lambda$:
\bea
\Theta_{\Lambda^*}(\tau)=(-i\tau)^{n/2\, }\Theta_{\Lambda}(-1/\tau).
\eea
When the lattice is even, $\Theta_{\Lambda}(\tau)$ is trivially invariant under $\tau\rightarrow \tau+1$. For an even self-dual lattice $\Theta_{\Lambda}(\tau)$ changes covariantly under the two generators of the modular group ${\rm PSL}(2,\Z)$, 
\bea
\tau\rightarrow \tau+1,\qquad \tau\rightarrow -1/\tau, 
\eea
and therefore $\Theta_{\Lambda}(\tau)$ is a modular form of weight $n/2$.

For a lattice obtained via Construction A, the theta function can be evaluated as follows. We split the sum in \eqref{Thetadef} into a sum over codewords, and for each codeword $c\in \C$ we sum over vectors  
\bea
\vec{v}={\vec{c}+2 \vec{a}\over \sqrt{2}}\in \Lambda(\C),\quad \vec{a}\in \Z^n.
\label{family}
\eea
The sum over each component $a_i\in \Z$ can be performed independently in terms of Jacobi theta-functions
\bea
\theta_{3}(q):=\sum_{n=-\infty}^\infty q^{n^2/2},\quad \theta_{2}(q):=\sum_{n=-\infty}^\infty q^{(n+1/2)^2/2},\quad q=e^{2\pi i \tau}.
\eea
Conventionally, Jacobi theta functions are understood as functions of $\tau$. We define them as functions of $q$ to emphasize that their algebraic combinations can be expanded as power series in $q$.
We  find that
\bea
\label{WtoT}
\Theta_{\Lambda(\C)}(\tau)=W_{\C}(\theta_3(q^2),\theta_2(q^2)).
\eea
Standard identities for the Jacobi theta functions imply that under $\tau\rightarrow \tau+1$ the function $\theta_3(q^2)$ remains invariant while $\theta_2(q^2)\rightarrow i\theta_2(q^2)$, and  under $\tau\rightarrow -1/\tau$ they change as follows 
\bea
\label{sshift1}
\theta_2(q^2)&\rightarrow &\theta_2(\tilde{q}^2)=\sqrt{-i\tau}\, { \theta_3(q^2)-\theta_2(q^2)\over \sqrt{2}},\\
\theta_3(q^2)&\rightarrow & \theta_3(\tilde{q}^2)=\sqrt{-i\tau}\,  { \theta_3(q^2)+\theta_2(q^2)\over \sqrt{2}},\quad \tilde{q}=e^{-2\pi i /\tau}.\label{sshift2}
\eea
These transformations match $y\rightarrow i y$ and \eqref{binaryMacWilliams}, confirming the modular properties of the theta function associated with the lattice $\Lambda(\C)$ of a doubly-even self-dual code $\C$.

\subsection*{Example: theta function of the $E_8$ root lattice}
The root lattice $E_8$ is also the Construction A lattice of the $e_8$ code. Therefore, its theta function is given by 
\bea
\Theta_{E_8}(\tau)=W_{e_8}(\theta_3(q^2),\theta_2(q^2)),\qquad W_{e_8}=x^8 + 14 x^4 y^4 + y^8.
\eea
The corresponding code is doubly-even self-dual (so the lattice is even self-dual), and therefore  $\Theta_{E_8}(\tau)$ is a modular form of weight $4$. There is a unique modular form of weight $4$, the Eisenstein series $E_4(\tau)$, and therefore $\Theta_{E_8}(\tau)=E_4$. The overall coefficient can be fixed by noting that both $\Theta_{E_8}(\tau)$ and $E_4$  for small $q$ behave as $1+O(q)$. In fact $E_4=1+240 q+O(q^2)$,  indicating that the $E_8$ lattice has $240$ roots. There are many ways $\Theta_{E_8}(\tau)=E_4$ can be expressed in terms of Jacobi theta-functions. It is customary to introduce 
\bea
\theta_4(q):=\sum_{n=-\infty}^\infty (-1) q^{n^2/2}, \label{abc}
\eea
and $a=\theta_2(q), b=\theta_3(q), c=\theta_4(q)$. They satisfy $a^4+c^4=b^4$. Then 
\bea
E_4(\tau)=\frac{a^8+b^8+c^8}{2}.
\eea\\

The analog of Gleason's theorem for theta functions of even unimodular lattices is the following. 
All theta functions for even self-dual lattices are polynomials in theta functions for the $E_8$ lattice and the Construction A lattice of the Golay code. This formulation is the direct analog of \eqref{Wi2-binary}, but it is not completely conventional in the choice of generators.  Upon substitution $x\rightarrow \theta_3(q)$, $y\rightarrow \theta_2(q)$, the combination $(xy(x^4-y^4))^4$ becomes 
\bea
\label{eta}
(xy(x^4-y^4))^4\rightarrow 16\, \eta^{24}={a^8 b^8 c^8\over 16}.
\eea
Correspondingly the theta series of any even self-dual lattice can be written as a polynomial in $E_4$ and $\eta^{24}$. This is of course a well-known result in the theory of modular forms. Since $1728 \eta^{24}=E_4^3-E_6^2$, 
this is simply a consequence of the statement that all modular forms of weight $n/2$ for $n$ divisible by $8$ are polynomials in $E_4$ and $E_6^2$. Another conventional choice of generators is provided by $E_4$ and the theta series of the Leech lattice introduced below. 

There is a close relation between codes and sphere packing. An optimal lattice sphere packing requires a lattice with a fundamental cell of unit volume and a shortest vector of maximum possible length. Codes of maximal Hamming distance for a given $n$ naturally lead to such lattices. Thinking of codewords as points on the unit cube, such codes maximize the distance  from the origin to all codewords. Via Construction A this should lead to a good  lattice sphere packing, and indeed the $E_8$ lattice is the optimal sphere packing in eight dimensions \cite{viazovska2017sphere}. The discussion above is intuitive but it has a serious flaw: all lattices obtained via Construction A include vectors of the form $\sqrt{2}\vec{a}$ with arbitrary $\vec{a}\in \Z^n$. Thus, no matter how good a code might be, the corresponding lattice $\Lambda(\C)$ would necessarily have vectors of length $\ell^2=2$. It so happens that $\ell^2=2$ is the largest possible  length of the shortest vector   in eight dimensions, but this observation renders Construction A an unsuitable approach for finding good lattice sphere packings in higher dimensions. To design good lattice packings starting from a good code with $n>8$, it is desirable to leave lattice vectors of the form $\vec{c}/\sqrt{2}$, $c\in \C$, intact because they have sufficient length, but remove short vectors of the form $\sqrt{2}(\pm 1,0,\dots,0)$, $\sqrt{2}(0,\pm 1,\dots,0)$, \dots.  There are several different ways (constructions) to achieve that result. Here we focus on a construction of particular physical relevance. 

Let us start with an even self-dual lattice $\Lambda$ and consider a vector  $\delta$ such that $2\delta \in \Lambda$. We demand that $\delta^2$ be an integer. The lattice $\Lambda$ can be represented as the disjoint union of two sets 
\bea
\Lambda_0&=&\{v\, |\, 2\,\delta\cdot v=0\, \, {\rm mod}\, \,2,\quad v\in \Lambda\},\\
\Lambda_1&=&\{v\, |\, 2\, \delta\cdot v=1\, \, {\rm mod}\, \,2,\quad v\in \Lambda\}.
\eea
Since the original lattice $\Lambda$ is integral,  $2\delta\cdot v$ is an integer and therefore $\Lambda=\Lambda_0 \cup \Lambda_1$. It is easy to see that $\Lambda_0$ is  closed under addition and therefore it is a lattice, while $\Lambda_1$ is not. We now shift all vectors in $\Lambda_1$ by $\delta$,
\bea
\Lambda'_1=\Lambda_1+\delta\equiv \{v+\delta\, |\, v\in \Lambda_1\}, \label{sumdelta}
\eea
and define a new lattice via 
\bea
\Lambda'=\Lambda_0\cup \Lambda'_1. \label{shift}
\eea
It is easy to check that $\Lambda'$ is a lattice: the sum of two vectors in $ \Lambda'$ belongs to $\Lambda'$. Furthermore if $\delta^2$ is odd, the lattice is even and self-dual. 
This procedure can also be applied to odd self-dual lattices, yielding a new odd self-dual lattice, in which case the condition that $\delta^2$ be odd is not necessary and is replaced by  $\vec{\delta}\notin\Lambda$. 

We will call the above construction of a new lattice  $\Lambda'$ a ``twist'' (by a half-lattice vector), following the nomenclature adopted in the context of 2d conformal theories \cite{nair1987compactification}.
The  twist can be used to construct new lattices with longer shortest vectors than the original ones. Since any self-dual code includes the codeword $\vec{1}$, any Construction A lattice includes the vector $2\vec{\delta}=\vec{1}/\sqrt{2}$. Choosing this $\delta$ removes the vectors $\sqrt{2}(\pm 1,0,\dots,0)$ from the lattice; they are instead replaced by $(-3/2,1/2,\dots,1/2)\sqrt{2}$ and $(5/2,1/2,\dots,1/2)\sqrt{2})$ which have length $\ell^2\geq 1+n/8$. The ``codeword'' vectors $\vec{c}/\sqrt{2}, c\in\C$, still belong to the lattice provided $\w(c)$ is divisible by four.  

The theta function of the new lattice, obtained from the Construction A lattice by the twist with $\vec{\delta}=\vec{1}/(2\sqrt{2})$, can be calculated in full generality. Because of permutation symmetry, the contribution of all vectors associated with a given codeword depends only on $\w(c)$. There are several cases to consider: $\w(c)=0$, $\w(c)/2$ is odd, and $\w(c)/2$ is positive and even. We spare the reader the details and simply present the answer, 
\bea
\label{twistedbinary}
\Theta_{\Lambda'(\C)}={\Theta_{\Lambda(\C)}+(ab)^{n/2}+(bc)^{n/2}-2^{-n/2}W_\C(1,i)(ac)^{n/2}\over 2}.
\eea
If the code is doubly even, $2^{-n/2}W_\C(1,i)=1$. 
Under the modular transformation $\tau\rightarrow \tau+1$, the functions $a,b,c$ change as follows: 
$a\rightarrow i^{1/2}a$, $b\leftrightarrow c$. Under $\tau\rightarrow -1/\tau$ they change as  
$a\rightarrow \sqrt{-i\tau}c, c\rightarrow \sqrt{-i\tau}a$, and $b\rightarrow \sqrt{-i\tau} b$. Therefore 
$\Theta_{\Lambda'(\C)}$ always changes covariantly under $\tau\rightarrow -1/\tau$, reflecting that the twist does not affect self-duality; however, modular invariance under $\tau\rightarrow \tau+1$ requires the code to be doubly-even and $n/8$ to be odd, to ensure that $\Lambda(\C)$ is even,  $\delta^2$ is odd, and therefore that $\Lambda'(\C)$ is even as well.

\subsection*{Example: twist of the $E_8$ lattice}
The Construction A lattice of $e_8$  is invariant under the twist by $\vec{\delta}=\vec{1}/(2\sqrt{2})$ in the sense that the new lattice is isomorphic to the old one. As a consistency check one can verify  using identity $a^4+c^4=b^4$ that the theta functions of the original and new lattices are equal  
\bea
E_4=\frac{a^8+b^8+c^8}{2} =(a b)^4+(b c)^4-(a c)^4.
\eea

\subsection*{Example: extended Golay $[24,12,8]$ code and Leech lattice}
The extended Golay $[24,12,8]$ code is the unique $n=24$, $d=8$ code (up to equivalences). It is denoted $g_{24}$. In the canonical form \eqref{canonical} it is specified by the matrix $\rm B$ given in \eqref{BGolay}.

It is a matter of a few minutes of computer algebra to verify that ${\rm B}\, {\rm B}^T={\rm I}$, confirming that the code is self-dual, and to evaluate its enumerator polynomial $W_{g_{24}}$  \eqref{Wg24}. The explicit form of $W_{g_{24}}$ confirms that $g_{24}$ is doubly-even. 
The theta function of $\Lambda(g_{24})$ is given by 
\bea
\Theta_{\Lambda(g_{24})}=E_4^3-672\, \eta^{24},
\eea 
which follows from the explicit form of $W_{g_{24}}$ and \eqref{eta}.

Applying the twist $\vec{\delta}=\vec{1}/(2\sqrt{2})$ to $\Lambda(g_{24})$ produces the Leech lattice, with theta function 
\bea
\Theta_{\rm Leech}=E_4^3-720\, \eta^{24}={\Theta_{\Lambda(g_{24})}+(ab)^{12}+(bc)^{12}-(ac)^{12}\over 2}.
\eea
Its small $q$ expansion reads 
\bea
\Theta_{\rm Leech}=1+196560 q^2+16773120 q^3+O(q^8),
\eea
which indicates that the Leech lattice (famously) has no roots -- vectors of length $0<\ell^2\leq 2$ -- and its shortest vector has length $\ell^2=4$.

The analog of Gleason's theorem for modular forms from above guarantees that $\Theta_{\rm Leech}$ can be expressed as an ``enumerator polynomial,'' i.e.~a polynomial in $x,y$ invariant under \eqref{binaryMacWilliams} and $y\rightarrow i y$,
\bea
\nonumber
W_{\rm Leech}&=&x^{24}-3 x^{20} y^4+771 x^{16} y^8+2558 x^{12} y^{12}+771 x^8 y^{16}-3 x^4 y^{20}+y^{24},\\
\Theta_{\rm Leech}&=&W_{\rm Leech}(\theta_3(q^2),\theta_2(q^2)).
\eea
We emphasize that while some coefficients of $W_{\rm Leech}(x,y)$ are negative, all coefficients in the $q$-expansion of $W_{\rm Leech}(\theta_3(q^2),\theta_2(q^2))$  are positive integers. \\

To summarize, there is a close relation between codes and their enumerator polynomials, and lattices and their theta functions. A natural question to ask is whether this relation is exclusive. The answer is no. Enumerator polynomials characterize codes, but not in a unique way: inequivalent codes may share the same polynomial. Accordingly, different non-isomorphic lattices can be isospectral, {\it i.e.}~have the same theta series. There are also  invariant polynomials which are not enumerator polynomials of any code. Likewise, there are self-dual lattices not related to any code via Construction A, and so on. To see how this works we discuss the most restrictive case of doubly-even self-dual codes for $n=8,16,24$. For even but not doubly-even self-dual codes the situation is even more complex. 

For $n=8$, $e_8$ is the unique self-dual doubly-even code, and there is a unique invariant polynomial $W_{e_8}$. There is also a unique even self-dual  lattice in eight dimensions, the root lattice of $E_8$, which is related to $e_8$ via Construction A. The theta series of that lattice, $E_4$, is the unique modular form of weight $4$. Thus for $n=8$ the story is simple: there is a perfect correspondence between self-dual doubly-even codes, even self-dual lattices, and invariant polynomials. 
For $n=16$, there is still a unique invariant polynomial $W_{e_8}^2$, but there are two inequivalent self-dual doubly-even codes, a decomposable code $e_8 \oplus e_8$ and an indecomposable code $d_{16}^+$ \cite{pless1972classification,pless1975classification}. Construction A applied to the latter yields the even self-dual lattice $D_{16}^+=D_{16}\cup (D_{16}+\vec{1}/2)$, where  $D_{16}^+$ is the root lattice of $Spin(32)/Z_2$. The Construction A lattice of the former code is $E_8 \oplus E_8$, which is not isomorphic to $D_{16}^+$. Both codes have the same enumerator polynomial, and thus both lattices have the same theta function, the unique modular form of weight eight, $E_4^2$. We conclude that these two non-isomorphic lattices are isospectral, since their theta series coincide. This is Milnor's famous example of distinct compact spaces (the tori defined by these lattices) with equivalent Laplacian spectra. An excellent nontechnical discussion of this point can be found in J.~Conway's book \cite{conway1997sensual}. The isospectral lattices $E_8 \oplus E_8$ and $D_{16}^+$ define two different heterotic string theories, related by duality \cite{PhysRevD.35.648,NARAIN1987369,NARAIN198641}. The situation is even more nuanced for $n=24$. In this case there are $9$ doubly-even self-dual codes, and overall $24$ non-isomorphic even self-dual lattices. There is no simple way to assign each lattice to a particular code besides those nine obtained via Construction A. In our discussion, and historically, the Leech lattice is associated with the $g_{24}$ code, but other lattices are related to each other in similar manner \cite{conway2013sphere}. For $n=24$, all invariant polynomials can be written as $W_{e_8}^3+r(xy(x^4-y^4))^4$ with $r$ an integer $-42\leq r \leq 147 $ to ensure all coefficients are positive. (The Golay code $g_{24}$ corresponds to the smallest allowed value of $r=-42$.)
Most of these polynomials are not enumerator polynomials for any code. We refer to such code-less invariant polynomials as ``fake.''

The relations between codes and lattices can be extended to CFTs and their vertex operator algebras \cite{dolan1996conformal,miyamoto1996binary,dong1998framed,lam1999codes,hohn2003self,lam2007characterization,hohn2008conformal}. (We refer the reader interested in a quick summary of these relations to Table 1 of  \cite{dolan1996conformal}.) 
In particular, Euclidean even self-dual lattices can be used to define chiral CFTs, 
which play a prominent role in string theory and mathematical physics. The connection to codes provides a new angle to probe various aspects of 2d chiral theories. In particular, the fake enumerator polynomials mentioned above give rise to ``would-be'' CFT partition functions, modular invariant functions satisfying positive conditions, at least some of which are not partition functions of any theory.\footnote{We thank Xi Yin for a discussion on this point.}
Speaking colloquially, our paper extends the relations between codes, lattices, and CFTs to include quantum codes and non-chiral CFTs.

One of the central questions of code theory is to understand the maximum possible value of $d/n$  for fixed $k/n$ when $n$ goes to infinity. Analogous questions can be asked about lattices and sphere packing. While the optimal value of $d/n$ is not known there are various upper and lower bounds. For self-dual codes $n=2k$ the Hamming bound \eqref{Hammingbound} readily provides an upper bound,
\bea
{d\over n} \leq 2 p^*,\quad  n\rightarrow \infty, \qquad H(p^*)={\ln(2)\over 2},\quad p^*\approx 0.11,
\eea 
where $H(p)=-p\ln(p)-(1-p)\ln(1-p)$  is the Shannon entropy. This bound is suboptimal. One can derive stronger bounds using linear programming techniques. For instance, for even self-dual codes the space of invariant polynomials is the linear space of all polynomials in $W_{i_2},W_{e_8}$ subject to linear constraints and inequalities. If we additionally require that the hypothetical enumerator polynomial describes a code of Hamming distance $d$, that would impose additional linear constrains $\left.\partial_y^k W(x,y)\right|_{y=0}=0$ for $1\leq k<d$. If the corresponding discrete linear programming problem is infeasible, there is no such code and $d$ must be reduced. For small $n$, but not in general, the bounds obtained this way are tight: the largest $d$ for which the problem of finding invariant polynomials is feasible  is also achievable as the Hamming distance of a code. Codes for which $d$ saturates the linear programming bound are called extremal. The linear programming bounds are not constructive -- they may yield invariant polynomials, but most of these are fake, and reconstructing a code from a polynomial is algorithmically hard. Nevertheless  one can establish asymptotic bounds on $d/n$ in this way, which for type I and II self-dual codes read $d/n\leq n/5$ and $d/n\leq n/6$, respectively \cite{mallows1973upper,mallows1975upper,rains1998shadow}. These bounds can be further improved \cite{605587,rains2003new} and it is expected that additional systematic improvements are possible. The linear programming bounds for codes are parallel to linear programming bounds on the length of the shortest vector of a unimodular lattice \cite{mallows1975upper,rains1998shadow}. They can be thought of as simpler, more restricted versions of linear programming bounds on sphere packing \cite{cohn2003new,cohn2002new,viazovska2017sphere,cohn2017sphere,cohn2016conceptual}, which, remarkably, are related to modular bootstrap bounds \cite{hartman2019sphere,afkhami2020high,afkhami2020free}.

Besides upper bounds, there is a lower bound on $d/n$ known as the Gilbert-Varshamov bound, which is closely related to the Hamming bound \eqref{Hammingbound}. The idea is to fix $d$ and $n$ and put a bound on the number of codewords $K$. Since $d$ is the minimal distance between any two codewords, the ball of radius $d-1$ centered at a given codeword does not include any other codeword. We consider balls of radius $d-1$ centered at all $K$ codewords and ask if they cover the whole space. If they do not, $K$ can be increased. Thus for a linear code we find the maximal $d$ for which the following inequality is satisfied 
\bea
V(d-1,n)<2^{n-k}.
\eea 
This bound can be improved by noticing that for even codes  the sum over $l$  in \eqref{Hammingbound} should go only over even values. Furthermore, the full space of even codewords has volume $2^{n-1}$. Similar improvements are possible also for doubly-even codes. In the considerations above we disregarded self-duality, but a generalization to self-dual codes is possible \cite{pless1973self}.

There is a conceptually different way to obtain a Gilbert-Varshamov  bound, suitable for self-dual codes, which leads to essentially the same results. For even $n$ there are 
\bea
{1\over 2}\prod_{j=0}^{n/2-1}(2^j+1)
\eea
type I self-dual codes (if $n$ is divisible by $8$, this number includes type II codes) and 
\bea
\prod_{j=0}^{n/2-2}(2^j+1)
\eea
type II self-dual codes, when $n$ is divisible by $8$. One can calculate the enumerator polynomial averaged over all such codes \cite{pless1975classification,rains2002self,nebe2006self}
\bea
{\overline W}_{\rm I}(x,y)&=&{2^{n/2}(x^n+y^n)+(x+y)^n+(x-y)^n\over 2(2^{n/2-1}+1)},\\
\label{averagedW}
{\overline W}_{\rm II}(x,y)&=&{2^{n/2}(x^n+y^n)+(x-y)^n+(x+y)^n+(x+i y)^n+(x-i y)^n\over 4(2^{n/2-2}+1)}.
\eea
So if the sum of the coefficients of $x^{n-k}y^k$, for $1\leq k<d$,  is smaller than one, then there is a code with Hamming distance $d$. 

Asymptotically, the lower bound on $d/n$ is given by the value of $d$ for which the coefficient of $x^{n-d}y^d$ becomes of order one when $n\rightarrow\infty$, yielding $d/n\geq p^*\approx 0.11$. In Section \eqref{sec:GV} we will interpret \eqref{averagedW} as the averaged partition function of certain chiral CFTs. The value of the Gilbert-Varshamov bound  $p^*\approx 0.11$ would then define the spectral gap in a random CFT from that class. 


\subsection{Codes over ${\rm GF}(4)$}
\label{sec:GF4}
Analogously to binary codes, one can define codes over any field $F$. We are specifically interested in codes over $F={\rm GF}(4)$ -- the unique field with four elements -- because of their relevance to quantum codes. The Galois field  ${\rm GF}(4)$ consists of four elements $0,1,\omega, \bar{\omega}$ subject to the following relations 
\bea
\forall\,\, x\in F,\qquad 0+x=x,\quad 0\times x=0,\quad 1\times x=x,\quad x+x=2\, x=0,
\eea
and $\bar{\omega}=\omega^2=1+\omega$. There is a conjugation operation  which leaves $0,1$ invariant and exchanges $\omega \leftrightarrow \bar{\omega}$. With the exception of $2\,x=0$, all other relations are automatically satisfied if we take $\omega=e^{2\pi i/3}$ and $\bar\omega=e^{-2\pi i/3}$. For example 
$1+\bar{\omega}=-\omega=\omega-2\omega\rightarrow \omega$. To impose the condition $2x=0$ we first consider the triangular lattice in the complex plane
\bea
\Gamma_{\rm E}=A_2/\sqrt{2}=\{a+b\,\omega\, |\,  a,b\in \Z\}\subset {\mathbb C}.
\eea 
This is the root lattice $A_2$ rescaled by $1/\sqrt 2$, the lattice of the so-called Eisenstein integers. If we define new lattice $2\,\Gamma_{\rm E}$ by requiring that both $a$ and $b$ be even, then ${\rm GF}(4)=\Gamma_{\rm E}/(2\Gamma_{\rm E})$. In contrast to the binary case, $2\,\Gamma_{\rm E}$ is not dual to $\Gamma_E$, a fact which will have consequences later. 

Now we are ready to define codes over $F={\rm GF}(4)$. 
A code $\C\subseteq F^n$ is called additive if $\C$ is a vector space over $F$, meaning that the sum of two codewords is a codeword. 
Additive codes always include the trivial codeword consisting of $n$ zeros.
The Hamming weight $\w(c)$ for $c\in F^n$ is the number of non-zero elements of $c$. As before, the Hamming distance of a code  is the minimal Hamming weight of all non-trivial codewords, and a code with Hamming distance $d$ and size $K=4^k$ is said to be of type $[n,k,d]$.

A code is called linear if it is a vector space over $F$, which requires that for any codeword $c\in\C$, $c'=\omega c$ must also be a codeword $c'\in \C$. All linear codewords are additive but not vice versa. For binary codes, the two notions coincide, but not for other fields. The length of a linear code is always $K=4^k$ for some $k\le n$. 
An additive code with $K=4^k$ can be specified by an $n\times k$ generator matrix $G$ with all codewords given by 
\bea
c(x)=Gx\in F^n,\qquad x\in F^k.
\eea

For each additive code we can define a lattice in ${\mathbb C}^n=\R^{2n}$ as the pre-image of the code under the map $(\Gamma_{\rm E})^n/(2\Gamma_{\rm E})^n$, or explicitly
\bea
\label{ELC}
\Lambda(\C)=\{\vec{a}+\vec{b}\, \omega\, |\,  \vec{a},\vec{b}\in\Z^n,\, \,  \vec{a}+\vec{b}\,\omega\equiv c \, \, ({\rm mod}\, \,  2),\, \,  c\in\C\}\subset \Gamma_{\rm E}\subset \R^{2n}.\ \ \ \ 
\eea
This is the analog of Construction A for codes over ${\rm GF}(4)$. 

To define duality on the space of codes, we need to introduce a scalar product. There are several natural choices. 
First, the so-called Euclidean scalar product of $x,y\in F^n$  is $(x,y)=\sum_i x_i\, y_i$. There is also a Hermitian version, $(x,y)_{\rm H}=\sum_i \bar{x}_i\, y_i$. These two versions are homogeneous and can be used to define duality on the space of linear codes. There is a third, physically relevant scalar product, 
\bea
\label{H+}
x \cdot y=\sum_i \bar{x}_i\, y_i+x_i\, \bar{y}_i,\qquad x,y\in F^n.
\eea
In all cases the algebra is over $F$.
The dual code $\C^\perp$ consists of all codewords orthogonal (with respect to a given inner product) to all codewords of $\C$. The code is called self-orthogonal if $\C\subset \C^\perp$ and self-dual if $\C=\C^\perp$. Linear self-dual codes under the Euclidean $(\,\, ,\, )$ and Hermitian $(\,\, ,\, )_{\rm H}$ products form the code families known as  $4^{\rm E}$ and $4^{\rm H}$. Additive codes self-dual under the Hermitian product \eqref{H+} make up to family $4^{\rm H+}$. It is easy to see that $4^{\rm H} \subset 4^{\rm H+}$. 

The Construction A lattice of an additive self-dual code $4^{\rm H+}$ is an integral lattice in $\R^{2n}$. It is not self-dual because $\Gamma_{\rm E}$ is not self-orthogonal, and $(\Gamma_{\rm E})^*$ includes points outside of $\Gamma_{\rm E}$, although $2\,\Gamma_{\rm E} \subset (\Gamma_{\rm E})^*$. If the Hamming weight $\w(c)$ of all codewords $c\in \C$ is even, the code is called even. The Construction A lattice $\Lambda(\C)$ of an even code is even. Even self-dual codes are said to be of type II, and belong to the family $4^{\rm H+}_{\rm II}$. Otherwise, if some of the weights $\w(c)$ are  odd, the codes are referred to as odd, or type I, and are in the family $4^{\rm H+}_{\rm I}$.

For codes over ${\rm GF}(4)$ one can introduce an enumerator polynomial exactly as in the binary case via \eqref{enumerator}, with the Hamming weight ${\rm w}(c)$ defined above. The MacWilliams identity relates the weight enumerators of the original and dual additive codes; it takes the same form for $4^{\rm E},4^{\rm H}$, and $4^{\rm H+}$ \cite{macwilliams1978self,rains2002self,nebe2006self}: 
\bea
\label{MacW4}
W_{\C^\perp}(x,y)=2^{n-k}W_\C\left({x+3y\over 2},{x-y\over 2}\right).
\eea
Enumerator polynomials of self-dual codes are invariant under 
\bea
\label{Mac4}
x\rightarrow{x+3y\over 2},\quad y\rightarrow {x-y\over 2}
\eea
and therefore are in the polynomial ring ${\mathcal P}(W_1,W_2)$ generated by \cite{calderbank1998quantum}
\bea
W_1=x+y \quad {\rm and} \quad W_{i_2}=x^2+3y^2. \label{Wi2}
\eea
Here $W_1$ is easy to recognize as the enumerator polynomial of the simplest $[1,1,1]$ self-dual additive code with only one non-trivial codeword $c=(1)$, while $W_{i_2}$ is the enumerator polynomial of a linear ``repetition'' code over $F$ with the generator matrix $G^T=(1,1)$.

If the self-dual code is even, the enumerator polynomial is additionally invariant under $y\rightarrow -y$, in which case it is a polynomial in $W_{i_2}$ and 
\bea
\label{hexaW}
W_{h_6}=x^6+45 x^2 y^4+18 y^6.
\eea
Here $W_{h_6}$ is the weight enumerator of the hexacode, introduced below. 

Additive codes over ${\rm GF(4)}$ are defined to be equivalent if they are related to each other by a permutation of their ``letters'' (components of the codewords), conjugation of some ``letters'' $\omega \leftrightarrow {\bar \omega}$, and multiplication of some ``letters'' by $\omega$ or $\bar{\omega}$. The same operation should be applied to all codewords of the code. In terms of the corresponding lattice $\Lambda(C)$ these are isomorphisms which permute $\CC$ planes inside $\R^{2n}$, and within each plane permute $1, \omega=e^{2\pi i/3},\omega=e^{-2\pi i/3}$ in an arbitrary order.  There are a total of $3! n!$ elements in the equivalence group.

To calculate the theta series for $\Lambda(\C)$, it is sufficient to consider each $\CC$ plane inside $\R^{2n}$ individually and sum either over the triangular lattice $\{2(a+b\omega)\, |\, a,b,\in Z\}$ or over the triangular lattice shifted by half-vector $\{1+2(a+b\omega)\, |\, a,b,\in Z\}$,
\bea
\nonumber
\phi_0(\tau)&=&{\theta_3(q)\theta_3(q^3)+\theta_4(q)\theta_4(q^3)\over 2},\\
\phi_1(\tau)&=&{\theta_3(q)\theta_3(q^3)-\theta_4(q)\theta_4(q^3)\over 2},\\ \label{phi01}
\Theta_{\Lambda(\C)}(\tau)&=&W_\C(\phi_0,\phi_1),\qquad q=e^{2\pi i \tau}.
\eea
Under modular  transformation $\tau\rightarrow \tau+1$, $\phi_0$ is invariant while $\phi_1$ changes sign. Hence $\Theta_{\Lambda(\C)}$ is invariant under $\tau\rightarrow \tau+1$ if and only if the code (and corresponding lattice) is even. The functions $\phi_0,\phi_1$ also change covariantly under the modular transformation $\tau \rightarrow -1/\tau$,
\bea
\phi_0(-1/\tau)&=&{-i \tau\over \sqrt{3}}\, {\phi_0(\tau/3)+3\phi_1(\tau/3)\over 2},\\
\phi_1(-1/\tau)&=&{-i \tau\over \sqrt{3}}\,  {\phi_0(\tau/3)-\phi_1(\tau/3)\over 2}.
\eea
These transformations coincide with \eqref{Mac4} after rescaling $\tau \rightarrow \tau/\sqrt{3}$. Upon rescaling the argument to $t = \sqrt 3 \tau$, the theta function for a self-dual $\C$  would be covariant under 
\bea
\tilde{\Theta}(-1/t)=(-it)^n \tilde{\Theta}(t) \quad {\rm where} \quad  \tilde{\Theta}(t):=\Theta_{\Lambda(\C)}(t/\sqrt{3}),
\eea
which reflects that the rescaled lattice $\Lambda(\C)/3^{1/4}$ is isodual, 
{\it i.e.}~it is equal to its dual $(\Lambda(\C)/3^{1/4})^*$ after a rotation by $\pi/2$ in each $\CC$ plane. Alternatively, one can characterize $\Lambda(\C)$ of a self-dual code $\C$ as a $3$-modular lattice \cite{rains1998shadow}.

\subsection*{Hexacode and Coxeter-Todd lattice}
The hexacode is the unique linear even self-dual $[6,3,4]$ code of type $4^{\rm H}$ defined by the following generator matrix 
\bea
G^T=\left(\begin{array}{cccccc}
1\, & 0\, & 0\, & 1\, & \omega & \omega \\
0\, & 1\, & 0\, & \omega & 1\, &\omega\\
0\, & 0\, & 1\, & \omega & \omega & 1\,
\end{array}\right).  \label{hexacode}
\eea
As an additive self-dual code from $4^{\rm H+}$ it would be denoted as $[6,6,4]$. Later in the text we will refer to it as $h_6$. Its enumerator polynomial is given by \eqref{hexaW}. Since it is a linear code, all coefficients of $W_{h_6}$ except for the first one are divisible by three. 
 
The Construction A lattice $\Lambda(h_6)\subset \R^{12}$ is the Coxeter-Todd lattice $K_{12}$, an even lattice with no roots (vectors of length $0<\ell^2\leq 2$) in twelve dimensions. This follows from the theta series 
\bea
\label{K12}
\Theta_{K_{12}}(\tau)=\phi_0^6+45 \phi_0^2 \phi_1^4+18 \phi_1^6=1+756 q^2+4032 q^3+O\left(q^{4}\right).
\eea

There are many other results concerning codes over ${\rm GF}(4)$, analogous to results about binary codes, including a series of linear programming bounds. We will present these bounds later in the text after making the connection between classical codes over ${\rm GF}(4)$ and binary quantum stabilizer codes.

\section{Quantum error-correcting codes}\label{sec:3}
In this section we introduce quantum stabilizer codes and establish their relation to Lorentzian integer lattices. Subsection \ref{sec:3.1} is mostly pedagogical; it introduces quantum stabilizer codes and explains their relation to classical codes over ${\rm GF}(4)$. Only the very last part of this section, where we discuss real self-dual stabilizer codes and their refined enumerator polynomials, is original. Subsection~\ref{sec:newA} explains the relation of stabilizer codes to Lorentzian lattices, which is the central ingredient in our construction.

\subsection{Quantum additive codes}\label{sec:3.1}
Let us consider a system consisting of $n$  quantum spins, or qubits. Initially the system is in some state $\psi\in \H$. Because of unwanted interactions with the environment the system changes its quantum state $\psi\rightarrow \psi'$ in some unpredictable way. This is quantum error. We would like to devise a protocol to return the system to its original state. That would be quantum error correction. Clearly this can not be done in full generality, so we must restrict to quantum errors of a particular type. For a system consisting of $n$ distinct physical qubits, one usually assumes a random interaction with the environment that affects at most $t$ qubits at once. Furthermore, the correction of quantum errors is possible only for certain states that belong to a special code subspace $\psi\in\H_\C\subset \H$. 

Interactions with the environment can be described as linear operations acting on $\psi$. More accurately, one should speak of a quantum channel acting on a density matrix, but for simplicity we will assume the system always remains in a pure state. Operators describing interactions with the environment form a linear space. We can choose a basis $E_i$ for this space, a basis of quantum errors.  Crucially, to ensure reversibility of quantum errors due to an arbitrary linear combination of  the $E_i$, it is necessary and sufficient that each $E_i$, restricted to $\H_\C$, be nondegenerate (reversible) and that the images of $\H_\C$ do not overlap, 
\bea
E_i \H_\C \cap E_j \H_\C=0,\qquad i\neq j. \label{nonoverlap}
\eea
This is the Knill-Laflamme condition \cite{knill1997theory}.
The reduction of all possible errors to a handful of linear operators $E_i$ is called a discretization of quantum errors \cite{nielsen2002quantum}.

The Knill-Laflamme condition  has a classical counterpart: correctable errors of a linear classical code must produce different results. Consider two codewords of a binary classical code $c_1,c_2\in \C$, and assume they are subject to errors, $c_1'=c_1+e_i$, $c_2'=c_2+e_j$. For both errors $e_i,e_j$, to be correctable, $c_1'$ and $c'_2$ must always be distinct, which is the classical analog of \eqref{nonoverlap}. Indeed, if  $c_1'=c_2'$ (all algebra is mod 2),
\bea
e_i+e_j=c_1+c_2 \in \C.
\eea
In this case the error $e_i+e_j$ is annihilated by the parity check matrix, meaning that both $e_i$ and $e_j$ will yield the same error correction protocol, which will fail to undo at least one of errors. The similarity of the quantum case comes from the linearity of quantum mechanics, {\it i.e.}~the possibility to represent any quantum error as a linear combination of the $E_i$.

There is one important exception when \eqref{nonoverlap} does not have to apply -- when two distinct errors act identically on $\H_\C$. In the classical case that would mean that the errors are the same, but in the quantum case the errors could act differently on $\H\, \backslash\, \H_\C$. A code for which all correctable errors satisfy \eqref{nonoverlap} is called non-degenerate. 

The condition $c_1'\neq c_2'$ in Section \eqref{sec:binary} leads to the classical Hamming bound \eqref{Hammingbound}. There is a quantum version of the Hamming bound, which is as follows. The linear space of quantum errors which affect exactly $l$ qubits is spanned by $3^l$ tensor products of Pauli matrices (the identity being excluded, as we want all $l$ qubits to be affected). Hence the total number of errors $E_i$, including the trivial one $E_1=\mathcal I$, affecting up to $t$ qubits is 
\bea
\label{Vq}
V_q(t,n)=\sum_{l=0}^t {n!\over l!(n-l)!} 3^l.
\eea
Each error $E_i$ restricted to $\H_\C$ is reversible, and therefore ${\rm dim}(E_i \H_\C)={\rm dim}(\H_\C)$. Assuming the code is nondegenerate, the images $E_i \H_\C$ must not overlap. The total dimension of all images can not  exceed the dimension of full Hilbert space, yielding 
\bea
{\rm dim}(\H_\C) V_q(t,n)\leq {\rm dim}(\H)=2^n. \label{QHB}
\eea 
This is the quantum Hamming bound for codes correcting arbitrary quantum errors affecting  up to $t$ qubits  \cite{ekert1996quantum}. A code saturating this bound is called perfect. Often the code subspace $\H_\C$ will describe $k$ logical qubits, in which case ${\rm dim}(\H_\C)=2^k$. From \eqref{QHB} it follows that to encode $k=1$ logical qubit and to be able to recover the state after any quantum error affecting $t=1$ physical qubit, one needs $n=5$ physical qubits. 
For example, the 5-qubit protocol of Laflamme at el.~\cite{laflamme1996perfect} introduced below is a perfect quantum error-correcting code. 

The quantum Hamming distance $d$ is the minimal number of physical qubits which need to be affected to map a state from $\H_\C$ into $\H_\C$. A quantum error-correcting code characterized by $n,k$, and $d$ is denoted $[[n,k,d]]$. Such a code can correct for any error affecting up to $t=[(d-1)/2]$ qubits.

To illustrate how quantum error-correcting codes work, we consider an oversimplified situation in which $k$ logical qubits are implemented as $k$ physical qubits, which are isolated in a lab as part of a perfect noiseless quantum computer. We additionally consider $n-k$ auxiliary qubits located in a different lab in an imperfect environment.  States of all $k+(n-k)=n$ qubits can be represented in the conventional 
binary basis ($0 =$ spin up, $1 =$ spin down)
\bea
|a_1\dots a_n\rangle=|a_1\dots a_k\rangle \otimes |a_{k+1}\dots a_n\rangle \in \H,\qquad a_i=\{0,1\}.
\eea
Auxiliary qubits will be initialized in the state $|0^{n-k}\rangle$, and will be left intact, while the quantum computer performs unitary evolution of the first $k$ ``logical'' qubits 
\bea
\psi=\psi_l \otimes |0^{n-k}\rangle. \label{desired}
\eea
Because the auxiliary lab is imperfect, after some time the state of the last $n-k$ qubits will evolve to
\bea 
\psi=\psi_l \otimes \psi_a, \label{corrupted}
\eea
where $\psi_l$ is the desired result of unitary evolution produced by the quantum computer, while $\psi_a$ is some unknown random state resulting from interactions with the environment. This example may appear unrealistic because we have physically isolated the logical qubits from the environment, the systems are not entangled, and therefore state of the auxiliary qubits does not matter. All measurements performed in the first lab will be insensitive to $\psi_a$, and our insistence on including  the auxiliary qubits in our considerations is inconsequential. Nevertheless, it is instructive to ask a question: can one devise a protocol to bring the corrupted state \eqref{corrupted} to the desired form \eqref{desired}. This is easy to do: one simply needs to re-initialize the auxiliary system. This can be done by first measuring the state of auxiliary qubits in the computational 
up-down basis, which will project the total state onto  $\psi_l \otimes |a_{k+1}\dots a_n\rangle$, and then applying the recovery operator 
\bea
\label{correction}
R={\rm I }_{2^k \times 2^k} \otimes \left((\sigma_z)^{a_{k+1}}\otimes \dots \otimes (\sigma_z)^{a_n}\right), \qquad (\sigma_z)^0\equiv {\rm I}.
\eea
Measuring  $\psi_a$ in the computational basis, called syndrome measurement, and then applying $R$, is analogous to evaluating \eqref{y} in classical case and using it to reconstruct the original codeword. 

In our example, the code subspace $\H_\C$ includes all states of the form $\psi_l \otimes |0^{n-k}\rangle$, and has dimension $2^k$. It can be defined as the subspace invariant under the action of $\sigma_z$ acting on any of the auxiliary spins,
\bea
\g_i&=&\underbrace{{\rm I} \otimes \dots \otimes {\rm I}}_{k+i-1} \otimes\, \sigma_z \otimes\underbrace{{\rm I} \otimes \dots \otimes {\rm I}}_{n-k-i},\qquad 1\leq i\leq n-k,\\
\g_i \H_\C&=&\H_\C. 
\eea
We additionally notice that $\g_i$ are unitary, nilpotent, traceless, and commute with each other,
\bea
\g_i\, \g_j=\g_j\, \g_i,\qquad (g_i)^2={\rm I},\qquad (\g_i)^\dagger  \g_i={\rm I}, \qquad {\rm Tr}(\g_i)=0.
\eea
They form an abelian group, which acts trivially on  $\H_\C$. The group generated by the $\g_i$ is called the stabilizer of $\H_\C$.
Crucially, the generators $\g_i$  define the basis  $|a_{k+1}\dots a_n\rangle$ in the quotient $\H/\H_\C$ as the mutual eigenbasis of the $\g_i$ with eigenvalues $1-2a_{k+i}$.

The code described above is degenerate. Different nontrivial combinations of Pauli matrices acting on the $n-k$ auxiliary qubits may act trivially on $|0^{n-k}\rangle$. This can be seen differently. The dimension  of $\H_\C$ is $2^k$ while the total dimension is $2^n$. Thus naively only $2^{n-k}$ errors are correctable, while in fact all $\sum_{t=0}^{n-k}V_q(t,n-k)=4^{n-k}$ operators  acting on the auxiliary qubits are correctable. 

The discussion above applies to the trivial case when the logical qubits are isolated from the environment. Now we want to consider the situation when all $n$ physical qubits are subject to noise, and we want to use them to encode $k$ logical qubits with the possibility to recover at least some errors. To that end we perform a unitary transformation on $\H$, and define new stabilizer group via $\g_i \rightarrow U \g_i U^\dagger$. Our code subspace is an image of $\psi_l\otimes |0^{n-l}\rangle$ under $U$. If $U$ is nontrivial, all states in the code subspace will be highly entangled. To correct the error, we can perform projective measurements of the $\g_i$, identify corresponding eigenvalues $\lambda_i=1-2a_{k+i}$ and then act by 
\bea
R=\prod_{i=1}^{n-k} \g^{a_{k+i}}_i 
\eea
on the projected state. As a result of these operations we are guaranteed to obtain a state from the code subspace. We will discuss later  which errors can be corrected in this way. 

The class of quantum error-correcting codes 
known as additive or stabilizer codes exploits the idea outlined above with the following restriction. The generators of the stabilizer group $\g_i$ are chosen to be tensor products of Pauli operators and identity operators acting on individual spins 
\bea
\label{generator}
\g(\nu)=\epsilon \left( \sigma_{\nu_1}\otimes \dots \otimes  \sigma_{\nu_n}\right),\qquad \nu_i\in\{0,1,2,3\},\quad 1\leq i \leq n.
\eea
Here $\sigma_0$ is the identity matrix, $\sigma_{1,2,3}$ are Pauli matrices and $\epsilon=\pm 1$ or $\epsilon=\pm i$ to ensure $\g^2={\rm I}$.  With this definition all properties are automatically satisfied except for commutativity.
The form of \eqref{generator} can be understood as a restriction on the unitary transformation $U$. 
 It is customary to rewrite \eqref{generator} in a slightly different form, using two binary vectors $\alpha,\beta$ of length $n$,
\bea
\label{g}
\g(\alpha,\beta)= i^{\alpha \cdot \beta} \epsilon \left((\sigma_x)^{\alpha_1} \otimes \dots \otimes (\sigma_x)^{\alpha_n}\right)
\left((\sigma_z)^{\beta_1} \otimes \dots \otimes (\sigma_z)^{\beta_n}\right),\ \alpha,\beta\in (\Z_2)^n.
\eea
The coefficient $\epsilon=\pm 1$ can be chosen at will. To describe $k$ logical qubits we would need $n-k$ generators of the stabilizer group, or $n-k$ vectors $(\alpha_i,\beta_i)$ for $1\leq i \leq n-k$. Commutativity of a pair of generators requires 
\bea
\g_i\, \g_j=\g_j\, \g_i \quad \Leftrightarrow \quad \alpha_i \cdot \beta_j-\alpha_j \cdot \beta_i\, \equiv 0\, \, \, {\rm mod}\, \, 2. \label{commute}
\eea
Because we are working mod 2, the minus sign in front of the second term can be flipped. 
It is convenient to combine the vectors $(\alpha_i,\beta_i)$ into an $(n-k)\times n$ binary ``parity check'' matrix 
\bea
\label{QH}
H=\left(\begin{array}{c|c}
\alpha_1 & \beta_1 \\
\dots & \dots \\
\alpha_{n-k} & \beta_{n-k} 
\end{array}\right),
\eea
and  introduce a $2n\times 2n$ matrix 
\bea
g=\left(\begin{array}{c|c}
\, 0\,  &\, {\rm I}\, \\ \hline
{\rm I} & 0 
\end{array}\right). \label{Mink}
\eea
Then the commutativity condition \eqref{commute} is $H g H^T=0$. Multiplying $H$ by an invertible binary matrix from the left would not change the stabilizer group but only the choice of the generators $\g_i$. All operations with $H$ are to be understood  mod 2.

The operators  on $\H$ commuting with the full stabilizer group are operators acting on $\psi_l$ in our example above. These are ``logical operations'' -- they change states from the code subspace into other states in the code subspace. Considering operators of the  form  \eqref{g}, there are exactly $2k$ generators of such transformations corresponding to $2k$ linearly independent vectors $(\alpha,\beta)$. In the example above those would be operators $\sigma_z$ and $\sigma_x$ acting on individual logical qubits.  

We introduce the binary ``generator matrix'' $G$ as a matrix of maximal rank satisfying 
\bea
H\, g\, G\,\equiv 0 \, \, \, {\rm mod}\, \, 2.
\eea   Its transpose $G^T$  will have $n+k$ rows, $n-k$ of which span the same space as the rows of $H$, while the remaining $2k$ rows are generators of logical operations on $\H_\C$,  
\bea
G^T=\left(
\begin{array}{c|c}
\alpha_1 & \beta_1\\
\dots  & \dots \\
\alpha_{n+k} & \beta_{n+k}
\end{array}
\right). \label{Gm}
\eea
The similarity with classical codes is striking at this point. We can identify rows of $G^T$ as codewords $c\in \Z_2^{2n}$. Assuming algebra over $\Z_2$, acting on $G$ from the right by any invertible $(n+k)\times (n+k)$ binary matrix would not change the code, exactly as in the classical case. That is why we can always assume that the first $n-k$ rows of $G^T$ coincide with $H$.  

At this point we would like to introduce the quantum Hamming weight $\w(c)=\w(\alpha,\beta)$ as the number of qubits affected by $\g(\alpha,\beta)$. For binary vectors $(\alpha,\beta)$ this can be written as follows 
\bea
\w(\alpha,\beta)=\alpha^2+\beta^2-\alpha\cdot \beta. \label{quantumweight}
\eea 
In the classical case we would define the Hamming distance of the code as the minimal weight of all $2^{n+k}-1$ linear combinations  of the rows of $G^T$ understood mod 2, with the exception of the trivial one. In the quantum case the situation is more nuanced. The rows of $H$ and their linear combinations, understood in the sense of \eqref{g}, are elements of the stabilizer group. They do not introduce errors as they do not affect states from the code subspace. Therefore the Hamming distance of a quantum stabilizer code is defined as follows:
\bea
d=\min \w(c),\qquad c=G\, x,\, \,  x\in (\Z_2)^{n+k},\quad  c\notin \Im(H). 
\eea
A quantum stabilizer code of Hamming distance $d$ is said to be of type $[[n,k,d]]$. It will protect against any quantum error affecting at most $t=[(n-1)/2]$ qubits. Indeed for any two such errors $E_i,E_j$ their linear combination $E_i-E_j$ would affect strictly fewer than $d$ qubits and therefore either  \eqref{nonoverlap} will be satisfied or $E_i$ and $E_j$ will act identically on $\H_\C$.

Since the stabilizer generators are nilpotent, $\g^2={\rm I}$, summing them up yields a projector on $\H_\C$,
\bea
\mathds{P}=\prod_{i=1}^{n-k}{{\rm I}+\g_i\over 2}={1\over 2^{n-k}}\sum_{\vec{\rm x}\in  \Z_2^{n-k}}  \prod_{i=1}^{n-k} \g_i^{{\rm x}_i}. \label{Projector}
\eea
In practice, to find states from $\H_\C$ in terms of the computational basis, it suffices to act by $\mathds{P}$ on $|0^n\rangle$.

In the literature, stabilizer codes are often specified by writing down stabilizer generators as products of Pauli matrices, denoted simply as X,Y,Z, and the identity I. 

\subsection*{Example: perfect LMPZ $[[5,1,3]]$ code}
The $[[5,1,3]]$ code was introduced by  Laflamme, Miquel, Paz and Zurek \cite{laflamme1996perfect}. We use an equivalent representation from \cite{devitt2013quantum}, which  specifies the code through the following four stabilizers. 
\bea
\begin{array}{l  c c c c c} 
 \hline
 \g_1\,  & {\rm X} & {\rm Z} & {\rm Z} & {\rm X} & {\rm I} \\ 
 \g_2\,  & {\rm I} & {\rm X} & {\rm Z} & {\rm Z} & {\rm X} \\ 
\g_3\,   & {\rm X} & {\rm I} & {\rm X} & {\rm Z} & {\rm Z} \\ 
\g_4\,   & {\rm Z} & {\rm X} & {\rm I} & {\rm X} & {\rm Z} \\ 
 \hline
\end{array}
\eea
The corresponding parity check and generator matrices are
\bea
H=\left(
\begin{array}{ccccc | ccccc}
1 & 0 & 0 & 1 & 0\, &\, 0 & 1 & 1 & 0 & 0 \\
0& 1 & 0 & 0 & 1\, &\, 0 & 0 & 1 & 1 & 0 \\
1 & 0& 1 & 0 & 0\, &\, 0& 0 & 0 & 1 & 1 \\
0& 1 & 0& 1 & 0\,  &\, 1 & 0 & 0 & 0 & 1 \\
\end{array}
\right), \qquad G^T=\left(
\begin{array}{ccccc|ccccc}
0 & 0 & 0 & 0 & 0\, &\, 1 & 1 & 1 & 1 & 1 \\
1 & 1 & 1 & 1 & 1\, &\, 0 & 0 & 0 & 0 & 0 \\
. & .& . & .& . & .& . & .& .& .
\end{array}
\right).
\eea
For the generator matrix $G^T$, we only explicitly write the two additional rows linearly independent from $H$. Linear combinations of the rows of $G^T$ include many vectors of minimal Hamming weight $3$, e.g.~$(1,1,1,0,0\,|\, 0^5)$. 

The code subspace is spanned by two vectors $|0\rangle_l$ and $|1\rangle_l$, with two algebraically-independent logical operators ${\rm X}_l,{\rm Z}_l$ represented by ${\rm X}\otimes {\rm X}\otimes {\rm X}\otimes {\rm X}\otimes {\rm X}$ and ${\rm Z}\otimes {\rm Z}\otimes {\rm Z}\otimes {\rm Z} \otimes {\rm Z}$. In terms of the physical basis 
\begin{align}
\label{0}
|0\rangle_l={1\over 4}\left(\right.&|00000\rangle+|01010\rangle + |10100 \rangle - |11110\rangle \  \\
+ &|01001\rangle - |00011\rangle - |11101\rangle - |10111\rangle\, \nonumber  \\
+ &|10010\rangle - |11000\rangle - |00110\rangle - |01100\rangle\, \nonumber \\ \nonumber
   -&|11011\rangle - |10001\rangle - |01111\rangle + \left. |00101\rangle\right). 
\end{align}
It can be checked that all four generators $\g_i$ leave $|0\rangle_l$ invariant. 
The state $|1\rangle_l$ can be obtained by flipping all spins in \eqref{0}. Many other details, including the circuit representation of the recovery protocol, can be found in the pedagogical review \cite{devitt2013quantum}. \\

The formulation of stabilizer codes given above was developed in \cite{calderbank1997quantum} and is reviewed in the textbook by Nielsen and Chuang \cite{nielsen2002quantum}. It suggests a close relation between quantum stabilizer codes and classical linear codes. This relation was further developed in a seminal paper by Calderbank, Rains, Shor and Sloane \cite{calderbank1998quantum} who reformulated quantum binary stabilizer codes as classical additive self-orthogonal codes over ${\rm GF}(4)$. 
There is an isomorphism under addition between ${\rm GF}(4)$  and $\Z_2^2$, called the Gray map,
\bea
\nonumber
0\, \,  \leftrightarrow \, \,  (0,0),\qquad 1\, \,  \leftrightarrow  \, \, (1,1),\\
\omega \, \,  \leftrightarrow \, \,  (1,0),\qquad \bar\omega \, \,  \leftrightarrow \, \,  (0,1).
\label{Gray}
\eea
By combining the $i$-th component of $\alpha\in \Z_2^n$ and $\beta\in \Z_2^n$ we can rewrite $(\alpha,\beta)$ as a vector with $n$ components, $c\in {\rm GF}(4)^n$. Then a straightforward check confirms that
\bea
\alpha_1 \cdot \beta_2+\alpha_2\cdot \beta_1=0\, \, \,  \Leftrightarrow \, \, \, \bar{c}_1 \cdot c_2+c_1\cdot \bar{c}_2 =0,
\eea
where the first equation is understood  in terms of algebra over ${\rm GF}(2)=\Z_2$ while the second equation is over ${\rm GF}(4)$. Any vector $c\in {\rm GF}(4)^n$ is orthogonal to itself. Therefore stabilizer codes of the form (\ref{g},\ref{QH}) are in one-to-one correspondence with self-orthogonal additive codes  over ${\rm GF}(4)$ with the scalar product \eqref{H+}. This class of codes is denoted by $4^{\rm H+}$ in Section \ref{sec:GF4}.  A quantum $[[n,k,d]]$ code corresponds to a classical $[n,n-k,\tilde{d}]$ code, but the relation between $d$ and $\tilde{d}$ is nontrivial. The quantum Hamming distance $d$ is the smallest Hamming weight of the ${\rm GF}(4)$ code $[n,n+k,\dbtilde{d}]$, which is dual to $[n,n-k,\tilde{d}]$,  after removing all codewords of $[n,n-k,\tilde{d}]\subset [n,n+k,\dbtilde{d}]$ from consideration.

Self-dual classical codes $[n,n,d]$ over ${\rm GF}(4)$ are a special case. They correspond to stabilizer codes with $k=0$, which means that the code subspace is one-dimensional. In this case the quantum state $\psi_\C \in \H_\C$ contains no information and one can not speak of quantum error correction. Rather $k=0$ stabilizer codes should be interpreted as quantum error detection protocols: they can detect any error acting on up to $d-1$ qubits, where $d$ is the largest quantum Hamming weight of all linear combinations of $H$ (except the trivial one), 
\bea
d=\min \w(c),\quad c=G x,\quad x\in \Z_2^{n},\quad x\neq 0^n,\qquad G=H^T.
\eea
In this case the classical and quantum Hamming distances coincide and self-dual stabilizer $[[n,0,d]]$ codes are non-degenerate. They are in one-to-one correspondence with classical self-dual $[n,n,d]$ codes of type $4^{\rm H+}$. 

 
The equivalence group of classical codes over ${\rm GF}(4)$ can be understood quantum mechanically as  the group of unitary transformations acting on individual qubits in such a way that the stabilizer generators remain of the form \eqref{g}. This group of transformations is called the Clifford group or local Clifford group (LC). The generators of the Clifford group include permutations of qubits, cyclic permutations of Pauli operators (multiplication by $\omega$ in ${\rm GF(4)}$ language), and the exchange of $\sigma_x$ and $\sigma_z$ 
generated by the Hadamard matrix (conjugation $\omega \leftrightarrow \bar \omega$). Assuming that the physical qubits are subject to uncorrelated noise, these are natural symmetries, which define the group of equivalences of stabilizer codes. 

The connection to classical codes over ${\rm GF}(4)$ enables many results developed for classical codes to be applied to the quantum case, and vice versa. In what follows we mostly focus on self-dual stabilizer codes, or equivalently on self-dual $4^{{\rm H}+}$  classical codes  over ${\rm GF}(4)$. In addition to the (total) Hamming weight $\rm w$ introduced in Section \ref{sec:GF4} we can introduce weights that count the number of individual ``letters'' in the codeword. Instead of using the ${\rm GF}(4)$ ``alphabet'' $1,\omega,\bar\omega$ we will use the labels of the Pauli matrices $\sigma_{x,y,z}$ of the corresponding stabilizer element $\g(\alpha,\beta)$. Using the Gray map $c=(\alpha,\beta)$ \eqref{Gray} we can write down explicit formulas for the weights
\bea
{\rm w}_x(c)=\vec{1}\cdot \alpha,\quad  {\rm w}_y(c)=\alpha\cdot \beta,\quad {\rm w}_z(c)=\vec{1}\cdot \beta,
\eea
and ${\rm w}(c)={\rm w}_x(c)+{\rm w}_y(c)+{\rm w}_z(c)$. Then, in addition to the enumerator polynomial \eqref{enumerator},  we can define the refined enumerator polynomial (REP)
\bea
W_\C(x,y,z)=\sum_{c\in \C} x^{n-{\rm w}(c)} y^{{\rm w}_y(c)} z^{{\rm w}_x(c)+{\rm w}_z(c)}.\label{REP}
\eea 
One can also define the full enumerator polynomial $W_\C(t,x,y,z)=\sum_{c\in \C} t^{n-{\rm w}(c)} x^{{\rm w}_x(c)} y^{{\rm w}_y(c)} z^{{\rm w}_z(c)}$, but this will not play an important role in what follows.  Under a duality transformation, the refined enumerator polynomial of an $[n,k,d]$ code changes as follows: 
\bea
\label{MacW}
W_{{\mathcal C}^\perp}(x,y,z)=2^{n-k}\,W_{\mathcal C}\left (\frac{x+y+2 z}{2} ,\frac{x+y-2 z}{2},\frac{x-y}{2}\right). \label{dual}
\eea
Thus for self-dual codes the refined enumerator polynomial is invariant under 
\bea
\label{dual}
x\to \frac{1}{2} (x+y+2 z),\quad y\to \frac{1}{2} (x+y-2 z),\quad z\to \frac{x-y}{2}.
\eea
Setting $z= y$ reduces \eqref{MacW} and \eqref{dual} to \eqref{MacW4} and \eqref{Mac4}. Enumerator polynomials and the MacWilliams identity \eqref{MacW} can also be defined at the level of quantum stabilizer codes without any reference to codes over ${\rm GF}(4)$ \cite{shor1997quantum,rains1998quantum,rains1999quantum}.

Focusing on self-dual codes $4^{{\rm H}+}$, their total number is \cite{pless1975classification,calderbank1998quantum}
\bea
\prod_{j=1}^n (2^j+1). \label{n4h+}
\eea
In Section  \ref{sec:GF4} we introduced even codes $4^{{\rm H}+}_{\rm II}$,  those whose codewords all have even Hamming weight. They exist only when $n$ is even, and their total number is 
\bea
\prod_{j=0}^{n-1} (2^j+1),\quad n\equiv 0 \, \, \,({\rm mod}\, \,  2). \label{4HII}
\eea

Real codes make up another class of codes, which is central to our considerations. A stabilizer code is called real if all generators $\g(c),\, c\in \C$, of the stabilizer group are real. This nomenclature was introduced in the  context of quantum codes in \cite{rains1999quantum}, where it was shown that any stabilizer code has an equivalent real code. Equivalence is defined with  respect  to  the Clifford group defined above.  This result is crucial, because it shows that modulo  equivalences, real  codes encompass all codes. Since the Pauli matrices $\sigma_x,\sigma_z$ are real and $\sigma_y$ is purely imaginary, a code is real if and only if ${\rm w}_y(c)$ of all codewords is even. We denote the space of real self-dual codes over ${\rm GF}(4)$ by $4^{{\rm H}+}_{\rm R}$. Their total number is
\bea
\prod_{j=0}^{n-1} (2^j+1). \label{n4hr}
\eea
While this formula is the same as \eqref{4HII}, the spaces $4^{{\rm H}+}_{\rm II}$ and $4^{{\rm H}+}_{\rm R}$ are not isomorphic. The former is defined for even $n$, while the latter exists for any $n$. Refined enumerator polynomials of real self-dual codes, besides being invariant under \eqref{dual}, must also be invariant under $y\rightarrow -y$. They are polynomials in the ring ${\mathcal P}(W_1,W_2,W_3)$ generated by
\bea
\label{invpols}
W_1=x+z,\quad W_2=x^2+y^2+2z^2,\quad W_3=x^3+3 x z^2+3 y^2 z+z^3,
\nonumber
\eea
which satisfy $W(1,0,0)=1$ and have positive integer coefficients. The polynomials $W_1,W_2,W_3$ are refined enumerator polynomials of three particular codes introduced below in Section \ref{sec:firstn}. In practice, instead of $W_3$ it is convenient to use ${\cal R}=(x-z)(y^2-z^2)$. The rings of invariant refined enumerator polynomials for $4^{{\rm H}+}$ and $4^{{\rm H}+}_{\rm II}$  can similarly be described explicitly.

\subsection{New Construction A: Lorentzian lattices}\label{sec:newA}
The connection to classical codes over ${\rm GF}(4)$ provides a  way to associate a stabilizer $[[n,k,d]]$ code to an integral Euclidean lattice in $\R^{2n}$, as described in Section \ref{sec:GF4}. This lattice is not in general self-dual, even when the underlying code is self-dual, and for this reason it is not connected in any obvious way to a CFT.
To obtain a self-dual lattice from a self-dual code over ${\rm GF}(4)$, we introduce a new version of Construction A for $4^{\rm H+}$ codes. 

Starting from a code of type $4^{\rm H+}$ and rewriting its codewords as vectors $c=(\alpha,\beta)\in \C\subset \Z_2^{2n}$ using the Gray map, we define a corresponding lattice using Construction A for binary codes,
\bea
\Lambda(\C)=\{v/\sqrt{2}\, |\, v\in \Z^{2n},\, \,  v=c,\, \, \, \, ({\rm mod}\, \, 2),\ c\in \C\},\quad \Gamma^*  \subset \Lambda(\C)  \subset \Gamma=(\Z/\sqrt{2})^{2n}. \nonumber
\eea
The lattice $\Lambda(\C)$ should be understood as a lattice in Lorentzian space $\R^{n,n}$  with the  metric \eqref{Mink}. Then the following crucial results  follow. The lattice of a dual code $\Lambda(\C^\perp)$ is equal to the dual lattice, $\Lambda(\C^\perp)=\Lambda(\C)^*$.
If $\C$ is self-orthogonal, the lattice is integral, $\Lambda(\C)\subset \Lambda(\C)^*$. If $\C$ is self-dual, then $\Lambda(\C)$ is self-dual. There is a one-to-one correspondence between lattices $\Lambda(\C)\subset (\Z/\sqrt{2})^{2n}\subset \R^{n,n}$ and codes $\C$ of type $4^{\rm H+}$.

Stabilizer codes are self-orthogonal codes of type $4^{\rm H+}$ and therefore correspond to integral lattices.  This correspondence provides a geometric way to interpret various aspects of stabilizer codes. Suppose $\C$ is an Abelian stabilizer group generated by the rows of \eqref{QH} (or equivalently, $\C$ is a self-orthogonal code from $4^{\rm H+}$). The corresponding lattice $\Lambda(\C)\subset (\Z/\sqrt{2})^{2n}\subset \R^{n,n}$ is then integral. The number of logical qubits $k$ is  equal to half the number of generators in the abelian group $\Lambda(\C)^*/\Lambda(\C)$
({\it i.e.}~the number of dimensions of the torus $\Lambda(\C)^*/\Lambda(\C)$).
We define the quantum Hamming  norm on the space of vectors $\vec{v}=(\alpha,\beta)\in \R^{n,n}$ via \eqref{quantumweight},
$|v|^2_q=\alpha^2+\beta^2-\alpha\cdot \beta$. The quantum Hamming distance of the $[[n,k,d]]$ code $\C$ is 
\bea
d=\min_{v\in \Lambda(\C)^*/\Lambda(\C) \atop v\neq 0} |v|^2_q,
\eea
where the norm on the quotient is understood to be the minimal value of the norm on all elements in the preimage.  

A particular class of stabilizer codes,  the Calderbank-Shor-Steane (CSS) codes, can easily be understood geometrically. In this case  the Lorentzian lattice $\Lambda(\C)$ is the direct sum of two Euclidean lattices $\Lambda(\C)=\Lambda_2\oplus (\Lambda_1)^*$, which additionally satisfy $\Lambda_2\subset \Lambda_1$. Furthermore, $\Lambda(\C)$ is automatically integral, $\Lambda^*(\C)=\Lambda_1\oplus (\Lambda_2)^* \supset \Lambda(\C)$. 
We additionally require $(\sqrt{2}\Z)^n\subset \Lambda_1,\Lambda_2\subset (\Z/\sqrt{2})^n$
such that $\Lambda_{1,2}$ are the Construction A lattices of some classical binary codes  $[n,k_{1,2},d_{1,2}]$. The number of logical qubits $k=k_1-k_2$ is equal to the dimension of the torus $\Lambda_1^*/\Lambda_2$ and the quantum Hamming distance of the CSS stabilizer code is the smallest length of any nontrivial vector in 
$\Lambda_1/\Lambda_2$  or $(\Lambda_2)^*/(\Lambda_1)^*$ calculated with the Euclidean metric. It is in any case not smaller than $\min(d_1,d^*_2)$, where $d^*_2$ is the Hamming distance of $\C_2^\perp$.

\subsection*{Example: Steane $[[7,1,3]]$ code}
The Steane $[[7,1,3]]$ quantum stabilizer code is a CSS code with $\C_1$ being the Hamming $[7,4,3]$ code and $\C_2=\C_1^\perp$ being  its dual, the Hamming $[7,3,4]$ code. In this case $k_1=4$, $k_2=3$, $d=3$, and the quantum code protects $k=1$ logical qubit against any one-qubit errors. Geometrically, $\Lambda_1/\Lambda_2=(\Lambda_2)^*/(\Lambda_1)^*$ is the quotient of the weight lattice of $E_7$  over the root lattice of $E_7$. As a group, the quotient is $\Z_2$, which corresponds to  Pauli operators $\sigma_{x,z}$ acting on the single logical qubit.\\

A Lorentzian lattice $\Lambda\subset\R^{n,n}$ with metric $|v|^2=p_L^2-p_R^2$ for $\vec{v}=(\vec{k}_L,\vec{k}_R)\in  \R^{n,n}$ can be characterized by the Siegel theta-function 
\bea
\label{Siegel}
\Theta_\Lambda(\tau,\bar\tau)=\sum_{v\in \Lambda} q^{\, p_L^2/ 2}\, \, {q'}^{\, p_R^2/2},\qquad q\equiv e^{2\pi i \tau},\quad q'\equiv e^{-2\pi i \tau'},
\eea
which is a holomorphic function of two complex variables $q,q'$, or, equivalently $\tau$, $\tau'$.  The implicit assumption in \eqref{Siegel} and in the rest of this paper is that the Lorentzian lattice is simultaneously equipped with the Euclidean metric. 
In  \eqref{Siegel} we assumed a diagonal Lorentzian metric, while previously it was given by \eqref{Mink}. The two metrics are related by the  following transformation which preserves the Euclidean metric on $\vec{v}=(\alpha,\beta)=(\vec{k}_L,\vec{k}_R)$,
\bea
\vec{k}_L={\alpha+\beta\over \sqrt{2}},\qquad \vec{k}_R={\alpha-\beta\over \sqrt{2}},
\eea
where we use different letters (Greek or Latin) and a vector arrow (or lack thereof) to imply the form of the metric tensor.

When the lattice is even, it is trivially invariant under the simultaneous shift $\tau\rightarrow \tau+1$, $\tau'\rightarrow \tau'+1$. When it is self-dual (unimodular) it is also covariant under $\tau\rightarrow -1/\tau$, $\tau'\rightarrow -1/\tau'$, changing as 
\bea
\Theta_\Lambda(-1/\tau,-1/\tau')=(\tau \tau')^{n/2}\, \Theta_\Lambda(\tau,\tau').
\eea 
Thus, the Siegel theta-function of an even self-dual Lorentzian lattice transforms covariantly under the full ${\rm PSL}(2,\Z)$ group acting simultaneously on $\tau$ and $\tau'$.

Similarly to the Euclidean lattices associated with  binary and ${\rm GF}(4)$ codes discussed in Sections \ref{sec:binary} and \ref{sec:GF4}, the Siegel theta function of a lattice $\Lambda(\C)$ associated with the stabilizer code $\C$  is determined in terms of its refined enumerator polynomial
\bea
\label{STL}
\Theta_{\Lambda(\C)}(\tau,\bar\tau)=2^{-n}W_\C\left({b\,b'+c\,c'},{b\,b'-c\, c'},{a\, a'}\right),\qquad {a'},{b'},{c'}=a,b,c(\tau').
\eea
The theta functions $a,b,c$ are defined in the text after \eqref{abc}.
The invariance of $\Theta_{\Lambda(\C)}(\tau,\tau')$  under $\tau\rightarrow \tau+1$, $\tau'\rightarrow \tau'+1$ and $\tau\rightarrow -1/\tau$, $\bar \tau\rightarrow -1/\tau'$ easily follows from the invariance of $W_\C(x,y,z)$ under $y\rightarrow -y$ and \eqref{dual} correspondingly. 

Comparing Construction A from Section \ref{sec:GF4} and the new Lorentzian Construction A defined above, we find that a classical self-orthogonal (self-dual) code $\C$ over ${\rm GF}(4)$ can be associated with both a Euclidean integral (integral) lattice and a Lorentzian integral (self-dual) lattice. These Euclidean and Lorentzian lattices are related to each other.  In terms  of the Euclidean vector $\vec{v}=(\alpha,\beta)=(\vec{x},\vec{y})$, the ${\rm GF}(4)$ Hamming weight, which is the same as the quantum Hamming weight, is 
\bea
|v|_q^2=\alpha^2+\beta^2-\alpha\cdot \beta= x^2+3y^2.
\eea
Therefore the Siegel theta function will become the theta function of the Euclidean lattice \eqref{ELC} upon the substitution $\tau'=-3\tau$, or $q'=q^3$. Indeed, it is straightforward to  check  using Jacobi theta function identities that in this case 
\bea
{b\,b'+c\,c'\over 2}\rightarrow \phi_0(\tau),\qquad 
{b\,b'-c\,c'\over 2},{a\,a'\over 2}\rightarrow \phi_1(\tau),
\eea
where $\phi_{0,1}$ were given in \eqref{phi01}. We postpone giving explicit examples until Sections \ref{sec:n=2},  \ref{sec:n6}.

Besides the Euclidean metric $x^2+3y^2$ associated with the Hamming weight of a ${\rm GF}(4)$ code, the conventional  Euclidean metric $x^2+y^2$ may also be considered. It is associated with the binary Hamming distance $d_{\rm b}(c)={\rm w}_x(c)+2{\rm w}_y(c)+{\rm w}_z(c)$, with respect to which the original self-dual ${\rm GF}(4)$ code $\C$ via the Gray map is interpreted as an isodual binary code.  Given that the generator matrix $\Uplambda$ of the Lorentzian $\Lambda(\C)$ satisfies
\bea
\Uplambda^T g\, \Uplambda\in {\rm GL}(2n,\Z),
\eea
and $g$ \eqref{Mink} can be interpreted as an orthogonal rotation of $\R^{2n}$, we immediately conclude that $\Lambda(\C)$, understood as a Euclidean lattice with metric $x^2+y^2$, is isodual. 
The theta function of this Euclidean lattice follows from $\Theta_\C(\tau,\tau')$ upon the substitution $\tau'=-\tau$.

Looking ahead, the theta function $\Theta_\C(\tau,-\tau)$ will turn out to count dimensions of the CFT operators. 
In full analogy with the case of classical binary codes, for which we defined a lattice and sought to maximize the norm of its shortest vector, in the quantum case we also want to maximize the Euclidean norm of the shortest nontrivial vector of $\Lambda(\C)$, which  defines the spectral gap of the theory. The same problem we encountered in the case of binary codes also appears here: Construction A Lorentzian lattices $\Lambda(\C)$ necessarily have vectors of the form $(\alpha,\beta)=(2,0^{2n-1})/\sqrt{2}$, resulting in operators of conformal dimension $\Delta=(p_L^2+p_R^2)/2=1$. To  partially resolve this problem we employ the procedure of twisting by a half-vector, which will yield a new even self-dual Lorentzian lattice $\Lambda'$ starting from the original lattice $\Lambda$, given a vector $2\vec{\delta}\in \Lambda$ with odd norm $\delta^2$. The procedure is identical to the one described in Section \ref{sec:binary}, with the only change that scalar product is now understood to be defined by the Lorentzian metric.

In the context of Construction A code lattices $\Lambda(\C)$, we can take the same vector $\vec{\delta}=\vec{1}/(2\sqrt{2})$ as in the binary case. Notice that for any $\vec{u}=(\alpha,\beta)/\sqrt{2}\in \Lambda(\C)$, the scalar product with $\vec{\delta}$ gives
\bea
2\,\vec{\delta}\cdot\vec{u}={\vec{1}\cdot \alpha+\vec{1}\cdot \beta\over 2}={{\rm w}_x(c)+{\rm w}_z(c)\over 2},\quad c=(\alpha,\beta)\in \C,
\eea
and therefore $2\vec{\delta}\in \Lambda(\C)$ for a  real  self-dual $\C$ if and only if the code $\C$ is also even, $\C \in 4^{{\rm H}+}_{\rm II}\cap 4^{{\rm H}+}_{\rm R}$. Additionally, for $\delta^2$ to be odd, $n\, \, {\rm mod}\, \, 4$ should  be odd. Provided both of these conditions are satisfied, the Siegel theta-function of the new lattice $\Lambda'(\C)$ will be  given by
\bea
\nonumber
\Theta_{\Lambda'(\C)}(\tau,\tau')={2^{n}\,\Theta_{\Lambda(\C)}+W_\C\left({bc'+cb'},{cb'-bc'},0\right)+W_\C(a b',a b',b a')-W_\C(a c',-a c',i c a') \over 2^{n+1}}.\\ \label{twisted}
\eea
When the code is self-dual and real  $W_\C\left(\mu,\pm \mu,\nu \right)=W_\C(\nu,\pm \nu,\mu)=W_\C(\mu+\nu,\pm \mu\mp \nu,0)$ for any $\mu,\nu$. From here it follows that $\Theta_{\Lambda'(\C)}(\tau,\tau')$ is real, i.e.~it is invariant under complex conjugation, accompanied by $\tau\leftrightarrow \tau'$ when $n$ is divisible by four.
Similarly, \eqref{twisted} changes covariantly under modular transformation $\Theta_{\Lambda'(\C)}(-1/\tau,-1/\tau')=(\tau \tau')^{n/2} \Theta_{\Lambda'(\C)}(\tau,\tau')$, provided $\C$ is also even. 
When additionally $n\, \, {\rm mod}\, \, 4$ is odd, which is the condition for $\delta^2$ to be odd, the Lorentzian lattice $\Lambda'(\C)$ is even, and accordingly the theta function $\Theta_{\Lambda'(\C)}(\tau,\tau')$ is invariant under $\tau\rightarrow \tau+1$, $\tau'\rightarrow \tau'+1$.
Upon substituting ${a'},{b'},{c'}\rightarrow a,b,c$, expression \eqref{twisted} reduces to \eqref{twistedbinary}.

\section{Codes and CFTs}\label{sec:4}
\subsection{Narain CFTs}\label{NarainCFTs}
We start with a brief review of toroidal compactifications of string theory and the moduli space of Narain CFTs. This topic is discussed in many textbooks including \cite{polchinski1998string,becker2006string}.

As a warm-up we consider a particle of unit mass on a circle of radius $R$, 
\bea
S=\int dt\,  {\dot{x}^2\over 2}.
\eea
The classical EOM can be easily solved: $x(t)=x(0)+p(0)\,t$, where $p=\dot{x}$ is the momentum. At  the quantum mechanical level, the wavefunction $\psi(x)$ must  be periodic, $\psi(x)=\psi(x+2\pi R)$, which results in the quantum-mechanical momentum operator $\hat{p}$ being quantized, $p=n/R$, $n\in \Z$ where $p$ now stands for the eigenvalue of $\hat{p}$. The solution for $x(t)$ should now be understood as the solution of the EOM in the Heisenberg picture, $\hat{x}(t)=\hat{x}(0)+\hat{p}(0)\,t$, with the caveat that the operator $\hat{x}$ is not well-defined. Since the points $x$ and $x+2\pi R$ are equivalent, the algebra of  operators consists of  $\hat{p}(0)$ with eigenvalues $n/R$, and operators $e^{i k \hat{x}(0)}$ for $k=m/R$, $m\in \Z$.

Next let us consider a two-dimensional classical worldsheet theory describing the motion of $n$ bosons $X^I$ on a  torus  $\R^n/(2\pi\Gamma)$, where $\Gamma \subset \R^n$ is a lattice and $2\pi \Gamma$ stands for that lattice rescaled by $2\pi$, 
\bea
S={1\over 4\pi \alpha'}\int dt\, \int_0^{2\pi} d\sigma \left(\dot{X}^2-X'^2-2 B_{IJ} \dot{X}^I X'^J\right).
\eea
As in the previous example, $\vec{X}(t,\sigma)$ and $\vec{X}(t,\sigma)+2\pi \vec{e}$ must be physically equivalent for any $\vec{e}\in \Gamma$ . The worldsheet spatial variable $\sigma$ is periodic, and therefore $\vec{X}(t,\sigma+2\pi)=\vec{X}(t,\sigma)+2\pi \vec{e}$. The antisymmetric $B$-field does not enter into the EOM, nor into the solution 
\bea
\vec{X}(t,\sigma)=\vec{X}(0,0)+\vec{V}\, t+{\vec{e}}\,\sigma+{i\over 2}\sum_{n\neq 0} {a_n\over n} e^{-i n(t+\sigma)}+{b_n\over n} e^{-i n(t-\sigma)}, \label{solEOM}
\eea
but it affects the relation between the center-of-mass velocity and the total momentum 
\bea
\label{P}
\alpha' P_I={1\over 2\pi}\int_0^{2\pi}d\sigma\left(\dot{X}^I -B_{IJ} X'^J\right)=V^I-B_{IJ} e^J.
\eea
Going back to \eqref{solEOM}, we can represent $\vec X(t,\sigma)$ as a sum of left and right-moving components, 
\bea
\vec{X}(t,\sigma)&=&\vec{X}_L(t+\sigma)+\vec{X}_R(t-\sigma),\\
\vec{X}_L(t+\sigma)&=&{\vec{X}(0,0)\over 2}+\alpha'{\vec{p}_L\over 2} (t+\sigma)+ {i\over 2}\sum_{n\neq 0} {a_n\over n} e^{-i n(t+\sigma)},\\
\vec{X}_R(t-\sigma)&=&{\vec{X}(0,0)\over 2}+\alpha'{\vec{p}_R\over 2} (t-\sigma)+ {i\over 2}\sum_{n\neq 0} {b_n\over n} e^{-i n(t-\sigma)}.
\eea
At the classical level, the vector $\alpha'(\vec{p}_L-\vec{p}_R)/2=\vec{e}\in \Gamma$, while $\alpha'(\vec{p}_L+\vec{p}_R)/2=\vec{V}$. Quantum mechanically, since $X(t,\sigma)$ is only defined up to an arbitrary shift by $2\pi \vec{e}\in  \Gamma$,
the total momentum   \eqref{P} must be a vector from the dual lattice, $\vec{P}\in  \Gamma^*$, such that $\vec{V}=\alpha'\vec{P}+B\vec{e}$, and
\bea
\label{tc}
\vec{p}_L={\alpha'\vec{P}+(B+{\rm I})\vec{e}\over \alpha'},\qquad \vec{p}_R={\alpha'\vec{P}+(B-{\rm I})\vec{e}\over \alpha'},
\quad \vec{e}\in  \Gamma,\quad \vec{P}\in  \Gamma^*.
\eea
The set of vectors $\vec{v}=(\vec{p}_L,\vec{p}_R)$ for all possible $\vec{e}\in  \Gamma, \vec{P}\in  \Gamma^*$ forms a lattice $\Lambda$  in $\R^{n,n}$. To render $\vec{p}_L,\vec{p}_R$ dimensionless,  we use the conventional choice $\alpha'=2$, in which case $\Lambda$ becomes even and self-dual. 
To  verify first property we calculate
\bea
|\vec{v}|^2\equiv {p}_L^2-{p}_R^2= {4\over \alpha'} \vec{P}\cdot \vec{e}\in 2\,\Z.
\eea
To verify self-duality it is convenient to perform a linear change of variables and represent vectors $\vec{v}=(\vec{p}_L,\vec{p}_R)$ as follows
\bea
v=(\alpha,\beta),\qquad \alpha={p_L+p_R\over \sqrt{2}},\quad \beta={p_L-p_R\over \sqrt{2}}. \label{ab}
\eea 
In $(\alpha,\beta)$-coordinates the metric is given by \eqref{Mink}. In  this representation, and taking $\alpha'=2$, the generator matrix of $\Lambda$ is 
\bea
\Uplambda=\left(\begin{array}{c|c}
2\gamma^* & B \gamma \\
\hline
0 & \gamma
\end{array}\right)/\sqrt{2}, \label{TCN}
\eea
where $\gamma$ and $\gamma^*=(\gamma^{-1})^T$ are the generator matrices of $\Gamma$ and $\Gamma^*$ correspondingly.  Then it is straightforward to check that 
\bea
\Uplambda^T g\, \Uplambda=g\in {\rm GL}(2d,\Z),
\eea
and therefore $\Lambda$ is self-dual.


The primary vertex operators of the $U(1)^d\times U(1)^d$\  CFT are indexed by elements of the lattice $\Lambda$, 
\bea
V_{p_L,p_R}=\ :e^{i\vec{p}_L \cdot \vec{X}_L(z)  +i\vec{p}_R \cdot \vec{X}_R({\bar z}) }:,\qquad (\vec{p}_L,\vec{p}_R)\in \Lambda.
\eea
The partition function of this theory on the Euclidean worldsheet torus $\tau$ is 
\bea
Z(\tau,\bar\tau)={1\over |\eta(\tau)|^{2d}}\sum_{(\vec{p}_L,\, \vec{p}_R)\in \Lambda} q^{\, p_L^2/2}\, \bar{q}^{\, p_R^2/2},\qquad q=e^{2\pi i\tau},\quad \bar{q}=e^{-2\pi i\bar{\tau}}. \label{ZCFT}
\eea 
If $\Lambda$ is an even self-dual lattice, $Z(\tau,\bar\tau)$ is modular invariant and the CFT is well-defined. We emphasize that in constrast to \eqref{Siegel}, where $\tau$ and $\tau'$ were two independent holomorphic variables, in  \eqref{ZCFT} $\tau$ and $\bar\tau$ are related by complex conjugation.

\subsection*{Example: twist of a compact boson on a circle}
The simplest example of a theory of the type described above consists of a single boson $X$ compactified on a circle of radius $R$.  The lattice $\Gamma$ consists of points $\vec{e}=m R\in \R^1$ for any integer $m$, and the vectors of the dual lattice are $\vec{P}=n/R\in \R^1$. Since there is only one boson $X$, the B-field  is trivial.  We immediately find
\bea
p_L={n\over R}+{m R\over 2},\quad p_R={n\over R}-{m R\over 2}. \label{circle}
\eea
The set of vectors $\vec{v}=(\vec{p}_L,\vec{p}_R)$ for all possible $n,m\in\Z$ defines an even self-dual lattice $\Lambda$ in $\R^{1,1}$. It is much simpler to represent $\Lambda$ in the coordinates \eqref{ab} 
\bea
v(n,m)=\left({n\sqrt{2} \over R},{ m R\over \sqrt{2}}\right)\in \R^{1,1},\qquad n,m\in \Z.
\label{vnm}
\eea
We would like to apply to this lattice the twist procedure \eqref{shift}. We start with a vector
$\delta=(2a /R,b R)/(2\sqrt{2})$ with some integers $a,b$, which automatically satisfies $2\vec{\delta}\in \Lambda$. For $\delta^2$ to be odd, we must require $ab/2\, \, {\rm mod}\, \,  2=1$. Therefore either $a=2(2k+1)$ is even, but not doubly-even, while $b=2l+1$ is odd, $k,l\in \Z$, or the other way around. In the first scenario we represent $\Lambda=\{v(n,m)|n,m\in\Z\}$ as the union of two sublattices
\bea
\Lambda&=&\Lambda_0 \cup \Lambda_1,\\
\Lambda_0&=&\{v(2n,m)\, |\, n,m\in Z\},\quad \Lambda_1=\{v(2n+1,m)\, |\, n,m\in Z\},
\eea
where $2\delta \cdot v(2n,m)$ is even and $2\delta \cdot v(2n+1,m+1)$ is odd. 
 Then
\bea
\Lambda'_1=\Lambda_1+\delta=\{v(2n+1,m)+v(2k+1,l+1/2)\, |\, n,m,k,l\in\Z\}.
\eea
The disjoint union of $\Lambda_0$ and $\Lambda'_1$ gives
\bea
\Lambda'=\{v(2n,m/2)\, |\, n,m\in\Z\}.
\eea
In other words $\Lambda'$ is the lattice of the single boson compactified on a circle of radius $R'=R/2$. In the second scenario $\Lambda'$ is given by all vectors of the form $v(n/2,2m),\, \, n,m\in\Z$, or $R'=2R$. Thus, starting from  some radius $R$ and applying the twist procedure repeatedly, we arrive at the lattice of a boson compactified on  a circle of radius $R'=2^k R $ for any  integer $k$.  \\  

In the example above, it was obvious that starting from a toroidal CFT \eqref{ZCFT} and twisting by a vector $\delta$ yields  
another CFT of the same type.  In fact, this construction can be applied to general CFTs of this type, and can be extended to CFTs with fermionic degrees of freedom as well \cite{Dixon:1986iz,nair1987compactification}.

The CFT partition function \eqref{ZCFT} is manifestly invariant under orthogonal transformations ${\rm O}(d)$ acting independently on $\vec{p}_L\in \R^d$ and  $\vec{p}_R\in \R^d$. They form a group ${\rm O}(d)_L\times {\rm O}(d)_R$ of symmetries of the CFT, which is a subgroup of T-duality transformations. 
Thus, from the CFT point of view, two even self-dual lattices $\Lambda$ and $\Lambda'$ related by an ${\rm O}(d)\times {\rm O}(d)$ transformation are equivalent. 

As we will see shortly, any even self-dual lattice $\Lambda\subset \R^{d,d}$ defines a CFT, with the partition function given by \eqref{ZCFT}. CFTs of this kind are called Narain CFTs.  A central mathematical result, which provides a description of the moduli space of all Narain theories, is that all even self-dual lattices $\Lambda\subset \R^{d,d}$ are related to each other by boost transformations in ${\rm O}(d,d)$. 
At the level of the generator matrix, and working in coordinates \eqref{ab} such that metric is given by \eqref{Mink},
\bea
\Uplambda={\cal O}\, {\rm I}, \label{gm}
\eea
where ${\cal O}\in {\rm O}(d,d)$ and the identity matrix ${\rm I}$ is the generator matrix of a particular  even self-dual lattice in $\R^{d,d}$. The lattice generated by ${\rm I}$ has an obvious symmetry: any element from ${O}(d,d,\Z)$ maps  it into itself.\footnote{Here and in what follows ${O}(d,d,\Z)$ is defined as a group of integer $2d\times 2d$ matrices ${\cal F}$ satisfying ${\cal F}^T g\, {\cal F}=g$.} Therefore the full Narain moduli space is given by 
\bea
{{\rm O}(d,d)\over {\rm O}(d)\times {\rm O}(d) \times {\rm O}(d,d,\Z)}, \label{Narain}
\eea
where the denominator represents the group of T-dualities -- symmetries of the two-dimensional CFT.
The first two factors in the denominator act from the left and relate physically equivalent lattices to each other. The last factor acts from the right. It is a symmetry of a particular lattice, which maps different lattice points into each other. 

Thus, CFTs with momentum lattices of the form \eqref{tc} cover all possible Narain CFTs.  In other words, any Lorentzian even self-dual lattice with generator matrix \eqref{gm} can be brought into the form \eqref{tc} by means of an  appropriate ${\rm O}(d)\times {\rm O}(d)$ transformation. 
We demonstrate this explicitly in Appendix~\ref{sec:ToroidalC}.

\subsection{Code CFTs}\label{sec:codeCFTs}
In Section \eqref{sec:newA} we established that real self-dual codes $\C$ are in one-to-one correspondence with even self-dual lattices $(\sqrt{2}\Z)^{2n}\subset \Lambda(\C)\subset \R^{n,n}$. In the previous section we saw that any even self-dual lattice in $\R^{n,n}$ defines a Narain CFT. We therefore arrive at the main point of this paper: real self-dual quantum stabilizer codes (or alternatively classical self-dual codes of type $4^{{\rm H}+}_{\rm R}$) define a family of Narain CFTs, which we will call code theories.  The partition function of a code theory  is given by the Siegel theta-function $\Theta_\C$ of $\Lambda(\C)$ divided by $|\eta(\tau)|^{2n}$, where  $\Theta_\C$ is given in terms of the refined enumerator polynomial of $\C$ via \eqref{STL},
\bea
Z(\tau,\bar\tau)={W_\C\left({b\,{\bar b}+c\,{\bar c}},{b\,{\bar b}-c\, {\bar c}},{a\, {\bar a}}\right)\over 2^n |\eta(\tau)|^{2n}}, \label{ZcodeCFT}
\eea
\bea
a=\theta_2(q),\, \, b=\theta_3(q),\, \, c=\theta_4(q), \, \,{\bar a}=\theta_2({\bar q}),\, \, {\bar b}=\theta_3({\bar q}), \, \,{\bar c}=\theta_4({\bar q}). \nonumber
\eea

The code equivalence group (Clifford group) includes arbitrary permutations of codeword components and conjugations of the $i$-th component $\omega \leftrightarrow \bar \omega$ (exchange of $\sigma^i_z$ and $\sigma^i_x$) for arbitrary $1\leq i\leq n$. The permutations  are orthogonal transformations ${\cal O}_p \in {\rm O}(n,\Z)$ which are diagonally embedded in the group  ${\rm O}(n)\times {\rm O}(n)$ of T-duality transformations. The exchange of $\sigma^i_z$ and $\sigma^i_x$ is also an element of ${\rm O}(n)\times {\rm O}(n)$ represented by ${\rm I}\times {\cal O}_i $,  where ${\cal O}_i \in {\rm O}(n,\Z)$ flips the sign of the $i$-th coordinate. In other words, the subgroup of the equivalence group generated by these transformations is a subgroup of T-duality transformations, which leave all physical properties of the CFT invariant. In what follows we will simply refer to these equivalence transformations  as T-equivalences. It should be immediately noted that the full equivalence group also includes cyclic permutations $\sigma_x^i \rightarrow \sigma_y^i \rightarrow \sigma_z^i  \rightarrow \sigma_x^i$, which are not T-equivalences. Therefore, two code CFTs associated with equivalent codes are not necessarily equivalent as CFTs and may have different physical properties. We will see many such examples below. 

It is important to ask whether any other T-duality transformations from ${\rm O}(n)\times {\rm O}(n)$,  besides those mentioned above, can  map  a code theory into a  theory based on an inequivalent code. We show in Appendix \ref{sec:TD} that this is not the case, and therefore any pair of T-dual code theories are equivalent also in the code sense.

In the $c=(\alpha,\beta)$ representation of codewords,  T-equivalences are generated by simultaneous permutations of the components of $\alpha$ and $\beta$ and by exchanges of the $i$-th component of $\alpha$ with the $i$-th component of $\beta$. Using T-equivalences we can bring the code generator matrix \eqref{Gm} to the following simple form. (We also need to  perform linear operations mod 2 with the codewords that change the generator matrix, but do not change the code or the lattice.) First, by using linear operations and permutations, we can bring the $n\times n$ matrix formed by the $\alpha_i$ to the form 
\bea
\left(\begin{array}{c}\alpha_1\\ \dots \\  \alpha_n\end{array}\right) =
\renewcommand{\arraystretch}{1.4}
\left(\begin{array}{c|c}
{\rm I}_{m\times m} & a \\  \hline 
0_{(n-m)\times m}\, & \, 0_{(n-m)\times (n-m)}\end{array}
\right) \label{alphai}
\eea 
where $a$ is some $m \times (n-m)$ binary matrix. The codewords which form the rows of the matrix
\bea
\left(0_{(n-m) \times n },-a^T,{\rm I}_{(n-m)\times (n-m)}\right) \label{nmb}
\eea
are orthogonal to $(\alpha_i,\beta_i)$ with $\alpha_i$ given by \eqref{alphai} and arbitrary $\beta_i$, and therefore they belong to this code. In other words, by an appropriate linear transformation in the algebra mod 2, the last $n-m$ rows of $G^T$ can be brought to the form \eqref{nmb}.
After exchanging the last $(n-m)$ components of $\alpha_i$ with $\beta_i$, and using the last $n-m$ rows to eliminate the last $n-m$ components of the first $m$ rows, we finally transform \eqref{alphai} into an identity matrix, yielding a generator matrix of the form $G^T=\left(\, {\rm I}\, |\, {\rm B}\, \right)$.
This is the ``canonical'' form of the generator matrix, analogous to \eqref{canonical}. 
For a real self-dual code the binary matrix $\rm B$  has zeros on the diagonal and is symmetric. Otherwise it is arbitrary. 
For notational convenience, we prefer to exchange all components of $\alpha$ and $\beta$ to bring the generator matrix to the form 
\bea
G^T=\left(\, {\rm B}\, |\, {\rm I}\,\right). \label{Bform}
\eea 
We will call codes whose generator matrix is of the form \eqref{Bform}, up to multiplication from the left by a non-degenerate $n\times n$ binary matrix,  B-form codes.  
There are (compare with \eqref{n4hr})
\bea
\prod_{j=0}^{n-1} 2^j=2^{n(n-1)/2}
\eea
distinct B-form codes in total,  and any real code has at least one T-equivalent B-form code.

Since any self-dual code is equivalent to a real code, and the generator matrix of any real code can be brought to the form  \eqref{Bform} using additional equivalence transformations, we conclude that any code of type $4^{{\rm H}+}$ is equivalent to a B-form code. This result  has been established in a  different way  in \cite{van2004graphical}.

The generator matrix of $\Lambda(\C)$ associated with the B-form code \eqref{Bform} is
\bea
\Uplambda=\left(\begin{array}{c|c}
2\, {\rm I}\,  &\, \, B\, \,  \\ \hline
0\, &\,  {\rm I}\end{array}\right)/\sqrt{2}, \label{TC}
\eea
where $B_{ij}\in\{0,\pm 1\}$, such that 
\bea
{\rm  B}=B\, \, {\rm mod}\, \, 2.
\eea
There is an ambiguity is choosing signs of $B_{ij}$, but any choice results in the same  lattice. 
Let us choose $B_{ij}$ to be antisymmetric, reducing the sign ambiguity to the simultaneous flips 
$B_{ij}\rightarrow -B_{ij}$, $B_{ji}\rightarrow -B_{ji}$. All such generator matrices are related by 
\bea
\Uplambda \rightarrow \Uplambda\,  {\cal G},\qquad {\cal G}=\left(\begin{array}{c|c}
{\rm I}\,  &\, X \\ \hline
0\, &\,  {\rm I}\end{array}\right)\in {\rm O}(n,n,\Z)\subset {\rm GL}(2n,\Z),
\eea
where $X_{ij}\in\{0,\pm 1\}$, $X=-X^T$. For all  $\Uplambda$ related in this way, the lattice  remains the same.

Comparing \eqref{TC} with \eqref{TCN}, we find that the code theories are toroidal compactifications on the cube of ``unit'' size $2\pi$ with quantized $B$-field flux, as well as their T-duals. Different $B$-fields corresponding  to the same ${\rm B}$ are related by T-duality transformations in ${\rm O}(n,n,\Z)$ (from the denominator of \eqref{Narain}) which preserve the lattice. 

The T-duality transformations which map a code to another code, permutations ${\cal O}_p$ and sign flips ${\cal O}_i$, are the following elements of ${\rm O}(n,n,\Z)$:
\bea
{\cal O}_p  \rightarrow \left(\begin{array}{c|c}
{\cal O}_p & 0\\
\hline 
0 &\, {\cal O}_p\end{array}\right)\in {\rm O}(n,n,\Z),\qquad 
{\cal O}_i\rightarrow \left(\begin{array}{c|c}
{\rm I} -1_{ii} & 1_{ii}\\
\hline 
1_{ii} &\, {\rm I} -1_{ii}\end{array}\right)\in {\rm O}(n,n,\Z), \label{Oi}
\eea  
where $1_{ii}$ is a diagonal matrix with all elements being zero, except for $ii$-th element, which is $1$. Finally, the generator matrix \eqref{TC} can be obtained from the matrix with $B=0$ by a transformation from $ {\rm O}(n,n,\Z)$, 
\bea
\left(\begin{array}{c|c}
2\,{\rm I}\,  &\, B \\ \hline
0\, &\,  {\rm I}\end{array}\right)/\sqrt{2}={\cal F}\left(\begin{array}{c|c}
2\,{\rm I}\,  &\, 0 \\ \hline
0\, &\,  {\rm I}\end{array}\right)/\sqrt{2},\qquad {\cal F}=\left(\begin{array}{c|c}
{\rm I}\,  &\, B \\ \hline
0\, &\,  {\rm I}\end{array}\right)\in {\rm O}(n,n,\Z).
\eea 
Therefore all Construction A lattices $\Lambda(\C)$ can be obtained 
by the action of ${\rm O}(n,n,\Z)$, with the generator matrix  being
\bea
\Uplambda({\cal F})={\cal F}\left(\begin{array}{c|c}
2\,{\rm I}\,  &\, \,  0\,  \\ \hline
0\, &\,  \, {\rm I}\, \end{array}\right)/\sqrt{2} . \label{cosete}
\eea
For any $\cal F$, $\Uplambda({\cal F})$ defines an even self-dual lattice, a sublattice of $(\Z/\sqrt{2})^{2n}$ within $\R^{n,n}$. 

Different ${\cal F}\in {\rm O}(n,n,\Z)$ may result in the same lattice, defining an equivalence class 
within ${\rm O}(n,n,\Z)$:
\bea
{\cal F} \sim {\cal F}\, {\cal H},\quad {\cal H}=\left(\begin{array}{c|c}
{\cal A}\,  &\, {\cal B} \\ \hline
{\cal C}\, &\,  {\cal D}\end{array}\right)\in{\rm O}(n,n,\Z),\quad \quad {\cal B}\,\, {\rm mod}\, \,  2=0.
\eea
The equivalence of the lattices $\Uplambda(\cal F)$ and $\Uplambda(\cal F\, H)$ follows from 
\bea
{\cal H}\left(\begin{array}{c|c}
2\,{\rm I}\,  &\, \,  0\,  \\ \hline
0\, &\,  \, {\rm I}\, \end{array}\right)=\left(\begin{array}{c|c}
2\,{\rm I}\,  &\, \,  0\,  \\ \hline
0\, &\,  \, {\rm I}\, \end{array}\right) {\cal H}',\qquad {\cal H}'=
\left(\begin{array}{c|c}
{\cal A}\,  &\, {\cal B}/2 \\ \hline
2\, {\cal C}\, &\,  {\cal D}\end{array}\right) \in {\rm O}(n,n,\Z).
\eea
Such matrices $\cal H$ form a congruence subgroup within $ {\rm O}(n,n,\Z)$, which we denote ${\rm O}_2(n,n,\Z)$. 
Accordingly, all real codes can be described as a coset 
\bea
{{\rm O}(n,n,\Z)\over {\rm O}_2(n,n,\Z)}. \label{CN}
\eea
The coset description  of the real self-dual codes \eqref{CN} is the analog for code  theories of the full Narain moduli space \eqref{Narain}.

There is a similar coset construction for all self-dual codes, {\it i.e.}~including non-real codes (odd self-dual lattices), 
\bea
{{\rm O}(n,n,\Z_2)\over {\rm O}_2(n,n,\Z_2)}. \label{CN2}
\eea
Here the subgroup $ {\rm O}_2$ includes all binary orthogonal matrices with zero $\cal B$. Equivalence classes \eqref{CN} are mapped into \eqref{CN2}  by reducing mod 2. 
We illustrate the coset construction in case of $n=1$ and $n=2$ codes in Sections \ref{sec:n=1} and \ref{sec:n=2}. 


Most T-duality transformations (we only discuss those which map code theories to code theories) do not preserve the B-form of $\Uplambda$ \eqref{TC}, but  there is a particular set of transformations which do. They
include permutations of $\rm B$,
\bea
\label{Bpermutation}
{\rm B} \rightarrow {\cal O}_p\, {\rm B}\, {\cal O}_p^T,
\eea 
and ``genuine'' T-duality transformations 
\bea
\label{TDB}
\renewcommand{\arraystretch}{1.4}
{\rm B}=\left(
\begin{array}{c|c}
b_{11} &\, b_{12} \\ \hline
b_{12}^T\, &\, \, b_{22}
\end{array}
\right)\quad \rightarrow 
\quad {\rm B}'=\left(
\begin{array}{c|c}
b_{11}^{-1}\, & -b_{11}^{-1} b_{12} \\ \hline
b_{12}^T b_{11}^{-1}\, &\, b_{22}-b_{12}^T\, b_{11}^{-1}\, b^{}_{12}
\end{array}
\right),  \label{elc}
\eea
where $b_{11}$ is an arbitrary nondegenerate submatrix.  It can be written as 
\bea
{\rm B} \rightarrow \left((D+{\rm I}){\rm B}+D\right)\left(D\, {\rm B}+D+{\rm I}\right)^{-1},\label{elcp}
\eea 
where all algebra is mod 2, and $D$ is the diagonal matrix with ones in the diagonal entries associated with $b_{11}$ in \eqref{TDB} and zeros elsewhere.  We  note that the composition of two transformations \eqref{elcp} parameterized by $D_1$ and $D_2$ is again a transformation of the form \eqref{elcp} with $D=D_1+D_2$, consistent with the consecutive action of ${\cal O}_i$ \eqref{Oi} with different $i$. 

\subsection*{Consistency check}
The T-duality transformation \eqref{elc}  of a code theory does not change the CFT partition function, and therefore it should leave the refined enumerator polynomial invariant. For the code $\C$ associated with $\rm B$ it can be written as 
\bea
W_\C=\sum_{\alpha_i} x^{n-{\rm w}(\alpha)} y^{{\rm w}_y(\alpha) } z^{{\rm w}(\alpha)-{\rm w}_y(\alpha) },
\eea
where the sum goes over all values of binary variables $\alpha_i\in\{0,1\}$,  we have introduced auxiliary binary variables $\beta_i$ via $\beta={\rm B}\, \alpha$, and
\bea
{\rm w}=\sum_i (\alpha_i+\beta_i)-{\rm w}_y,\qquad {\rm w}_y=\sum_{i} \alpha_i\, \beta_i. \label{wxy}
\eea
We recognize the transformation of $b_{22}$ in \eqref{elc} as the Schur complement, with the property that the new matrix ${\rm B}'$ satisfies  $(\beta_1,\dots,\beta_m,\alpha_{m+1},\dots, \alpha_n)=(\alpha_1,\dots,\alpha_m,\beta_{m+1},\dots, \beta_n){\rm B'}$, where we have assumed that $b_{11}$ is $m\times m$. In other words, the transformation ${\rm B}\rightarrow {\rm B}'$ swaps $\alpha_i \leftrightarrow \beta_i$  for $i\leq m$, while \eqref{wxy} remains invariant. \\

The binary symmetric matrix ${\rm B}$ with zeros on the diagonal can be interpreted as the adjacency matrix of a graph on $n$ nodes. In this way all B-form codes (and code theories) correspond uniquely to graphs. Exchanging all $\sigma_z$ and $\sigma_x$ maps B-form codes into canonical form, and the stabilizer generators in this case are 
\bea
\g_i=\sigma_x^i \prod_{j=1}^n (\sigma_z^j)^{{\rm B}_{ij}}. \label{graphcode} 
\eea
Stabilizer codes with generators of the form \eqref{graphcode} are called graph codes \cite{schlingemann2001quantum,mackay2004sparse}.
That we can always bring a stabilizer code by means of equivalence transformations (unitary transformations from the Clifford group) to the canonical form with stabilizer generators of the form \eqref{graphcode} is in a nutshell the statement that any stabilizer code is equivalent to a graph code \cite{schlingemann2001stabilizer}. We should also mention that non-self-dual codes, i.e.~$[[n,k,d]]$ codes with $k>0$, also can be represented as graphs with labeled nodes \cite{schlingemann2001quantum}. Returning to self-dual codes, 
the one-dimensional code subspace ${\cal H}_\C$, defined as the state $\psi_\C$ invariant under the action of $\g_i$, {\it i.e.} $g_i\psi_\C=\psi_C$ for all $i$ (see \eqref{Projector}),
\bea
\psi_\C={1\over 2^n} \sum_{\alpha_i=0,1} (-1)^{f(\alpha_i)} |\alpha_1,\dots,\alpha_n\rangle,\qquad f(\alpha_i)={\sum_{i>j} \alpha_i {\rm B}_{ij} \alpha_j}. \label{codesubspace}
\eea
is the so-called graph state \cite{dur2003multiparticle,casati2006quantum}. Many aspects of code theory, including the action of the equivalence group (Clifford transformations), have been discussed in the literature in the context of graph states \cite{van2004graphical, hein2006entanglement}. An alternative language, also used in the literature, is that  of boolean functions $f(\alpha_i)$ \cite{riera2006pivot}.

In terms  of graphs, the permutation \eqref{Bpermutation} is simply the graph isomorphism which relabels the nodes, while \eqref{elcp} describes  
 all possible compositions of edge local complementation \cite{van2005edge}.  Local complementation of a graph $\rm B$ (we associate the graph with the adjacency matrix) with respect to  the node $i$, denoted ${\rm B} * i$,  is a new graph defined as follows. We define the ``neighborhood'' of $i$ as
a subgraph consisting of all nodes $j$ connected to $i$, {\it i.e.}~such that ${\rm B}_{ij}=1$, and the edges between them. The complementation procedure, applied to a (sub)graph, removes all existing links, and connects all pairs of nodes which were not previously connected. At the level of the adjacency matrix this is simply $B_{kl}\rightarrow B_{kl}+1\, \, {\rm mod}\, \, 2$. Local complementation ${\rm B}* i$ is a new graph defined as complementation applied to the neighborhood of $i$. In terms of ${\rm B}$ it can be written as 
\bea
{\rm B}_{kl}\rightarrow {\rm B}_{kl}+{\rm B}_{ik}{\rm B}_{il}+{\rm B}_{ik}\delta_{kl}\, \, {\rm mod}\, \, 2,\qquad k,l\neq i,
\eea
while ${\rm B}_{ii}$ and ${\rm B}_{ij}$ remain unchanged.
Edge local complementation is defined with respect to an edge -- a pair of vertices $(i,j)$ -- as a repeated application of local complementation 
\bea
{\rm B}\rightarrow (({\rm B}* i)*j)*i=(({\rm B}* j)*i)*j. \label{LC}
\eea
Two graphs related to each other by a sequence of isomorphisms and edge local complementation are said to be edge local equivalent. The edge local complementation (ELC) equivalence classes of graphs are therefore in one-to-one correspondence with the classes of physically  equivalent  B-form code theories, i.e.~those related to each other by T-duality.  ELC equivalence classes (which are defined to include isomorphic graphs) have been studied in \cite{danielsen2008edge} as a means to  classify equivalence classes of classical binary codes. Their connection with stabilizer codes and Clifford transformations has also been discussed in \cite{van2005edge,riera2006pivot}.
The number of classes of ELC equivalent  graphs $t_n^{\rm ELC}$ on $n$ nodes is known as the OEIS integer sequence \href{https://oeis.org/A156801}{\tt A156801}, see Table~\ref{table:ELC}. 
\begin{table}
\begin{center}
\begin{tabular}{c|cccccccccccc}
$n$ & 1 &  2 & 3 & 4 &5 & 6 &7 &8 & 9 & 10 & 11 &12\\  \hline  \\[-13pt]
$t_n^{\rm ELC}$ & 1 & 2& 4& 9& 21& 64& 218& 1068& 8038& 114188& 3493965& 235176097\\  \hline \\[-13pt]
$i_n^{\rm ELC}$ & 1 & 1& 2& 4& 10& 35& 134& 777& 6702& 104825& 3370317& 231557290
\end{tabular}
\end{center}
\caption{Number of equivalence classes of graphs under  edge local   complementation (ELC) $t_n^{\rm ELC}$,   for $n\leq 12$. Number of  ELC equivalence classes of  indecomposable graphs $i_n^{\rm ELC}$.}
\label{table:ELC}
\end{table}
As the number of inequivalent graphs grows rapidly, it is more convenient to keep track of indecomposable graphs/codes. The number of  classes of edge local equivalent indecomposable graphs $i_n^{\rm ELC}$  is related  to the full number of equivalence classes $t_n^{\rm ELC}$ via the Euler transform. (We note that the code CFT for a decomposable code is the tensor product of the CFTs associated with the indecomposable codes into which the original code factors.) 

To summarize, B-form codes (graph codes) are in one to one correspondence with graphs, and we will use both languages interchangeably. The T-dualities that transform a code theory into another code theory are necessarily code equivalences; we call them T-equivalences.  If we consider their action restricted to the space of B-form codes, then at the level of graphs, T-dualities are generated by permutations of nodes (graph isomorphisms) and edge local complementations.  In what follows we will simply say that graphs (or B-form codes) are T-equivalent if they belong to the same ELC equivalence class. 

As was mentioned above, T-duality  leaves the refined enumerator polynomial invariant; $W_{\cal C}$ is the same for all codes associated with edge local equivalent graphs. There is another homogeneous polynomial with this same property, {\it i.e.}~it is the same for all graphs belonging to the same ELC equivalence class. This is the interlace polynomial \cite{aigner2004interlace,van2005edge} defined via\footnote{The conventional definition is related to our definition via $q(x)=Q(1,x-1)$.} 
\bea
Q(x,y)=\sum_{w\subseteq \{1,\dots,n\}}   x^{n-s({\rm B}[w])} y^{s({\rm B}[w])},\qquad s(X):={\rm dim}({\rm Ker}(X)).
\eea
Here ${\rm B}[w]$ denotes the submatrix of ${\rm B}_{ij}$ for $i,j\in w$. The kernel ${\rm Ker}(X)$ of a binary matrix is understood to be with respect to mod 2 algebra.

Since any code theory can be brought  to the B-form using T-duality, and the interlace polynomial would be the same no matter which B-form representative we choose, the interlace  polynomial is a  proper characteristic of the code CFT.  Explicit examples of the interlace polynomial will be given in Section 
\ref{sec:firstn}. 

In the beginning of this section we mentioned that cyclic permutations of $\sigma_{x,y,z}$ (multiplying by $\omega$ in the language of ${\rm GF}(4)$ codes) are not T-dualities. 
For simplicity we consider $n=1$ and multiplication by $\omega$. The action of $\omega$ on $(\alpha,\beta)$  can be written as  
\bea
(\alpha,\beta) \rightarrow (\alpha,\beta)\left(\begin{array}{cc}
0 & 1\\
1 & 1
\end{array}\right),\label{cyclic}
\eea
where algebra is mod 2. This action automatically extends to the code lattice $\Lambda(\C)$.   Provided that we start with an even self-dual Lorentzian $\Lambda(\C)$,  the new lattice will be self-dual but may not be even. Thus, cyclic permutations, in general, do not preserve the property of codes being real.  Combining cyclic permutations of different components with exchanges of the $i$-th components of $\alpha$ and $\beta$ generated by ${\cal O}_i$ \eqref{Oi}, one can occasionally find transformations which transform  a B-form code into another  B-form code. The orbit of all B-form codes related to a given one via cyclic permutations of $\sigma_{x,y,z}$ and exchanges $\sigma_x\leftrightarrow \sigma_z$ is equivalent, at the level of graphs, to the orbit with respect to consecutive actions of local complementation (LC) \eqref{LC} \cite{van2004graphical,glynn2004geometry,hein2006entanglement}. 
Two graphs related to each other by a sequence of isomorphisms and local complementations are called LC equivalent. Local complementation equivalence classes of graphs are therefore in one-to-one correspondence with classes of  equivalent codes, {\it i.e.}~those related to each other by the Clifford group (also called the local Clifford group in the literature). LC equivalence classes (which are defined to include isomorphic graphs) have been studied in \cite{bouchet1993recognizing,hohn2003self,glynn2004geometry,hein2006entanglement,danielsen2006classification}, in particular to  classify equivalence classes of self-dual quantum stabilizer codes (or, equivalently, graph states).  The number of classes of LC-equivalent  graphs $t_n^{\rm LC}$  on $n$ nodes is known as the OEIS integer sequence \href{https://oeis.org/A094927}{\tt A094927}, see Table~\ref{table:LC}, \cite{danielsen2006classification}. 
\begin{table}[t]
\begin{center}
\begin{tabular}{c|cccccccccccc}
$n$ & 1 &  2 & 3 & 4 &5 & 6 &7 &8 & 9 & 10 & 11 &12\\  \hline  \\[-13pt]
$t_n^{\rm LC}$ & 1 & 2 &3 & 6 &11 & 26 & 59 & 182 & 675 & 3990 & 45144 & 1323363\\  \hline \\[-13pt]
$i_n^{\rm LC}$ & 1 & 1 & 1& 2& 4& 11& 26& 101& 440& 3132 & 40457 & 1274068
\end{tabular}
\end{center}
\caption{Number of classes of graphs equivalent under local complementation (LC) $t_n^{\rm LC}$,  for $n\leq 12$. Number of LC equivalence classes of indecomposable graphs $i_n^{\rm LC}$.}
\label{table:LC}
\end{table}

If two codes are equivalent in the code equivalence sense, but not related by T-equivalence (T-duality transformations), we call them ``C-equivalent,'' where the C stands for cyclic permutations of $\sigma_{x,y,z}$. C-equivalent codes necessarily share the same enumerator polynomial but usually have different refined enumerators.  The code CFTs associated with C-equivalent codes are generally physically distinct. 
At the level of graphs, C-equivalent B-form codes correspond to the graphs related by LC, but not by ELC.

The role played by ELC graph equivalence in determining the physical equivalence of the corresponding code theories 
motivates us to classify ELC classes within the LC classes of graphs that correspond to equivalent codes. To our knowledge such a classification has not previously been performed.  We provide a full classification for graphs on up to $n\leq 8$ nodes, obtained with help of computer algebra, in  Appendix~\ref{sec:graphs}.

The relation between quantum stabilizer codes and 2d CFTs outlined in this section is only one particular aspect of what is likely a much richer story. Given the role classical codes play in the context of chiral CFTs, we can essentially take for granted that quantum codes can be used to define non-chiral vertex operator algebras, a subject we leave for future investigation. Here we only briefly comment on the recent work \cite{harvey2020moonshine}, which establishes a  relation between the Hexacode, understood as the quantum stabilizer code, and  a particular SCFT. The SCFT in question, 
the GTVW theory \cite{gaberdiel2014k3}, has chiral vertex operators of dimension $3/2$ parametrized by vectors $\vec{k}\in \R^6$ with all components being half-integer, $k_i=\pm 1/2$. These vertex operators can be associated with the ket vectors of the Hilbert space of the Hexacode, $\vec{k} \rightarrow |(k_1+1/2)\dots (k_6+1/2)\rangle$, such that any linear combination in the Hilbert space is mapped to a linear combination of vertex operators. Harvey and Moore show that the code subspace $\psi_\C$, defined via an analog of   \eqref{codesubspace} (\cite{harvey2020moonshine} uses a code equivalent to  Hexacode \eqref{hexacode}, and therefore the analog of \eqref{codesubspace}  includes imaginary coefficients), is mapped to the special vertex operator, the ${\mathcal N}=1$ supercurrent. 
They conjecture that other ${\mathcal N}=1$ SCFTs are related to other stabilizer codes. 

There is a particular technical aspect emphasized in \cite{harvey2020moonshine}. The expression for the code state $\psi_\C=\mathds{P}|0\rangle$ with $\mathds{P}$ given by \eqref{Projector} exists for any stabilizer code, but it depends on the choice of $n$ generators $\g_i$. Choosing different combinations of the $\g_i$ as generators may result in a different  $\psi_\C$. This is because 
$\g(c)$ \eqref{g} understood as a map from codewords $c=(\alpha,\beta) \in {\rm GF}(4)^n$ to generators
is not a representation, but a projective representation 
\bea
\g(c_1)\g(c_2)=\epsilon(c_1,c_2)\g(c_1+c_2),\qquad c_1,c_2 \in {\rm GF}(4)^n.
\eea
The cocycle $\epsilon(c_1,c_2)=\pm 1$ is in general nontrivial, but in the example considered in 
\cite{harvey2020moonshine} it vanishes,  $\epsilon(c_1,c_2)=1$. Here we point out this is not a unique situation, and in fact other codes also have a vanishing cocycle, with an appropriate choice of the map from ${\rm GF}(4)$ to the group of Pauli matrices. Let us choose 
\begin{eqnarray}
c&=&(0,0) \rightarrow {\rm I},\quad \qquad \, \, c=(1,0) \rightarrow i^p \sigma_x,\\
c&=&(0,1) \rightarrow i^q \sigma_z,\qquad c=(1,1)\rightarrow i^{r} \sigma_y,
\end{eqnarray}
where $p,q,r$ are integer numbers between $0$ and $3$. Then the
coefficient $\epsilon(c)$ in \eqref{g} is equal to $\epsilon=i^{p\, {\rm w}_x(c)+q\, {\rm w}_z(c)+r\,  {\rm w}_y(c)}$. It should be real, which is a consistency condition on the code and $p,q,r$. For the cocycle to vanish, $\g(c_1)\g(c_2)=\g(c_1+c2)$, 
\bea
\nonumber
p\, ({\rm w}_x(c_1)+{\rm w}_x(c_2)-{\rm w}_x(c))+q\, ({\rm w}_z(c_1)+{\rm w}_z(c_2)-{\rm w}_z(c))+\\ r\, ({\rm w}_y(c_1)+{\rm w}_y(c_2)-{\rm w}_y(c)) =2 \beta_2\, \alpha_1\, \, {\rm mod}\, \, 4,
\label{nococycle}
\eea
where $c=(c_1+c_2) \, \, {\rm mod}\, \, 2$.  This condition is symmetric under $c_1 \leftrightarrow c_2$ because of \eqref{commute}, and should hold for any two codewords $c_1,c_2\in \C$.
When it holds the stabilizer group is a genuine representation of the code $\C\subset {\rm GF}(4)^n$, understood as an abelian group under addition. In general this condition is not invariant under code equivalence transformations.  Focusing on real self-dual codes, we have verified that all codes with $n=2,3$ satisfy this condition for some $p,q,r$. For $n=3$, $24$ out of $30$ codes, and for $n=4$, $103$ out of $270$ codes satisfy \eqref{nococycle}.

\section{Bounds, averaging over codes, and holography}\label{sec:GV}
One of the central questions of coding theory is how well one can protect (quantum) information when the number of qubits $n$ goes to infinity. In the case of self-dual stabilizer codes this is the question of determining the largest possible ratio $d/n$ in the limit $n\rightarrow \infty$, where $d$ is the maximal achievable Hamming distance.  

The quantum Hamming bound \eqref{QHB} readily provides an  upper limit\footnote{Self-dual codes are detection codes, so they are automatically non-degenerate.} 
\bea
\label{qhbp}
{d\over n}\leq 2p_q^*,\quad n\rightarrow \infty, \qquad H(p^*)=\ln(2)-p_q^* \ln(3),\quad p_q^*\approx  0.1893,
\eea
but it is known to be conservative. A stronger upper bound was found by  Rains in \cite{rains1999quantum} by analytically treating the linear programming constraints,
\bea
\label{Rains}
d\leq 2\left[{n\over 6}\right]+2+\delta_{n\equiv 5\, \, ({\rm mod}\, \, 6)},\qquad \Rightarrow\qquad  {d\over n}\lesssim 0.33,\quad n\rightarrow \infty.
\eea
Further improvements in the asymptotic bound for $d/n$ are possible \cite{rains2003new}.

\begin{figure}
\includegraphics[width=\textwidth]{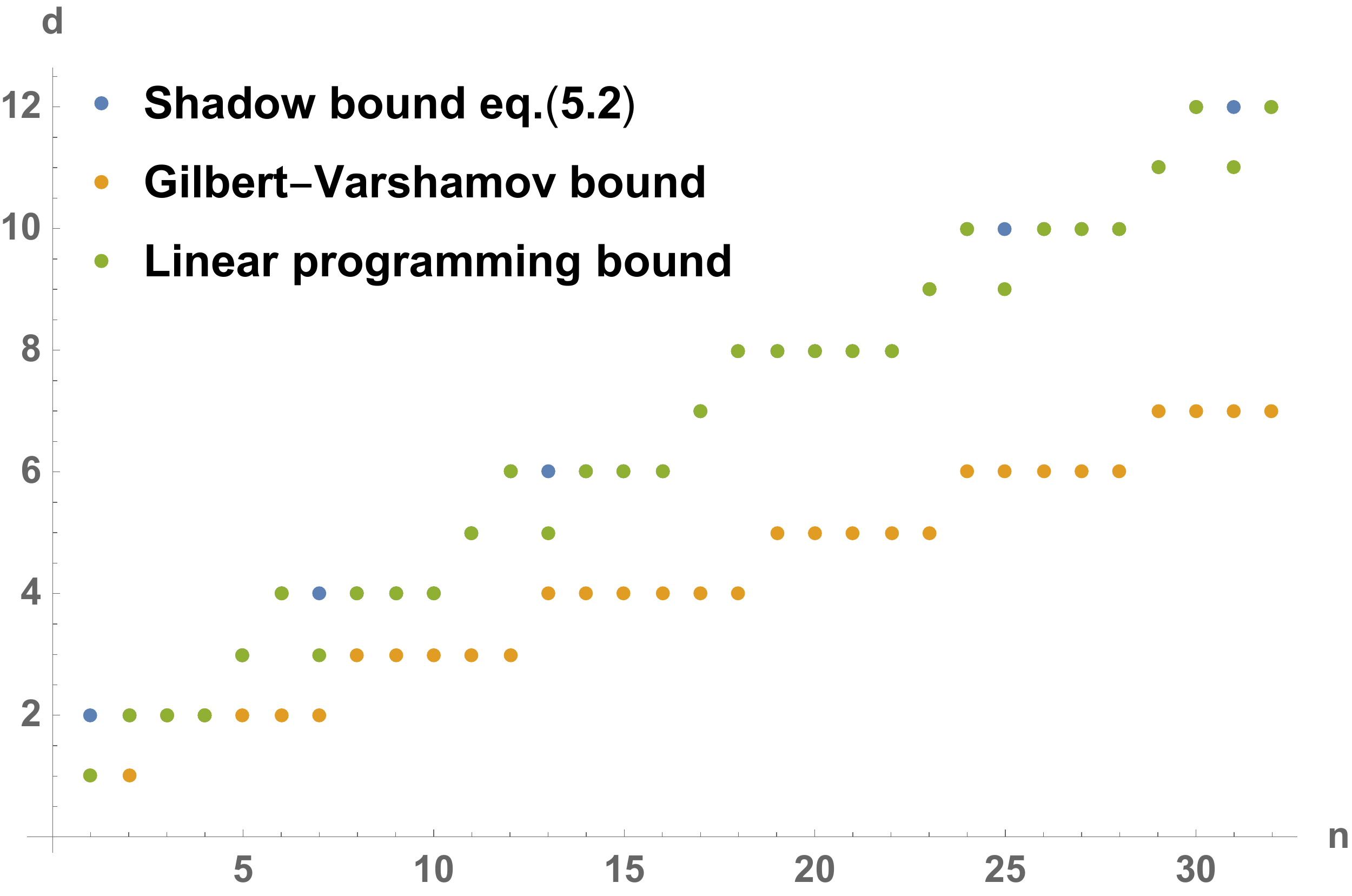}
\label{lpbound}
\caption{Upper and lower bounds on the maximal Hamming distance for self-dual stabilizer  codes. Blue: ``shadow'' upper bound  \eqref{Rains} found by Rains \cite{rains1999quantum}. It  is  an analytic constraint following  from the linear programming bound. Green:  linear programming upper bound solved numerically for $n\leq 32$. For $n\leq 18$ it coincides with the actual maximal $d$ for the given $n$, but it is known to be conservative for $n=19$.
Yellow:  Gilbert-Varshamov lower bound  obtained from the refined enumerator polynomial averaged over all $\rm B$-form codes \eqref{REPaverage}.}
\end{figure}

Our first task in this section will be to obtain linear programming bounds on $d$ numerically, for $n\le 32$.  
To illustrate the main idea of our approach we consider the following problem:  to find a homogeneous polynomial $W(x,y,z)$ of degree $n$, invariant under the duality transformation \eqref{dual}, with all coefficients being integer and non-negative, and satisfying $W(1,0,0)=1$. We additionally want to maximize $d$ over the set of such polynomials, which can be formulated as the linear programming optimization (or feasibility) problem
\bea
\left.\sum_{l=1}^{d-1}  {\partial^l W(1,y,y)\over (\partial y)^l}\right|_{y=0}=0. \label{lpc}
\eea
This is a slight modification of the linear programming bound considered in \cite{calderbank1998quantum}, where the feasibility of enumerator polynomials $W(x,y)$  was considered.\footnote{One of the important ingredients in the analysis of \cite{calderbank1998quantum} and  \cite{rains1999quantum} was the additional condition that the coefficients of the so-called shadow enumerator be integer and positive. This condition is automatically satisfied for real self-dual codes, and therefore we do not discuss it here.} We find that considering the feasibility of refined enumerators somewhat  strengthens the bound, which mostly follows \eqref{Rains} except for certain values of $n\, \, {\rm mod}\, \,  6=1$. For $n\leq 32$ the  results are shown in Fig.~\ref{lpbound}. Comparing with known results for $n\leq 30$ \cite{rains2002self,nebe2006self}, we find that the linear programming  bound is mostly tight, meaning that the maximal  $d$ for which the linear programming problem is feasible is also achievable by a code (or potentially many codes) with that value  of $d$, with at least one known exception when $n=19$. In the latter case our linear programming  bound gives $d\leq 8$, while no self-dual codes with $d=8$ exist.\footnote{We thank E.~Rains for a discussion on this point.} Even when extremal codes, {\it i.e.}~codes with $d$ saturating the bound exist, there are usually many other ``fake'' refined enumerator polynomials which satisfy  the linear programing optimization constraints and have the same $d$. 

The linear programming bound discussed above can be thought of as a toy version of the conformal modular bootstrap
\cite{Hellerman,hellerman2011bounds,keller2013modular,friedan2013constraints,qualls2014bounds,
hartman2014universal,qualls2015universal,ashrafi2016non,kim2016reflections,lin20172,anous2018parity,collier2018modular,afkhami2019fast,
cho2019genus,hartman2019sphere,afkhami2020high,afkhami2020free,gliozzi2020modular},
which aims to establish universal bounds on the spectral gap and other similar properties of the 2d theories. Here we are essentially restricting our analysis to the subset of code theories defined in the previous section. Then the partition function is fully specified by the refined enumerator, which reduces the nontrivial modular bootstrap analysis to a simple linear programing problem in the space of invariant polynomials. 

Linear programming bounds are not constructive. In practice, they may be used to produce invariant polynomials with the desired properties, but verifying if there is an actual code associated with this polynomial is a difficult task.  While various shortcuts are possible, the only universal way to make sure the polynomial is not ``fake'' is to  construct the code, which is exponentially difficult.  

There is also a nonconstructive Gilbert-Varshamov bound for quantum codes, which bounds  maximal $d$ from below. Similarly to the classical case, to obtain the bound we calculate the refined enumerator polynomial averaged over all $\rm B$-form codes, i.e.~codes specified by the generator matrix \eqref{Bform} with all $2^{n(n-1)/2}$ possible matrices $\rm B$, 
\bea
\label{REPaverage}
\overline{W}(x,y,z)=x^n+{(x+y+2z)^n\over 2^n}+{(x-y+2z)^n\over 2^n}-2{(x+z)^n\over 2^n}.
\eea
$\overline{W}$ is manifestly invariant under the duality transformation \eqref{dual}, as well as $y\rightarrow -y$. Taking $z=y$ reduces it to the averaged enumerator polynomial, from which we immediately conclude that for $n\rightarrow \infty$ the maximal $d$ is bounded from below by $d/n\geq p_q^*$, which is twice smaller than the upper bound \eqref{qhbp}, as expected. For a  general $n$ we find the maximal $d$ to be equal to or larger than the maximal value $d_{\rm GV}$ for which the  following constraint is satisfied,
\bea
\sum_{l=1}^{d_{\rm GV}-1}\, {3^l-1\over 2^n}{n!\over l! (n-l)!} <1. 
\eea
This is a somewhat stronger bound than the conventional Gilbert-Varshamov bound which would naively follow  from \eqref{Vq}. Similar lower bounds can in principle be obtained by averaging over any class of codes for which this averaging is feasible.  Averaging over $B$-form codes proves to be a convenient choice. We plot the numerical values $d_{\rm GV}(n)$  for  $n\leq 32$ in Fig.~\ref{lpbound}. Even though the Gilbert-Varshamov bound is asymptotically weaker than the linear  programming upper bound (and presumably the actual maximal value of $d$), there is no known systematic construction for producing codes with $d\geq d_{\rm GV}$  for arbitrarily large $n$. 

The averaged refined enumerator polynomial, via \eqref{ZcodeCFT}, can be interpreted as the averaged  partition function of all $\rm B$-code theories. In light of recent results relating the average over Narain CFTs to $U(1)^n\times U(1)^n$ Chern-Simons theory in AdS$_3$ \cite{afkhami2020free,maloney2020averaging}, it is natural to ask if the averaged code  theory may have a holographic interpretation. To see if a weakly coupled bulk description is possible, we would like to calculate the spectral gap of $U(1)^n$ primaries. Following \cite{hartman2014universal} we define the spectral gap as the value of $\Delta$ for which the density of primary states $\rho(\Delta)$ assumes its asymptotic form. This might be different from the dimension of the lightest nontrivial primary. Primaries of Narain CFTs correspond to vectors in the Lorentzian lattice, and their dimension is proportional to the Euclidean norm-squared $\Delta=\ell^2/2$. 
For vectors associated with codewords, $\ell^2=d_{\rm b}/2$, where the binary Hamming weight is $d_{\rm b}(c):={\rm w}_z(c)+2{\rm w}_y(c)$. The binary Hamming distance of a code (minimal weight of all non-trivial codewords) is the conventional Hamming distance of a classical binary $[2n,n,d_{\rm b}]$ isodual code defined from $\C$ via the Gray map. 
For large $n$ and sufficiently large $\Delta \gg n$, the density of vectors of a unimodular lattice $\Lambda(\C)\subset \R^{2n}$ is given by the volume of a $(2n-1)$-dimensional sphere, yielding
\bea
\label{density}
\rho(\Delta)d\Delta ={(2\pi)^n \Delta^{n-1}\over \Gamma(n)} d\Delta.
\eea 
For a lattice $\Lambda(\C)$ associated with a stabilizer code, all points of the form $\sqrt{2}(a,b)$,  $a,b\in \Z^n$ belong to the lattice and for sufficiently large $\Delta$ their contribution to the density  is given by \eqref{density} divided by $1/2^n$.
When $\Delta$ increases such that the sphere of radius $\ell^2=\Delta/2$ includes new codewords, the overall coefficient grows with each codeword eventually contributing $1/2^n$, until the overall coefficient saturates at one for $\Delta \gg n$. 
The spectral gap can be defined as the value of $\Delta$, for which the coefficient in front of \eqref{density} becomes of order one. This is the value of $d_{\rm b}$ for which coefficients  of the averaged enumerator polynomial $\overline{W}_\C(x^2,y^2,xy)$ become of order one, 
\bea
4{\Delta\over n}\equiv{d_{\rm b}\over n}=p^*,\qquad n\rightarrow \infty. \label{spectralgap}
\eea
This is exactly the Gilbert-Varshamov bound for binary self-dual codes, see Section \ref{sec:binary}. This suggests the following interpretation. The class of binary self-dual codes obtained through the Gray map from $\rm B$-form codes is a good representation, in the statistical sense, of all self-dual codes.  At the same time, isodual codes obtained from $\rm B$-form codes either share a similar distribution of Hamming distances with self-dual codes, or their overall number is much smaller than the number of self-dual codes.  

Since the spectral gap \eqref{spectralgap} scales linearly with the central charge $n$, we expect the corresponding averaged theory to be holographic. We leave the task of understanding the gravity dual theory for the future, while here we consider the simpler case of chiral theories and speculate about their possible gravity dual description. 
The chiral analogue of the average over the Narain lattices would be the average over even self-dual Euclidean lattices. The averaged theta-function is known to be given  by the Eisenstein series $E_{n/2}(\tau)$ \cite{nebe2006self}, such that the averaged partition function of the corresponding chiral CFTs is 
\bea
\overline{Z}_{\rm CFT}(\tau)={E_{n/2}(\tau)\over \eta^n(\tau)}. \label{ZCFTaveraged}
\eea  
For $n$ divisible by  $24$, \eqref{ZCFTaveraged} is  modular invariant. Otherwise, since $n$ is divisible by $8$,  it is invariant under the subgroup of ${\rm PSL}(2,\Z)$ generated by  $\tau\rightarrow \tau+3$ and $\tau\rightarrow -1/\tau$. Using the conventional representation for the Eisenstein series we can rewrite \eqref{ZCFTaveraged} as
\bea
\overline{Z}_{\rm CFT}(\tau)=\sum_{\gamma\in \Gamma_{\infty}\backslash {\rm SL}(2,\Z)} {1\over \eta^n(\gamma \tau)},  \label{holography}
\eea
for $n$ divisible by $24$ 
and interpret this sum as a sum over handlebodies, with $1/\eta^n(\tau)$ being the partition function of $U(1)^n$ Chern-Simons on thermal AdS$_3$ geometry,  parametrized  by  the modular parameter $\tau$ of the boundary torus. This  holographic interpretation is schematic, and similarly to  the  non-chiral case \cite{afkhami2020free,maloney2020averaging} requires further checks and clarifications. Furthermore, if $n$ is not divisible by $24$, additional degrees of freedom in the bulk, possibly  in the form of a $\Z_3$ gauge field, would be necessary to make the sum in \eqref{holography} well defined. We overlook  these important nuances, as our goal here is to understand how averaging specifically over code CFTs would change the story.  In  the chiral case, we would consider even self-dual lattices $\Lambda(\C)$ associated with doubly-even self-dual binary codes. Their averaged theta-function is given by (\ref{averagedW},\ref{WtoT}), which differs from $E_{n/2}(\tau)$ by an appropriate modular form. For simplicity we consider  the case $n=24$, for which   
\bea
\overline{Z}_{\rm codes}={a_{12} E_{12}(\tau)+a_{6^2} E_6^2(\tau) \over \eta^{24}(\tau)},\qquad a_{12}+a_{6^2}=1, \label{Zc}
\eea
with the values of $a_{12}$ and $a_{6^2}$ being unimportant. This  expression can be represented in a way similar  to \eqref{holography},
\bea
\overline{Z}_{\rm codes}=\sum_{\gamma\in \Gamma_{\infty}\backslash {\rm SL}(2,\Z)} {a_{12}+a_{6^2} E_6(\gamma\tau)\over \eta^{24}(\gamma \tau)}.  \label{holography2}
\eea
Again, the term $1/\eta^{24}(\tau)$ can be interpreted as the holographic contribution of $U(1)^n$ gauge fields, while $E_6$  in the numerator suggests presence of  non-abelian gauge fields in the bulk, which  are known to produce  certain combinations of Jacobi theta-functions \cite{maloney2020averaging}. 

Our considerations  do not prove that the averaged code theories have weakly coupled holographic duals, but they indicate that such an interpretation may be possible. We end  this discussion with a few concrete questions. First,  the representation of a modular form as a polynomial in terms of $E_4, E_6$ is not unique.  Instead of  $a_{12} E_{12}+a_{6^2} E_6^2$ in \eqref{Zc} we can write the most general expression $a_{12} E_{12}+a_{6^2} E_6^2+ a_{4^3} E_4^3$, and correspondingly the numerator  in \eqref{holography2} will become $a_{12}+a_{6^2} E_6+b_{4} E_4+b_{4^2} E_4^2$ with  arbitrary $b_{4}, b_{4^2}$ satisfying $b_{4}+b_{4^2}=a_{4^3}$. Thus the holographic partition function for $n=24$ can be written in many different ways, suggesting there are different microscopic descriptions in the bulk, related by  dualities. For larger $n$ there would be more representations and potentially a larger duality  web  in the bulk.  

Even more intriguing is the chiral case of $n=8$, for which there is a unique even self-dual lattice $E_8$. Modulo the important subtlety that to make the sum over ${\rm SL}(2,\Z)$ well-defined, additional degrees of freedom in the bulk would be needed, the partition function $E_8/\eta^8(\tau)$ similarly admits a holographic interpretation. It therefore provides a setting to study from the CFT side interpretation of the Euclidean wormhole geometries when the boundary is not connected.

\begin{figure}
\includegraphics[width=0.95\textwidth]{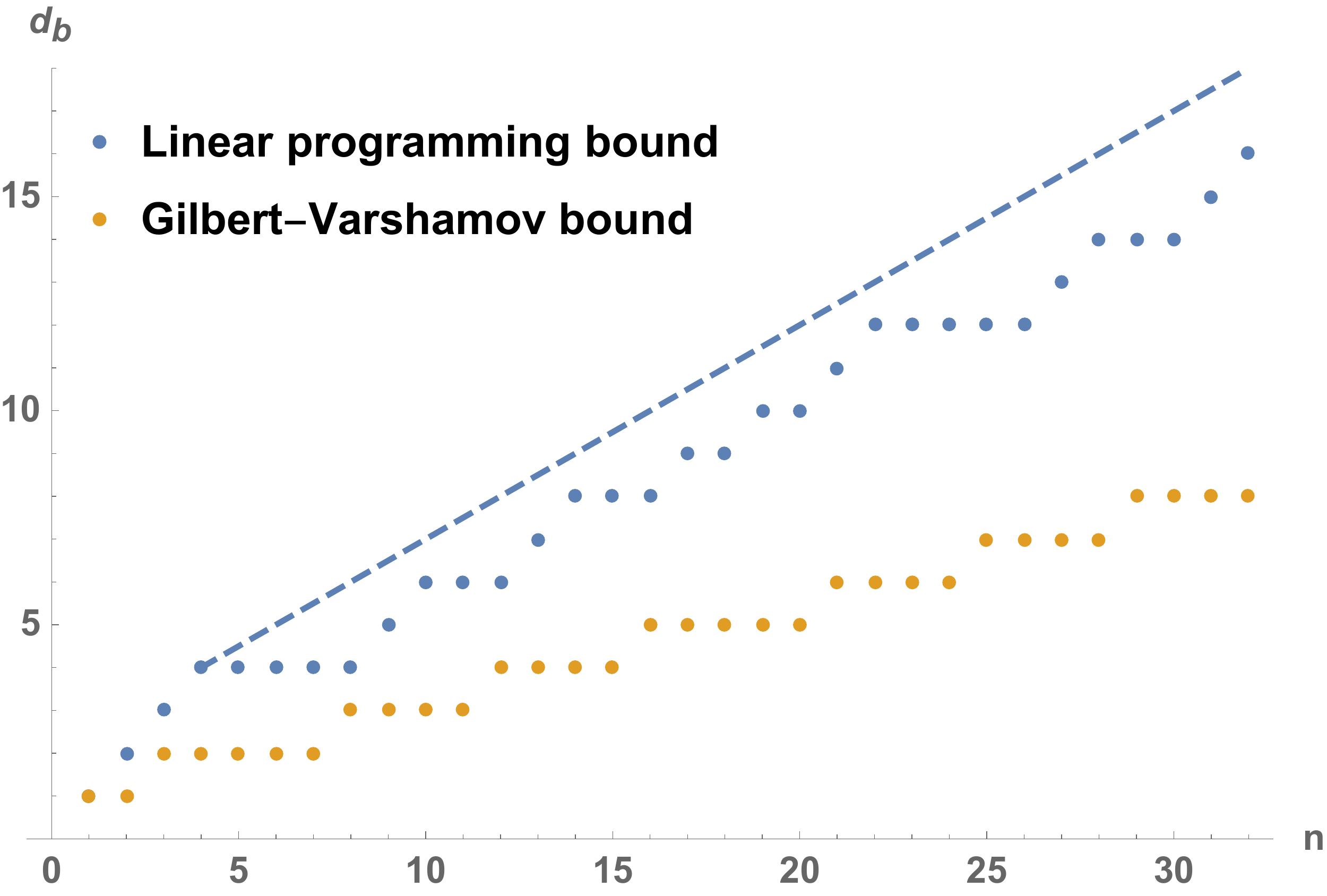}
\caption{Blue points: linear programming upper bound on binary Hamming distance $d_{\rm b}$ of real self-dual codes. It is tight for small but presumably not all $n$. Dashed blue line: $n/2+2$ is an approximate theoretical fit of the numerical bootstrap constraint for the maximal value of spectral gap, measured in units of $d_{\rm b}=4\Delta$ \cite{afkhami2020free}.  Yellow points: Gilbert-Varshamov  lower bound  on binary Hamming distance obtained from \eqref{REPaverage}.}
\label{lpbound-binary}
\end{figure}

To conclude this section, we return back to the question of the spectral gap, which we now define strictly as the dimension of the lightest non-trivial primary. It can be equivalently defined as the length-squared of the shortest non-trivial vector of the Narain lattice, and in this way the maximal value of the spectral gap is related to the efficiency of lattice sphere packing in a given dimension. This question has been recently studied numerically in \cite{hartman2019sphere,afkhami2020high,afkhami2020free}. Similarly to classical binary codes, which give rise to optimal sphere packings in $8$ and $24$ dimensions and provide valuable insight about scaling for large $n$, one may expect quantum codes to occasionally saturate the spectral gap bounds and inform the large-$n$ behavior.  In terms of real stabilizer codes, maximizing the spectral gap is equivalent to maximizing the binary Hamming distance for a given $n$. The question of finding a quantum code with maximal $d_{\rm b}$ for a given value of $n$ is very similar to the conventional question of finding the ``best'' code with largest $d$, but to our knowledge it was not previously discussed in the literature.
In Fig.~\ref{lpbound-binary} we plot the linear programming bound on $d_{\rm b}$ obtained by imposing constraints similar to \eqref{lpc}. The bound is tight at least for $n\leq 8$. Superimposed with the approximate theoretical fit of the numerical spectral gap bound \cite{afkhami2020free},  translated into units of 
$d_{\rm b}=4\Delta\lesssim n/2+2$  (dashed line in Fig.~\ref{lpbound-binary}), we find that both exhibit approximately linear growth with a similar slope. We leave it as an open problem to find the analytic analogue of \eqref{Rains} for the binary Hamming distance, or at least its asymptotic behavior for large $n$. 

By comparison with the case of classical codes and Euclidean lattices, we may expect quantum codes to yield Narain CFTs with the maximal spectral gap for certain special values of $n$, in particular for $n=4$ and $n=12$. This is indeed the case for $n=4$, as is evident from Fig.~\ref{lpbound-binary} and discussed in Section \ref{sec:n=4}. For $n=12$ code theories fall short of  saturating the spectral gap bound, which is discussed in more detail in Section \ref{sec:n=12}.

In the discussion above we identified the spectral gap (length of the shortest vector) with the binary Hamming distance of a code. This is correct for $d_{\rm b}\leq 4$, but for larger $d_{\rm b}$ there are always lattice vectors of the form $\sqrt{2}(\pm 1,0^{2n-1})$ which are shorter. This is the same problem, discussed in Section \ref{sec:binary}, which precludes classical binary codes from yielding efficient sphere packings in large dimensions, at least directly via Construction A. For small $n$ this problem can be partially solved by applying the shift procedure to the Construction A lattice, which will remove unwanted short vectors $\sqrt{2}(\pm 1,0^{2n-1})$. A similar strategy can be employed in the quantum case,
making it possible to relate codes with larger $d_b$ to CFTs with larger spectral gap. We consider an explicit example of a lattice with shortest vector controlled by $d_{\rm b}> 4$ in Section \ref{sec:n=12}.

\section{Enumeration of self-dual codes with small $n$}
\label{sec:firstn}
In this section we discuss many explicit examples of self-dual stabilizer codes for $n\leq 12$. As the number of codes rapidly grows with $n$, we emphasize different points for different $n$. For $n=1$ we discuss all codes in detail. For $n=2$ we  focus on real codes, discuss them in detail and then illustrate the coset construction. Starting from $n=3$ we restrict our attention to B-form codes, and for $n=3,4$  go over all classes of T-equivalent codes. For $n=3,4$ we also explicitly write down all ``fake'' enumerators. 
Starting from $n\geq  5$  we only consider codes with maximal values of $d$ and/or $d_{\rm b}$. For $n=7,8$ we give explicit examples of non-equivalent codes with the same refined enumerator polynomials, giving rise to groups of physically distinct isospectral code CFTs. For $n=4,6,12$ we give examples of codes related to special lattices.

\subsection{$n=1$}\label{sec:n=1}
For $n=1$ there are  three codes, see \eqref{n4h+},  specified by the unique stabilizer generator
\bea
\g={\rm X,\, or\, \,  Y,\, or\, \,  Z}. \label{n=1}
\eea
Obviously all three codes are equivalent (in the sense of the code equivalence group).
Two codes, see \eqref{n4hr}, the first and third, are real and correspond to code CFTs. The first code $\g={\rm X}$ is a B-form code  \eqref{TC} with $\rm B=0$. This is the only B-form code for $n=1$. The corresponding graph is simply a graph consisting of one vertex.  The corresponding CFT is a boson on a circle of radius $R=1$. 
The third code $\g={\rm Z}$ is T-dual to the first one. This is a boson compactified on a circle of radius $R=2$. Its lattice generator matrix is $\Uplambda={\rm diag}(1,2)/\sqrt{2}$.  The refined enumerator polynomial of these two  codes is,  {\it cf.}~\eqref{invpols},
\bea
W_1=x+z.
\eea
Since all three codes \eqref{n=1} are equivalent, they share the same enumerator polynomial $W(x,y)=x+y$. The Hamming distance of all three codes is $d=1$, see Fig.~\ref{lpbound}.  The binary Hamming distance of real codes $d_{\rm b}=1$, in  agreement with the spectral gap of the compact scalar on a circle of radius $R=1$ (or $2$), $\Delta=d_{\rm b}/4=1/4$. The interlace polynomial of the graph with ${\rm B}=0$ is $Q_1=x+y$.

We would like to see how the coset description \eqref{CN} works in the case $n=1$. The group ${\rm O}(1,1,\Z)$ consists of four matrices
\bea
\pm \left(\begin{array}{cc}1 & 0 \\ 0 & 1
\end{array}\right),\quad \pm \left(\begin{array}{cc} 0 & 1 \\ 1 & 0
\end{array}\right).
\eea 
The first two matrices form the subgroup ${\rm O}_2(1,1,\Z)$. The coset \eqref{CN} includes two elements, with representatives 
\bea
{\cal F}=\left(\begin{array}{cc}1 & 0 \\ 0 & 1
\end{array}\right),\quad  \left(\begin{array}{cc} 0 & 1 \\ 1 & 0
\end{array}\right). \label{classes}
\eea
The resulting lattice generator matrices \eqref{cosete} correspond to two $n=1$ codes, with stabilizer generators $\g={\rm X}$ and $\g={\rm Z}$ correspondingly.

The coset description of all codes \eqref{CN2} is equally straightforward. The group ${\rm O}(1,1,\Z_2)$ includes six elements, 
\bea
\left(
\begin{array}{cc}
 1 & 0 \\
 0 & 1 \\
\end{array}
\right),\quad 
\left(
\begin{array}{cc}
 1 & 0 \\
 1 & 1 \\
\end{array}
\right),\quad 
\left(
\begin{array}{cc}
 1 & 1 \\
 0 & 1 \\
\end{array}
\right),\quad 
\left(
\begin{array}{cc}
 0 & 1 \\
 1 & 1 \\
\end{array}
\right),
\quad 
\left(
\begin{array}{cc}
 0 & 1 \\
 1 & 0 \\
\end{array}
\right),\quad 
\left(
\begin{array}{cc}
 1 & 1 \\
 1 & 0 \\
\end{array}
\right), \label{O2}
\eea
with the first two forming the subgroup ${\rm O}_2(1,1,\Z_2)$. Multiplication by the only non-trivial element of ${\rm O}_2(1,1,\Z_2)$ permutes the matrices in \eqref{O2} as follows
\bea
1\leftrightarrow 2,\quad 3\leftrightarrow 4,\quad 5\leftrightarrow 6.
\eea
Therefore matrices 1, 3, and 5 can be chosen as class representatives (cf.~\eqref{classes}). They correspond to $\g={\rm X}, {\rm Y}, {\rm Z}$.

\subsection{$n=2$}\label{sec:n=2}
\begin{figure}
\begin{minipage}{.5\textwidth}
\centering
\includegraphics[]{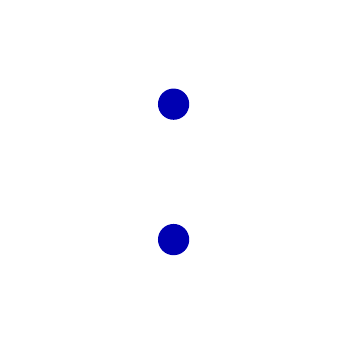}
\begin{eqnarray}
\nonumber
{\rm B}=\left(\begin{array}{cc}
0 & 0 \\
0 & 0
\end{array}\right)
\end{eqnarray}
\end{minipage}
\begin{minipage}{.5\textwidth}
\centering
\includegraphics[]{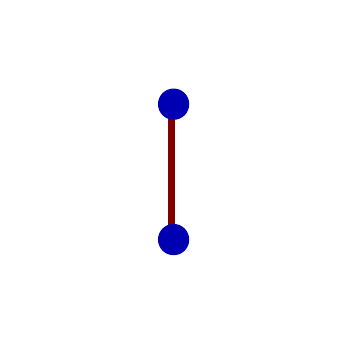}
\begin{eqnarray}
\nonumber
{\rm B}=\left(\begin{array}{cc}
0 & 1 \\
1 & 0
\end{array}\right)
\end{eqnarray}
\end{minipage}
\caption{Graphs and their adjacency matrix associated with the first code in \eqref{first} (left) and the first code in \eqref{second} (right).}
\label{Fig2}
\end{figure}

There are fifteen codes with $n=2$. Six of them, see \eqref{n4hr}, are real. We only list real codes, which are specified by a pair of stabilizer generators,
\bea
\label{first}
(\g_1,\g_2)&=& ({\rm X}\, {\rm I}, {\rm I}\, {\rm X}),\quad\, \, \, \, ({\rm X}\, {\rm I}, {\rm I}\, {\rm Z}),\quad \, \, \, \, ({\rm Z}\, {\rm I}, {\rm I}\, {\rm X}),\quad \, \, \, \, ({\rm Z}\, {\rm I}, {\rm I}\, {\rm Z}),\\
(\g_1,\g_2)&=&({\rm X}\, {\rm Z},\, {\rm Z}\, {\rm X}),\quad ({\rm X}\, {\rm X},\, {\rm Z}\, {\rm Z}).
\label{second}
\eea
The codes are split into two groups, with $4$ and $2$ elements. All codes within each group are T-dual to each other.  The first group consists of decomposable codes. The corresponding code CFTs are tensor products of two bosons compactified on circles or radius $R=1$ or $2$. The refined enumerator polynomial of these codes is $W=W_1^2$  and the (binary) Hamming distance is $d=d_{\rm b}=1$. Only the  first code in \eqref{first} is of B-form, with zero $2\times 2$ matrix $\rm B$. The corresponding graph is shown in Fig.~\ref{Fig2} left. The interlace polynomial of this graph is $Q=Q_1^2=(x+y)^2$.

Codes in the second group are indecomposable. They have $W=W_2=x^2+y^2+2z^2$, c.f.~with \eqref{invpols}, and (binary) Hamming distance $d=d_{\rm b}=2$. The first code in \eqref{second} is of B-form. The corresponding CFT consists of two compact bosons on circles of radius $R=1$ with one unit of $B$-flux, $B_{ij}=\epsilon_{ij}$. The corresponding graph  is shown in Fig.~\ref{Fig2} right, and the interlace polynomial is $Q_2=2x(x+y)$.  The second code in \eqref{second} is T-dual to the first one. The generating matrix of its lattice $\Lambda(\C)$ has the toroidal compactification form \eqref{TCN} with vanishing $B$-field and 
\bea
2\gamma^*=\gamma=\left(\begin{array}{cc}
1\, &\, \, \, \,1\\
1 & -1
\end{array}\right). \label{gamma}
\eea
The lattice generated by $\gamma$ is the square lattice with minimal length $\sqrt{2}$.
Therefore the corresponding CFT is the tensor product of two theories, each being a  boson compactified on a circle of self-dual radius $R=\sqrt{2}$. This is confirmed by the partition function
\bea
\label{Theta-i2}
|\eta(\tau)|^2\, Z_{R=\sqrt{2}}^2=\left({\theta_3(q^2)\theta_3({\bar q}^2)+\theta_4(q^2)\theta_4({\bar q}^2)\over 2}\right)^2=\left.{a^2 {a'}^2+b^2 {b'}^2+c^2 {c'}^2 \over 2}\right|_{\tau'=\bar \tau},
\eea 
where the right-hand-side follows from \eqref{ZcodeCFT}.
This is a curious situation because the code CFT is a tensor product of two theories, while the code itself is indecomposable. The spectral gap of this theory is $\Delta=d_{\rm b}/4=1/2$.

This  code has other interesting properties. It is in fact the repetition code $i_2$, the linear self-dual code of type $4^{{\rm H}}$ mentioned in Section \ref{sec:GF4}. Comparing with \eqref{Wi2}, we find $W_{i_2}(x,y)=W_2(x,y,y)$. The Lorentzian lattice in $\R^{2,2}$ generated by \eqref{TCN} with $\gamma$ as in \eqref{gamma}, and the Euclidean lattice in $\CC^2=\R^4$ associated with $i_2 \in 4^{{\rm H}}$ 
 are related to each other by a linear transformation 
\bea
\left(
\begin{array}{cc}
 -1 & -1 \\
 \sqrt{3} & -\sqrt{3} \\
\end{array}
\right)/\sqrt{2} \label{T}
\eea
in each $\R^2=\CC$ plane. Upon setting $\tau'=-3\tau$, the Siegel theta function \eqref{Theta-i2} reduces to the theta function of the Euclidean lattice $\phi_0^2+3\phi_1^2$, as expected. 

Using the Gray map, we can also interpret this code as a binary repetition code, such that the enumerator polynomial \eqref{Wi2-binary} is $W_{i_2}(x,y)=W_2(x^2,y^2,x y)$.   The theta function following from \eqref{Theta-i2},   reduces to $b^4$ after setting $\tau'=-\tau$, which is the correct theta function of a cubic lattice of size $\sqrt{2}$ in $\R^4$. 

To conclude the case of $n=2$, we describe the coset construction \eqref{CN} of real self-dual codes. The group ${\rm SO}(2,2,\R)$ is the product of two ${\rm SL}(2,\R)$ factors mod $\Z_2$,
\bea
S_i&=&\left(\begin{array}{cc}
a_i & b_i \\
c_i & d_i\end{array}\right),\quad a_i d_i-b_i c_i=1, \quad i=1,2,\\
S_1\times S_2 &=&\left(\begin{array}{c| c}
a_1 S_2 & b_1 S_2 \epsilon \\ \hline
-c_1 \epsilon S_2\, &\, d_1 (S_2^T)^{-1}
\end{array}\right)\in {\rm SO}(2,2,\R),\quad \epsilon=\left(\begin{array}{cc}
0 & 1\\
-1 & 0
\end{array}\right).
\eea
Elements of ${\rm O}(2,2,\Z)$ can be described in a similar way. They include products $S_1 \times S_2$ where $S_{1,2}\in {\rm SL}(2,\Z)$ (these are matrices of ${\rm det}=1$)  and 
$t\, (S_1\times S_2)$ (these are matrices of ${\rm det}=-1$), where
\bea
t=\left(
\begin{array}{cccc}
\, 0\, & 0\, & 1\, & 0\, \\
\, 0\, & 1\, & 0\, & 0\, \\
\, 1\, & 0\, & 0\, & 0\, \\
\, 0\, & 0\, & 0\, & 1\, \\
\end{array}
\right).
\eea
The subgroup ${\rm O}_2(2,2,\Z)$ includes only matrices with all elements of the  upper-right $2\times 2$ submatrix being even. This leaves $S_1\times S_2$ where $c_1\, \,{\rm mod}\, \, 2=0$ and $S_2$ is arbitrary. In other words 
\bea
{\rm O}_2(2,2,\Z)=\Gamma_0(2)\times {\rm SL}(2,\Z),
\eea
where $\Gamma_0(2)$ is the Hecke congruence subgroup of level $2$. The quotient ${\rm SL}(2,\Z)/\Gamma_0(2)$ includes three matrices,
\bea
s_1=\left(
\begin{array}{cc}
 1 & 0 \\
 0 & 1 \\
\end{array}
\right),\quad s_2=\left(
\begin{array}{cc}
 1 & 1 \\
 0 & 1 \\
\end{array}
\right),\quad s_3=\left(
\begin{array}{cc}
 0 & 1 \\
 -1 & 0 \\
\end{array}
\right).
\eea 
Six real self-dual codes, arranged the same way as (\ref{first},\ref{second}), are 
\bea
&& s_1\times {\rm I},\qquad t\, (s_1\times {\rm I}),\qquad t\, (s_3\times {\rm I}),\qquad s_3\times {\rm I},\\
&& s_2\times {\rm I}, \qquad t\, (s_2\times {\rm I}).
\eea

\subsection{$n=3$}
There are $30$ real codes for $n=3$, which split into $t_4^{\rm ELC}=4$ orbits under T-duality equivalences (T-equivalences). In terms of the B-field representations, these four orbits correspond to the four inequivalent graphs with three vertices, labeled by the number of  edges (links)  $0\leq l\leq 3$. The equivalence of codes within each orbit is obvious, as the graphs with the same number of links $l$ are isomorphic for $n=3$. 
When $l=0,1$, the graphs, and hence the codes, are decomposable. We will not discuss these cases in detail as their properties were discussed above. B-form codes with $l=2$ are T-dual to the code with lattice $\Lambda(\C)$ \eqref{TCN}, with vanishing $B$-field and 
\bea
\gamma=\left(
\begin{array}{ccc}
 0 & 1 & 1 \\
 1 & 0 & 1 \\
 1 & 1 & 0 \\
\end{array}
\right). \label{gn3}
\eea
B-form codes with $l=3$ are T-dual to the code lattice $\Lambda(\C)$ \eqref{TCN}, with the same $\gamma$  \eqref{gn3} and with non-trivial $B$-field 
\bea
B=\left(
\begin{array}{ccc}
 0 & 0 & 0 \\
 0 & 0 & 1 \\
 0 & -1 & 0 \\
\end{array}
\right).
\eea
In this latter case there is no T-dual theory with lattice of the form \eqref{TCN}, with some  $\gamma$, and vanishing $B$-field. This is the general situation: it is always possible to use T-duality to bring \eqref{TCN} to the form \eqref{TC} with $\gamma={\rm I}$ and some non-trivial $B$-field, but it is almost never possible to  get rid of $B$ by changing $\gamma$. 

There are $t_4^{\rm LC}=3$ inequivalent classes of codes for $n=3$. The classes of T-equivalent codes  with $l=2$ and $l=3$  are related to each other via C-equivalence.  Accordingly, their refined enumerator polynomials 
\bea
W_3=x^3 + 3 y^2 z + 3 x z^2 + z^3, \qquad {\rm and}\qquad \tilde{W}_3=x^3 + 3 x y^2 + 4 z^3,
\eea
yield the same enumerator polynomial $W_3(x,y,y)=\tilde{W}_3(x,y,y)=x^3+3 x y^2+ 4 y^3$. Both classes of codes have the same Hamming distance $d=2$ but different binary Hamming distances $d_{\rm b}=2$ and $d_{\rm b}=3$. The corresponding CFTs will have different partition functions, as well as different spectral gaps, $\Delta=1/2$ and $\Delta=3/4$ correspondingly. 
\begin{figure}
\begin{minipage}{.23\textwidth}
\centering
\includegraphics[width=0.7\textwidth]{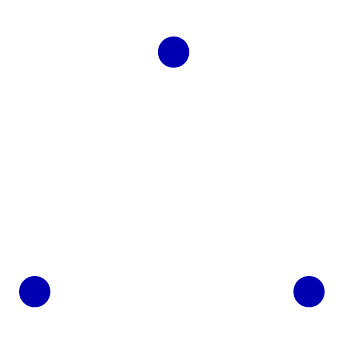}
\begin{eqnarray}
\nonumber
{\rm B}=\left(\begin{array}{ccc}
0 & 0 & 0 \\
0 & 0 & 0\\
0 & 0 & 0
\end{array}\right)
\end{eqnarray}
\begin{eqnarray}
W=W_1^3 \nonumber \\
Q=Q_1^3 \nonumber
\end{eqnarray}
\end{minipage}
\begin{minipage}{.23\textwidth}
\centering
\includegraphics[width=0.7\textwidth]{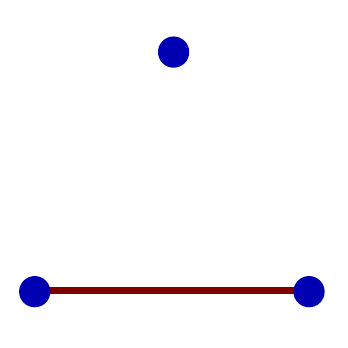}
\begin{eqnarray}
\nonumber
{\rm B}=\left(\begin{array}{ccc}
0 & 1 & 0 \\
1 & 0 & 0\\
0 & 0 & 0
\end{array}\right)
\end{eqnarray}
\begin{eqnarray}
W=W_1 W_2 \nonumber \\
Q=Q_1 Q_2 \nonumber
\end{eqnarray}
\end{minipage}
\begin{minipage}{.23\textwidth}
\centering
\includegraphics[width=0.7\textwidth]{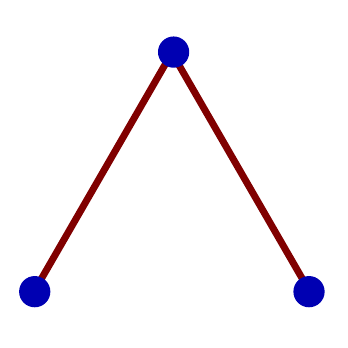}
\begin{eqnarray}
\nonumber
{\rm B}=\left(\begin{array}{ccc}
0 & 1 & 1 \\
1 & 0 & 0\\
1 & 0 & 0
\end{array}\right)
\end{eqnarray}
\begin{eqnarray}
W=W_3 \nonumber\\
Q= x(3x+y) Q_1 \nonumber
\end{eqnarray}
\end{minipage}
\begin{minipage}{.23\textwidth}
\centering
\includegraphics[width=0.7\textwidth]{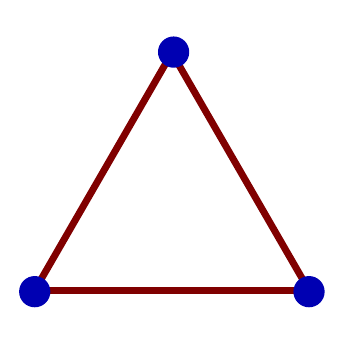}
\begin{eqnarray}
\nonumber
{\rm B}=\left(\begin{array}{ccc}
0 & 1 & 1 \\
1 & 0 & 1\\
1 & 1 & 0
\end{array}\right)\end{eqnarray}
\begin{align}
W=\tilde{W}_3  \nonumber \\
Q= 4 x^2 Q_1 \nonumber
\end{align}
\end{minipage}
\caption{Graphs, their adjacency matrices, refined and interlace polynomials, associated with the four T-dual classes of $n=3$ codes. The polynomials $W_1,W_2,W_3$ are defined in \eqref{invpols}, and $\tilde{W}_3\equiv x^3 + 3 x y^2 + 4 z^3=-W_3+3 W_1 W_2+W_1^3$.}
\label{Fig4}
\end{figure}

For $n=1$ and $n=2$, all polynomials $W(x,y,z)$ invariant under \eqref{dual} and $y\rightarrow -y$, and satisfying additional conditions, $W(1,0,0)=1$, all coefficients integer and positive, are actual refined enumerator polynomials of additive codes. For $n=3$, besides the four polynomials associated with the four classes of T-equivalent codes, see Fig.~\ref{Fig4}, there are another six ``fake'' polynomials, 
\bea
W&=&x^3+2 x^2 z+3 x z^2+y^2 z+z^3,\\
W&=&x^3+x^2 z+3 x z^2+2 y^2 z+z^3,\\
W&=&x^3+2 x^2 z+x y^2+2 x z^2+2 z^3,\\
W&=&x^3+x y^2+2 x z^2+2 y^2 z+2 z^3,\\
W&=&x^3+x^2 z+2 x y^2+x z^2+3 z^3,\\
W&=&x^3+2 x y^2+x z^2+y^2 z+3 z^3.
\eea
There are no additive self-dual codes for which these polynomials are refined enumerator polynomials, yet they satisfy all necessary properties, and the ``partition function'' defined via \eqref{ZcodeCFT} is modular invariant and satisfies other basic properties expected of the CFT partition function.  This poses the following question important in light of the  modular bootstrap program: do those would-be CFT partition functions correspond to actual theories?  Given that the number of ``fake'' polynomials increases rapidly with $n$, unless they correspond to actual CFTs, ``bootstrapping''  2d theories must yield a growing number of consistency  regions (occasionally taking the form of ``islands'') in the exclusion  plots, which are in fact empty, contradicting our experience so far. Assuming the opposite, that some of these ``fake'' polynomials correspond to actual CFTs, they likely can be identified as refined enumerator polynomials for non-additive codes.  In this case the scope of what we call code theories should be extended to include CFTs based on a wider class of codes. Continuing this logic further, the CFT partition  function is a much richer object than the code enumerator polynomial, and may satisfy additional non-trivial  conditions \cite{gaberdiel2007constraints,gaiotto2008monster}. An interesting scenario  would be if these additional conditions could be used to distinguish ``fake'' code enumerators from actual ones, thus introducing  a new string theoretic tool to code theory.

\subsection{$n=4$}\label{sec:n=4}
There are $t_4^{\rm ELC}=9$ classes of T-equivalent codes in this case, $i_4^{\rm ELC}=4$ of which correspond to indecomposable codes. We discuss only the indecomposable ones.
The first class includes B-form codes with the ${\rm B}$ matrices (graphs)
\bea
\label{n4g1label} 
{\rm B}=\left(
\begin{array}{cccc}
 0\, & 1\, & 0\, & 1\, \\
 1\, & 0\, & 1\, & 0\, \\
 0\, & 1\, & 0\, & 1\, \\
 1\, & 0\, & 1\, & 0\, \\
\end{array}
\right), \qquad \quad 
{\rm B}=\left(
\begin{array}{cccc}
 0\, & 1\, & 0\, & 0\, \\
 1\, & 0\, & 1\, & 0\, \\
 0\, & 1\, & 0\, & 1\, \\
 0\, & 0\, & 1\, & 0\,
\end{array}
\right), \quad 
\eea
\begin{center}
\, \, \includegraphics[width=0.55\textwidth]{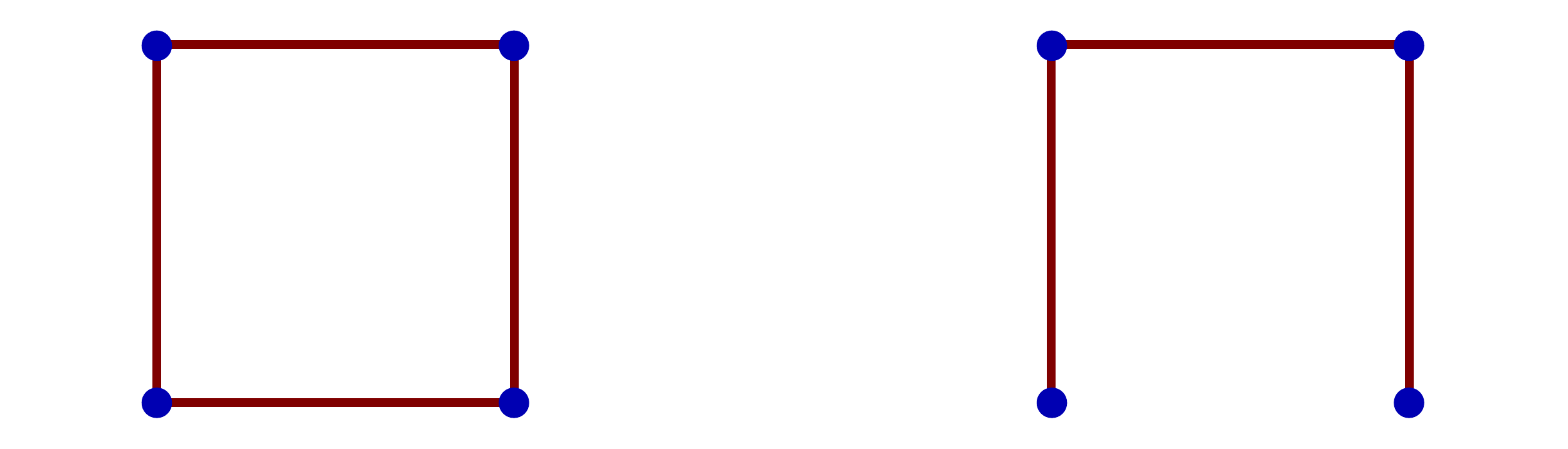}
\end{center}
and those isomorphic (permutation equivalent) to them. We explicitly write down two different ${\rm  B}$'s for the same  class of T-equivalent codes to emphasize that edge local complementation can change the number of links, topology, etc. The REP for this class is $W=x^4 + 4 x y^2 z + 2 x^2 z^2 + 4 y^2 z^2 + 4 x z^3 + z^4=2 W_1^2 W_2-2 W_1 {\cal R}-W_1^4$, and the interlace polynomial is $Q=5 x^4 + 8 x^3 y + 3 x^2 y^2$.

The second equivalence class include B-form codes with the ${\rm B}$ matrices (graphs)
\bea
\label{n4g2label} 
{\rm B}=\left(
\begin{array}{cccc}
 0\, & 0\, & 0\, & 1 \\
 0\, & 0\, & 1\, & 1 \\
 0\, & 1\, & 0\, & 1 \\
 1\, & 1\, & 1\, & 0 \\
\end{array}
\right), \qquad \quad 
{\rm B}=\left(
\begin{array}{cccc}
 0\, & 0\, & 1\, & 1 \\
 0\, & 0\, & 1\, & 1 \\
 1\, & 1\, & 0\, & 1 \\
 1\, & 1\, & 1\, & 0 \\
\end{array}
\right), \quad 
\eea
\begin{center}
\, \, \includegraphics[width=0.55\textwidth]{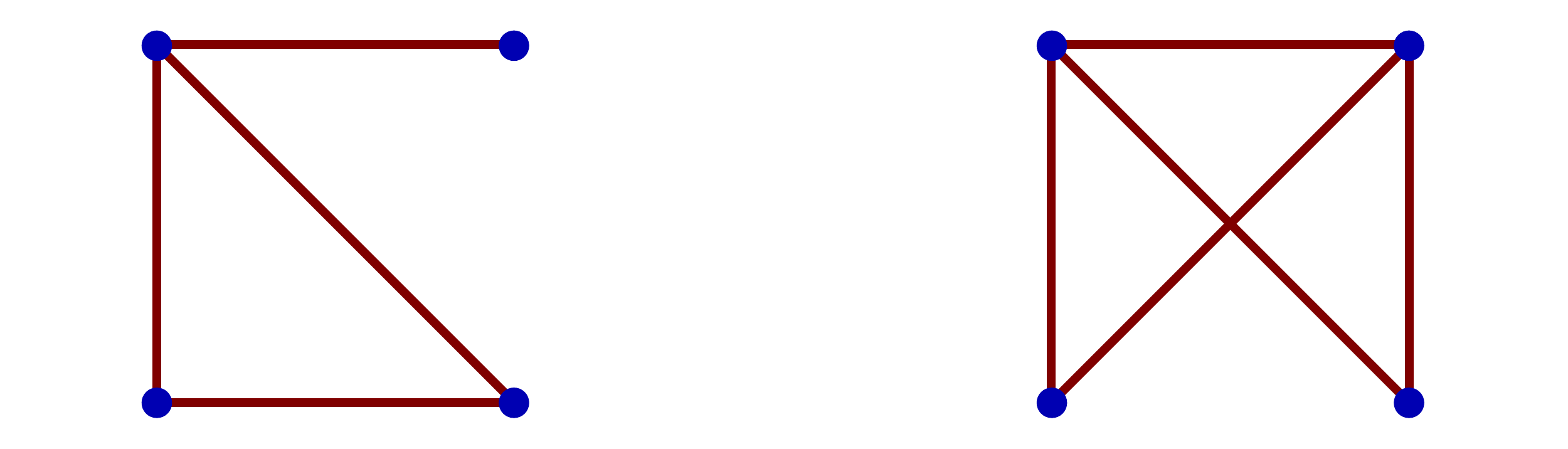}
\end{center}
and those isomorphic  to them.  The edge local complementarity of these two graphs is discussed as an example in \cite{danielsen2008edge}. The REP for this class is
$W=x^4+x^2 y^2+x^2 z^2+4 x y^2 z+4 x z^3+3 y^2 z^2+2 z^4=2 W_1^2 W_2-W_1 {\cal R}-W_1^4$, and the interlace polynomial is $Q=6 x^4+8 x^3 y+2 x^2 y^2$.

The two classes of T-equivalent codes  described above are related to each other via C-equivalence. One can easily check that the enumerator polynomial in both cases is the same by taking ${\cal R}\rightarrow 0$. Both classes of code  theories have the same value of $d=d_{\rm b}=2$ and spectral gap $\Delta=1/2$.

There are two more classes of T-equivalent codes for $n=3$ . B-form codes from the third class are given by 
\bea
{\rm B}=\left(
\begin{array}{cccc}
 0\, & 1\, & 1\, & 1\, \\
 1\, & 0\, & 0\, & 0\, \\
 1\, & 0\, & 0\, & 0\, \\
 1\, & 0\, & 0\, & 0\, 
\end{array}
\right),\quad \quad\quad \qquad \adjustimage{width=0.15\textwidth,valign=m}{n4-3}\qquad
\eea
%
and those isomorphic  to them. The  REP of this class is $W=x^4+6 x^2 z^2+y^4+6 y^2 z^2+2 z^4=W_2^2-2 W_1 {\cal R}$ and interlace polynomial is $Q=4 x^4+7 x^3 y+4 x^2 y^2+x y^3$.

Finally, the fourth class has a unique B-form representative with $B_{ij}=1-\delta_{ij}$ associated with the complete graph with four vertices
\bea
\label{qe8}
{\rm B}=\left(
\begin{array}{cccc}
 0\, & 1\, & 1\, & 1\, \\
 1\, & 0\, & 1\, & 1\, \\
 1\, & 1\, & 0\, & 1\, \\
 1\, & 1\, & 1\, & 0\, 
\end{array}
\right). \quad \qquad \qquad \adjustimage{width=0.15\textwidth,valign=m}{n4-4}\qquad
\eea
%
The REP of this class is $W_{qe_8}=x^4+6 x^2 y^2+y^4+8 z^4=W_2^2+4 W_1 {\cal R}$ and the interlace polynomial is $Q=8 x^3(x + y)$.

The third and fourth classes are related to each other via C-equivalence (thus there are two  classes  of indecomposable equivalent codes $i_4^{\rm LC}=2$). Accordingly, the enumerator polynomials of the third and fourth classes are the  same, $W(x,y)=W_2(x,y,y)^2$, and also the same as the enumerator polynomial of the decomposable code consisting of two $n=2$ codes. This is an example of a generic situation: enumerator polynomials are not unique and different codes may share the same enumerator polynomial. The same is also true for the refined enumerator polynomial, see Section \ref{sec:n=7}. Codes from the third and fourth classes have different  $d_{\rm b}$,  $2$ and $4$ correspondingly. This is similar to the case of $n=3$ and is a reflection of the general situation: C-equivalent codes must have the same $d$ but usually  have different $d_{\rm b}$.

Comparing ${\rm B}$ from  \eqref{qe8}  with \eqref{H8} we immediately recognize that the binary code obtained from this stabilizer code  via Gray map is not merely isodual, but in fact self-dual. It is the extended Hamming $[8,4,4]$ binary code, and the Lorentzian lattice $\Lambda(\C)$, understood as a lattice in the Euclidean space, is the even self-dual lattice $E_8$. In other words $E_8$ is an even self-dual lattice in both Lorentzian and Euclidean signatures!  As a consistency check one can easily verify that the enumerator polynomial of the resulting binary code $W_{qe_8}(x^2,y^2,xy)=W_{e_8}(x,y)$. Now we can recognize $d_{\rm b}/2=2$ as the length-squared of the shortest root of $E_8$ lattice $\ell^2=2$. The corresponding Narain CFT, which we will refer to as the non-chiral $E_8$ theory, has  spectral gap $\Delta=1$.

Modular bootstrap studies of the maximal spectral gap in $U(1)^4\times U(1)^4$ theories reveal with an astonishing precision that $\Delta=1$ is in fact the optimal (maximal possible) value \cite{afkhami2020free}. 
The theory of eight free Majorana fermions with diagonal GSO projection was identified in \cite{collier2018modular} as the CFT saturating the bound. Here we have found that the non-chiral $E_8$ Narain CFT  also saturates the bound. Using the explicit form of $W_{qe_8}$ and \eqref{ZcodeCFT} we readily find the  partition function of this theory
\bea
Z_{\rm E_8}(\tau,\bar \tau)={|a|^8+|b|^8+|c|^8\over 2|\eta(\tau)|^8},
\eea
which coincides with the partition function of eight fermions mentioned above \cite{collier2018modular}.  
In fact these are the same theory, which has other descriptions including as 
the $\widehat{{\rm SO}(8)}_1$ WZW model. The theory of eight fermions exhibits rich group of symmetries, known as triality, which has been recently discussed in \cite{tong2019notes}. While the description of this theory as a Narain CFT has been discussed previously, to our knowledge connection with the $E_8$ lattice has not been pointed out. We establish an explicit relation between the theory of eight Majorana fermions and the non-chiral Narain $E_8$ theory in the Appendix~\ref{appdx:e7e8}.


There are $11$ ``fake'' REPs for $n=4$, 
\bea
W&=&W_2^2+k\, W_1 {\cal R},\qquad k=-1,1,2,3,\\
W&=&W_1^2 W_2\pm W_1{\cal R},\\
W&=&2 W_1^2 W_2-W_1^4,\\
W&=& \frac{1}{2} W_2 W_1^2+\frac{W_1^4}{2}\pm\frac{{\cal R} W_1}{2},\\
W&=&\frac{3}{2} W_2 W_1^2-\frac{W_1^4}{2}\pm\frac{{\cal R} W_1}{2}.
\eea

\begin{figure}
\centering
\, \, \includegraphics[width=0.55\textwidth]{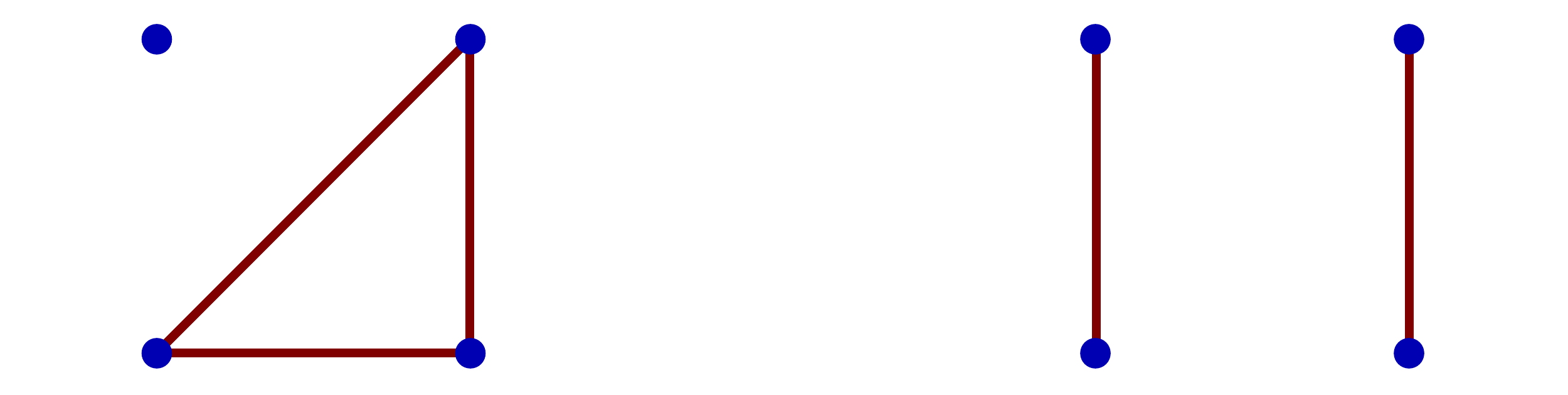}
\caption{Two  non-equivalent graphs  under edge local complementation, which  have the same interlace polynomial $Q=4 x^2 (x + y)^2$.}
\label{fig:Fig4}
\end{figure}

The interlace polynomial $Q(x)$ is a characteristic of the graph equivalence class under  edge local complementation (and isomorphisms), but  different classes may share the same $Q(x)$. This happens for the first time (meaning smallest $n$) for $n=4$, for the decomposable graphs shown in Fig.~\ref{fig:Fig4}. Edge local complementation acts on each disconnected subgraph individually, and therefore these graphs, which have different decompositions $1+3$ and $2+2$ can not be equivalent. 

\subsection{$n=5$}
For $n=5$ there are too many classes of codes ($t_5^{\rm ELC}=21,  i_5^{\rm ELC}=10$) to describe all of them in detail. From now on we will only focus  on codes maximizing the Hamming distance $d$ (the conventional  measure of quality for quantum codes) or the binary Hamming distance $d_{\rm b}$ (which determines, up  to some nuances, the spectral gap in the code CFT). For all $n\leq 4$, there was a unique class of T-equivalent codes with maximal $d_{\rm b}$, which also had maximal $d$ (note, there were other codes with the same $d$, but smaller $d_{\rm b}$).
For $n=5$, there is a unique class of codes with the maximal $d=3$ (and $d_{\rm b}=3$), the so-called shorter hexacode related via Construction A of Section~\ref{sec:GF4} to the shorter Coxeter-Todd lattice \cite{rains1998shadow}, and there is another unique class of codes with the maximal $d_{\rm b}=4$ (and $d=2$).

\begin{figure}
\centering
\, \, \includegraphics[width=0.7\textwidth]{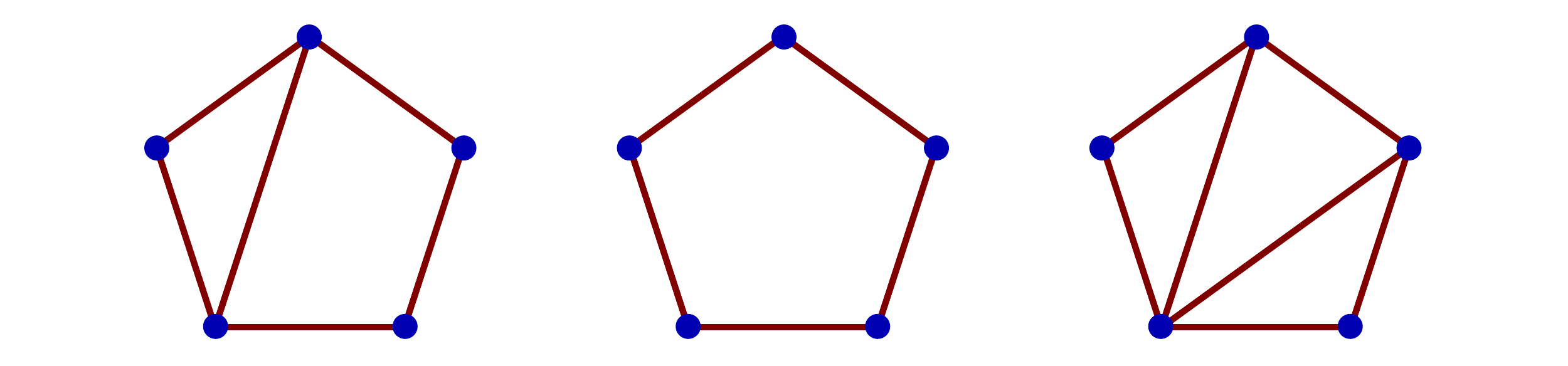}
\caption{ELC-equivalent (T-dual equivalent) graphs corresponding to  $n=5$ code with the largest $d=3$, $d_{\rm b}=3$, $W=x^5+5 x^2 y^2 z+5 x^2 z^3+10 x y^2 z^2+5 x z^4+5 y^2 z^3+z^5$ and $Q=x^3 \left(11 x^2+16 x y+5 y^2\right)$. }
\label{fig:Fig5-1}
\centering
\, \, \includegraphics[width=0.18\textwidth]{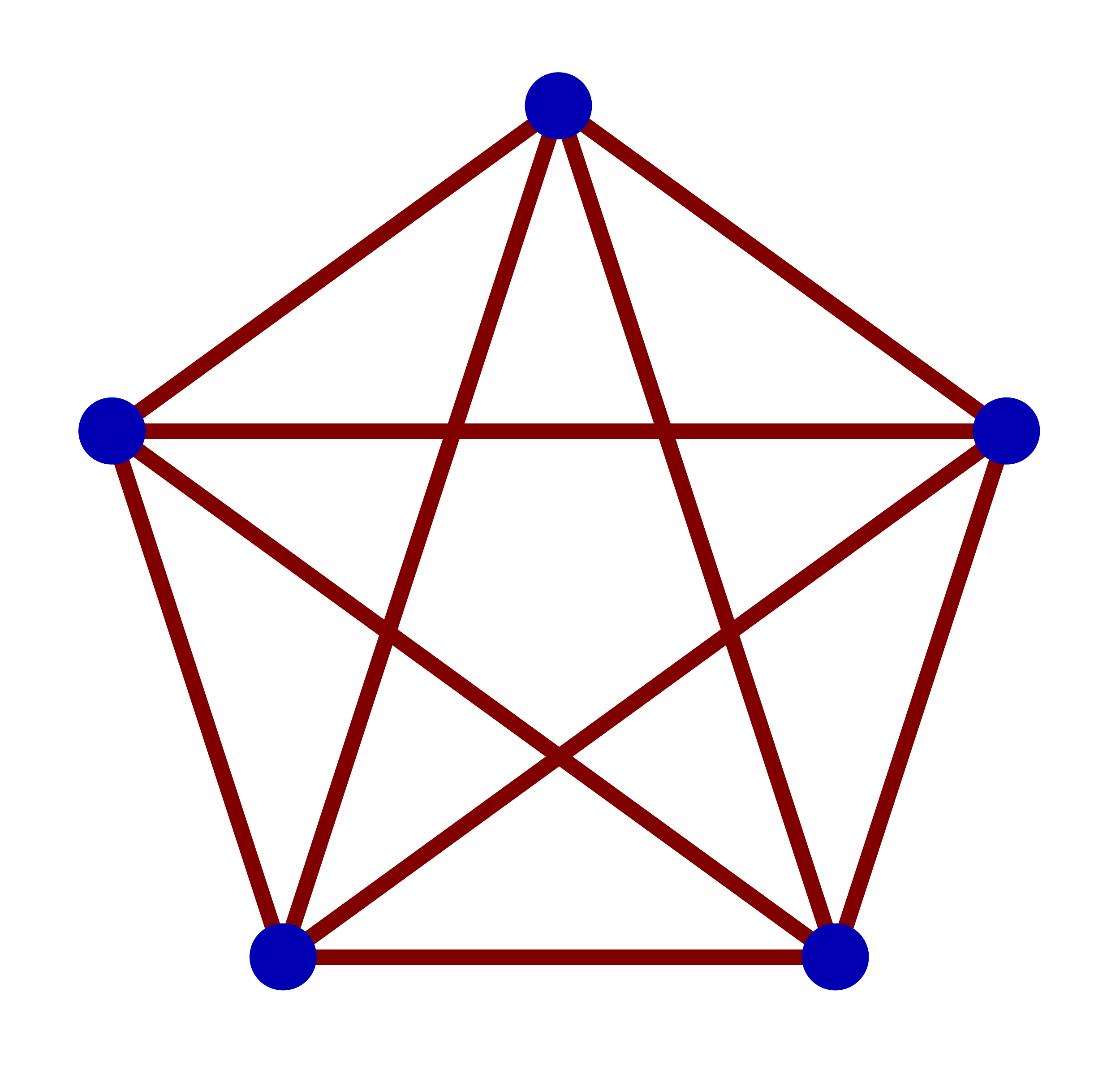}
\caption{Unique (up to isomorphisms) graph corresponding to $n=5$ class of codes with the largest $d_{\rm b}=4$, $d=2$,   $W=x^5+10 x^3 y^2+5 x y^4+16 z^5$ and $Q=16 x^4 (x + y)$. }
\label{fig:Fig5-2}
\end{figure}

There are three distinct graphs (up to isomorphisms), which correspond to the shorter hexacode  class, see Fig.~\ref{fig:Fig5-1}. And there is a unique graph (up to isomorphism), which corresponds to the second class with $d_{\rm b}=4$, Fig.~\ref{fig:Fig5-2}.

For $n=3$ and $n=4$ we saw examples where C-equivalence would relate two classes of T-equivalent codes. For $n=5$ there are already groups of $2$, $3$ and $4$ classes of codes related to each other by C-equivalences.

There are $128$ ``fake'' REPs for $n=5$. The number of ``fake'' REPs increases rapidly with $n$, $2835$ for $n=6$, $71164$ for $n=7$, 4012529 for $n = 8$ and so on. 

\subsection{$n=6$}\label{sec:n6}
\begin{figure}[t]
\centering
\includegraphics[width=0.55\textwidth]{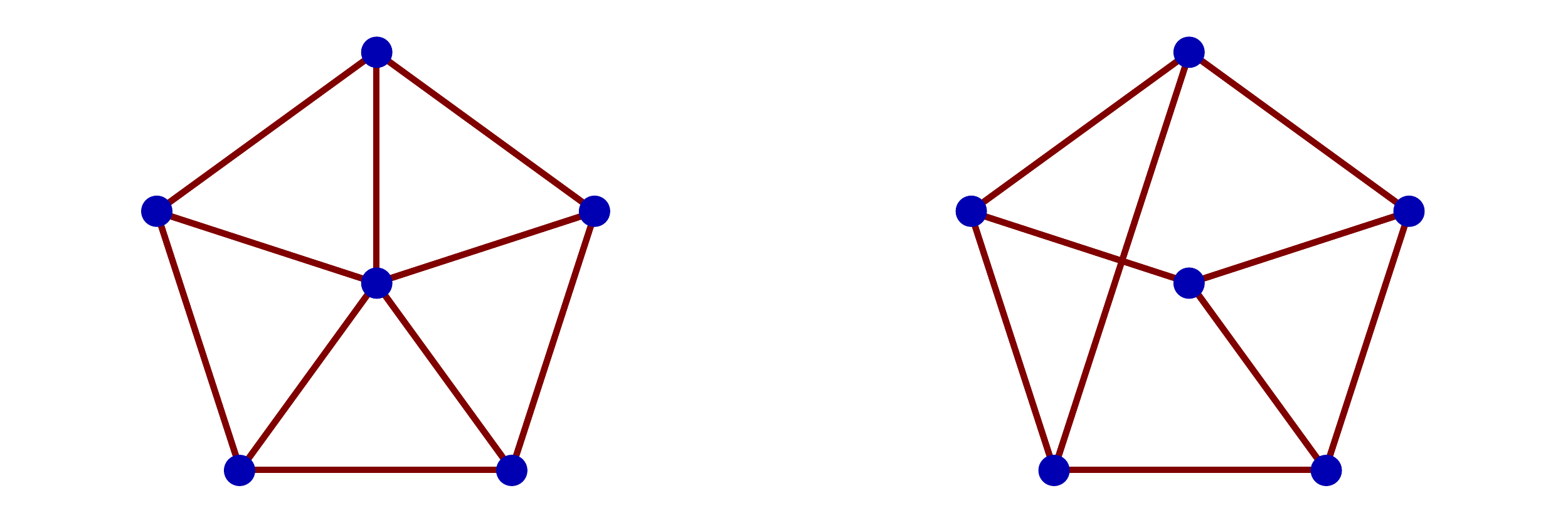}
\caption{ELC-equivalent (T-equivalent) graphs corresponding to the hexacode. These graphs have $Q=2 x^4 (11 x^2 + 16 x y + 5 y^2)$.}
\label{fig:hexa}
\end{figure}

There is a unique class of codes which achieves both maximal $d=4$ and maximal $d_{\rm b}=4$. This is the hexacode $h_6$, introduced in Section \ref{sec:GF4}. As can be easily seen from \eqref{hexacode}, the hexacode is a real code, and by using T-duality transformations it can be brought to the B-form. (We should note that there are other codes, C-equivalent to the hexacode \eqref{hexacode}, which are also called by this name in the literature, see \cite{conway2013sphere,harvey2020moonshine}. Those codes are not real.) There are two graphs shown in Fig.~\ref{fig:hexa}, which are associated with the class of T-equivalent codes that includes the hexacode. 
The refined enumerator polynomial of the hexacode is $W_{h_6}(x,y,z)=x^6+30 x^2 y^2 z^2+15 x^2 z^4+y^6+15 y^2 z^4+2 z^6$, which reduces to \eqref{hexaW} upon substituting $z\rightarrow y$. 

The Lorentzian lattice of the hexacode $\Lambda(h_6)$ is related to the Euclidean Coxeter-Todd lattice $K_{12}$ by the linear transformation \eqref{T} applied in each $\CC$ plane. Upon setting $\tau'=-3\tau$, the Siegel theta-function of  $\Lambda(h_6)$ reduces to the theta 
function of $K_{12}$ \eqref{K12}.  
 We should note that the Lorentzian lattice $\Lambda(h_6)$, although related to $K_{12}$, is not the same as the Coxeter-Todd lattice understood as a Lorentzian even self-dual lattice. The latter interpretation and the related Narain CFT was recently introduced in \cite{afkhami2020free}. That construction is analogous to our construction of $E_8$ as a Lorentzian even self-dual lattice, discussed in Section \ref{sec:n=4}.

There are two other classes of codes with maximal $d_{\rm b}=4$ and $d=2$, which  we do not discuss  here.

\subsection{$n=7$}\label{sec:n=7}
\begin{figure}[b]
\centering
\includegraphics[width=0.9\textwidth]{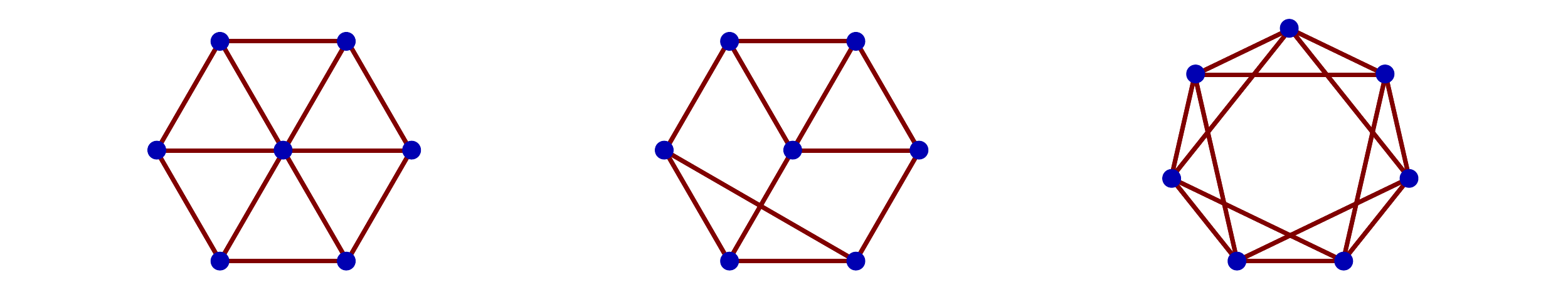}
\caption{Representatives from three distinct classes of T-equivalent codes,  which maximize both $d=3$ and $d_{\rm b}=4$. The interlace polynomials for these classes are $Q=40 x^7+62 x^6 y+24 x^5 y^2+2 x^4 y^3$, $Q=41 x^7 + 63 x^6 y + 23 x^5 y^2 + x^4 y^3$, and $Q=43 x^7 + 64 x^6 y + 21 x^5 y^2$.}
\label{fig:Fig7}
\end{figure}

There are $t_7^{\rm ELC}=218$ classes of T-equivalent  codes, with $18$ classes
attaining the maximal value of $d=3$, and $8$ classes with the maximal value of $d_{\rm b}=4$. Let us first focus on those $3$ classes which have both maximal $d$ and $d_{\rm b}$. The first class includes $3$ graphs (up to isomorphisms), the second $12$ and the third $6$. We only show one representative for each class of T-equivalent codes in figure~\ref{fig:Fig7}.
The REPs for these codes are (from left to right)
\bea
\nonumber
W&=&\frac{3 W_2 W_1^5}{2}+\frac{W_2^2 W_1^3}{4}+\frac{W_2^3 W_1}{2}+\frac{11{\cal R}^2 W_1}{4}-\frac{{\cal R} W_2^2}{2}-\frac{5 W_1^7}{4}-\frac{5{\cal R} W_1^4}{2}-\frac{{\cal R} W_2 W_1^2}{2},\\ \nonumber
W&=&\frac{3 W_2 W_1^5}{2}+\frac{W_2^2 W_1^3}{4}+\frac{W_2^3 W_1}{2}-\frac{{\cal R} W_2 W_1^2}{2}-\frac{{\cal R} W_2^2}{2}-\frac{5 W_1^7}{4}-\frac{{\cal R}^2 W_1}{4}-\frac{{\cal R} W_1^4}{2},\\ \nonumber
W&=&\frac{7{\cal R}^2 W_1}{4}-\frac{7{\cal R} W_2 W_1^2}{2} -\frac{3 W_1^7}{4}+\frac{7W_2^2 W_1^3}{4}.
\eea

\begin{figure}
\centering
\includegraphics[width=1\textwidth]{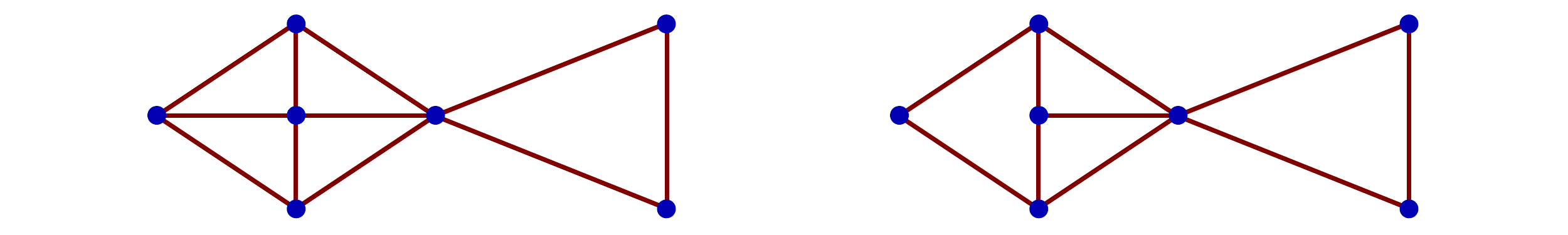}
\caption{``Fish'' graphs -- representatives from two classes of T-dual codes, which share the same refined enumerator polynomial \eqref{rep7}; they are C-equivalent but not T-dual to each other. The first class includes $10$ distinct non-isomorphic graphs; the second class includes $9$. We choose among them two representatives based on simplicity and aesthetics. These two classes of graphs also share the interlace polynomial $Q=30 x^7 + 58 x^6 y + 34 x^5 y^2 + 6 x^4 y^3.$}
\label{fig:Fig7fish}
\end{figure}

It turns out, there are exactly two distinct  classes of T-equivalent codes for $n=7$ which share the same REP,
\bea\nonumber
W_{\rm isospectral}&=&x^7 + x^5 y^2 + 5 x^4 y^2 z + 5 x^2 y^4 z + x^5 z^2 + 
 12 x^3 y^2 z^2 + 9 x y^4 z^2 + 4 x^4 z^3 +\\ 22 x^2 y^2 z^3 &+&    \label{rep7}
 4 y^4 z^3 + 5 x^3 z^4 + 25 x y^2 z^4 + 11 x^2 z^5 + 11 y^2 z^5 + 
 10 x z^6 + 2 z^7.
\eea
Two representative graphs (there are many others) associated with these two classes are shown in fig.~\ref{fig:Fig7fish}. 
(We note that the REP is unique for all classes of T-equivalent codes for $n\leq 6$). It should be noted that the shown graphs are related by LC, which means that the corresponding classes of codes are C-equivalent. And while in the general case C-equivalence can change the REP, in this case it does not.

Since these two classes of codes are not T-equivalent, corresponding code theories are not T-dual to each other, see Appendix~\ref{sec:TD}. Because they share the same REP, the corresponding code CFTs have the same partition function. In other words we have obtained an explicit example of two Narain CFTs, not related by T-duality, with the same spectrum. At the level of lattices, this is a pair of isospectral but not isomorphic even self-dual Lorentzian lattices in $\R^{7,7}$.
It is interesting to compare this example with the examples of isospectral but not isomorphic Euclidean lattices associated with inequivalent classical codes, in particular Milnor's example of $E_8\oplus E_8$ and $D_{16}^+$ even self-dual lattices in $\R^{16}$. In String Theory this example famously corresponds to the two possible isospectral  compactifications of 16 left-moving modes of the heterotic string. These two isospectral theories  are related by T-duality upon compactification \cite{PhysRevD.35.648,NARAIN198641,NARAIN1987369}. Our construction gives an example  of isospectral even self-dual lattices in the smallest number of dimensions, and in that dimension it is unique, much like the Milnor's example. (We note our analysis covers only code-related lattices. It is an open question if there are other even self-dual isospectral lattices in $\R^{n,n}$ for $n\leq 7$.) But it is also different from Milnor's example in several ways. First, Milnor's example related a decomposable code with an indecomposable one. Here we have two indecomposable codes. Second, at the level of CFTs it is an example of two isospectral non-chiral CFTs, not related in any simple way to chiral CFTs. Furthermore, there is no obvious symmetry which would make this example unique or special, raising doubts that these isospectral theories might be related by a duality. 

As we will see shortly there are many more examples of isospectral Narain CFTs with $n\geq 8$. Our finding  highlights a limitation of the modular bootstrap approach, which is incapable of differentiating isospectral theories. 



\subsection{$n=8$}
\label{n=8}
There are $t_8^{\rm ELC}=1068$ classes of T-equivalent codes with $n=8$. Fourteen classes 
achieve the maximal allowed value of $d=4$, and $d_{\rm b}=4$. They form 5 groups of C-equivalent codes \cite{danielsen2006classification}. There are other classes with maximal $d_{\rm b}=4$ but they have smaller $d$. 

Among the fourteen classes with the maximal $d=d_{\rm b}=4$ is the code with lattice $\Lambda(\C)=E_8 \oplus E_8$, understood with the metric \eqref{Mink}. Upon bringing it to B-form, its $\rm B$-matrix  is given by 
\bea
\label{qe16}
{\rm B}=\left(
\begin{array}{c|c}
 0 & {\rm B}_4 \\ \hline
{\rm B}_4 & 0 
\end{array}
\right), \quad \qquad \qquad \adjustimage{width=0.2\textwidth,valign=m}{e8e8}\qquad
\eea
where the $4\times 4$ matrix ${\rm B}_4$ is given by \eqref{qe8}. As can be seen from its graph, this code is indecomposable and its REP is 
\bea
W_{(qe_8)^2}=x^8+14 x^4 y^4+28 x^4 z^4+168 x^2 y^2 z^4+y^8+28 y^4 z^4+16 z^8.
\eea 
We should note right away that there is another decomposable code, which is a product of two $n=4$ codes  \eqref{qe8}. Its ${\rm B}$ matrix is block-diagonal, with each block equal to $B_4$. The lattice $\Lambda(\C)$ of that code is also $E_8 \oplus E_8$ but this time each $E_8$ is understood as a Lorentzian lattice, as in Section \ref{sec:n=4}. The REP of this code is $W_{qe_8}^2$ and $d=2, d_{\rm b}=4$. Both codes would be equivalent as binary codes, and in particular $W_{(qe_8)^2}(x^2,y^2,xy)=W_{qe_8}^2(x^2,y^2,xy)=W_{e_8}(x,y)^2$. As we have mentioned several times already, the binary $e_8\oplus e_8$ code is isospectral with $d_{16}^+$ (denoted $E_{16}$ in \cite{pless1975classification}). The latter can be brought to canonical form with the $\rm B$-matrix being symmetric and ${\rm B}_{ii}=0$. This means the binary self-dual code $d_{16}^+$ can be uplifted to the real self-dual stabilizer code with $\rm B$-matrix (one of many representatives from the T-equivalence class) and graph
\bea
\label{d16}
{\rm B}=\left(
\begin{array}{cccccccc}
 0\, & 1\, & 1\, & 1\, & 1\, & 1\, & 1\, & 1\, \\
 1 & 0 & 1 & 1 & 1 & 1 & 1 & 1 \\
 1 & 1 & 0 & 0 & 0 & 1 & 0 & 0 \\
 1 & 1 & 0 & 0 & 0 & 0 & 1 & 0 \\
 1 & 1 & 0 & 0 & 0 & 0 & 0 & 1 \\
 1 & 1 & 1 & 0 & 0 & 0 & 0 & 0 \\
 1 & 1 & 0 & 1 & 0 & 0 & 0 & 0 \\
 1 & 1 & 0 & 0 & 1 & 0 & 0 & 0 \\
\end{array}
\right),\qquad\adjustimage{width=0.32\textwidth,valign=m}{d16}\qquad
\eea
This code has $d=2$, $d_{\rm b}=4$ and REP
\bea
\nonumber
W_{d_{16}^+}=x^8 + 4 x^6 y^2 + 22 x^4 y^4 + 4 x^2 y^6 + y^8 + 24 x^4 z^4 + 
 144 x^2 y^2 z^4 + 24 y^4 z^4 + 32 z^8.
\eea
Of course $W_{d_{16}^+}(x^2,y^2,xy)=W_{e_8}(x,y)^2$, which means that all three code CFTs -- the one associated with \eqref{qe16}, the tensor product of two \eqref{qe8} theories, and the one associated with \eqref{d16} -- have partition functions which coincide along the diagonal $\bar \tau=-\tau$ (purely imaginary $\tau$), but are different otherwise. 

We just saw that Milnor's example of isospectral even self-dual lattices in Euclidean space $\R^{16}$ does not lead to isospectral Lorentzian lattices. This does not mean there is any lack of isospectral even self-dual lattices in 
$\R^{8,8}$. Among  $n=8$ real self-dual stabilizer codes there are 60 isospectral pairs (excluding the product of the $n=1$ code with the isospectral $n=7$ codes shown in Fig.~\ref{fig:Fig7fish}). Among these $60$ pairs two relate a decomposable code with an indecomposable one, while the other $58$ relate two indecomposable codes. Among the first two cases is the hexacode, see Fig.~\ref{fig:hexa}, combined together with the $n=2$ code shown in Fig.~\ref{Fig2} right, which is isospectral with the indecomposable $n=8$ code  associated with the graph (one of many representatives from the T-equivalence class)
\bea
{\rm B}=\left(
\begin{array}{cccccccc}
 0\, & 1\, & 1\, & 1\, & 1\, & 1\, & 0\, & 0\, \\
 1 & 0 & 1 & 1 & 1 & 1 & 0 & 0 \\
 1 & 1 & 0 & 1 & 1 & 0 & 1 & 0 \\
 1 & 1 & 1 & 0 & 0 & 1 & 0 & 1 \\
 1 & 1 & 1 & 0 & 0 & 0 & 0 & 0 \\
 1 & 1 & 0 & 1 & 0 & 0 & 0 & 0 \\
 0 & 0 & 1 & 0 & 0 & 0 & 0 & 0 \\
 0 & 0 & 0 & 1 & 0 & 0 & 0 & 0 \\
\end{array}
\right),\qquad\adjustimage{width=0.32\textwidth,valign=m}{isotoh}\qquad
\eea
One can easily find two codewords with $d_{\rm b}=2$ that are not orthogonal to other codewords in terms of the Euclidean metric. This means corresponding lattice is not decomposable into a  sum of two lattices, but is isospectral with a decomposable one. In this sense this example is similar to Milnor's example. 

\begin{figure}[t]
\includegraphics[width=\textwidth]{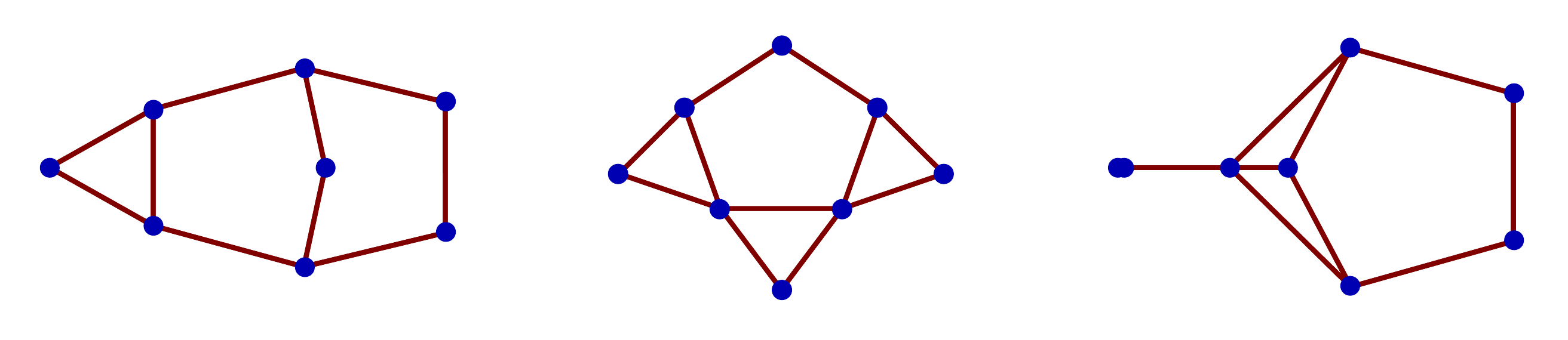}
\caption{Representatives from three different not C-equivalent classes of T-equivalent codes, which share the same REP $W=(x+z)^2(x^6-2 x^5 z+3 x^4 z^2+4 x^3 y^2 z+x^2 y^4+2 x^2 y^2 z^2+4 x^2 z^4+2 x y^4 z+24 x y^2 z^3+4 x z^5+9 y^4 z^2+10 y^2 z^4+2 z^6)$.}
\label{triple8}
\end{figure}

We will not discuss other examples of isospectral pairs in detail, but just mention that in 36 instances isospectral codes are C-equivalent, while in 24 instances they are not. Besides 60 isospectral pairs, there are 5 isospectral triples, when three different code CFTs are isospectral. Four triples include two C-equivalent codes, and another one, not C-equivalent. All three codes in the fifth triple are not C-equivalent. Representative graphs from the fifth triple are shown in Fig.~\ref{triple8}.

\subsection{$n=9-11$}\label{sec:n=9-11}
We have classified all graphs with up to $n\leq 8$ vertexes, see Appendix~\ref{sec:graphs}, and one can easily generate all corresponding refined enumerator polynomials and identify equivalent ones using computer algebra. We leave the task of classifying ELC classes of graphs (classes of T-equivalent codes) with larger $n$ for the future. There is a full classification  of LC classes (classes of equivalent codes) for $n\leq 12$ obtained in \cite{danielsen2006classification}, with the corresponding database available \href{http://www.ii.uib.no/~larsed/vncorbits/}{\tt online}. Going through 675 $n=9$, 3990 $n=10$  and 45144 $n=11$ codes available there  we confirm  there are more new examples of isospectral codes.  There are instances of pairs, triples, and quadruples of isospectral $n=10$ codes, and $k$-tuples of isospectral $n=11$ codes for all $k\leq 11$.

\subsection{$n=12$}\label{sec:n=12}
The theoretical fit of the numerical bootstrap constraint for the value of the spectral gap $\Delta\leq (n+4)/8$, depicted as a dashed line in Fig.~\ref{lpbound-binary}, seems to suggest the celebrated Leech lattice, the unique self-dual lattice in $d=24$ with no vectors shorter than $\ell^2=4$, will make an appearance when $n=12$, saturating the bound. But this is not the case. First, the numerical bound on  the spectral gap  is close, but is strictly smaller than $\Delta< 2$ \cite{afkhami2020free}. Second, the Leech lattice understood as a self-dual Lorentzian lattice is odd, see Appendix~\ref{appdx:g24}. That leaves  the possibility for the Leech lattice to define some special non-chiral fermionic CFT with large spectral gap, a question we leave for the future. 

The largest achievable binary Hamming distance for real self-dual codes with $n=12$ is $d_{\rm b}=6$.  It corresponds to spectral gap $\Delta=d_{\rm b}/4=3/2$.  As we have mentioned already, the Construction A lattice $\Lambda(\C)$ of any stabilizer code necessarily has vectors of length $\ell^2=2$, which limits the spectral gap to $\Delta\leq 1$. Nevertheless in certain cases one can apply a twist by a half lattice vector $\delta$ to attain larger spectral gaps. To turn an even lattice into an even lattice, $\delta^2$ should be odd. Assuming the vector $\vec{1}$ is one of the codewords, when $n/4$ is odd e.g.~for $n=12$, a twist by $\delta=\vec{1}/(2\sqrt{2})$ will yield a new even self-dual Lorentzian lattice, whose corresponding Narain CFT has spectral gap $\Delta=d_{\rm b}/4$. The Siegel theta-function and hence the partition function of the corresponding CFT is given by \eqref{twisted}. This procedure  is universal, and can be applied to any code  whose REP includes the term $y^{12}$.

\begin{figure}
\centering
\includegraphics[width=0.3\textwidth]{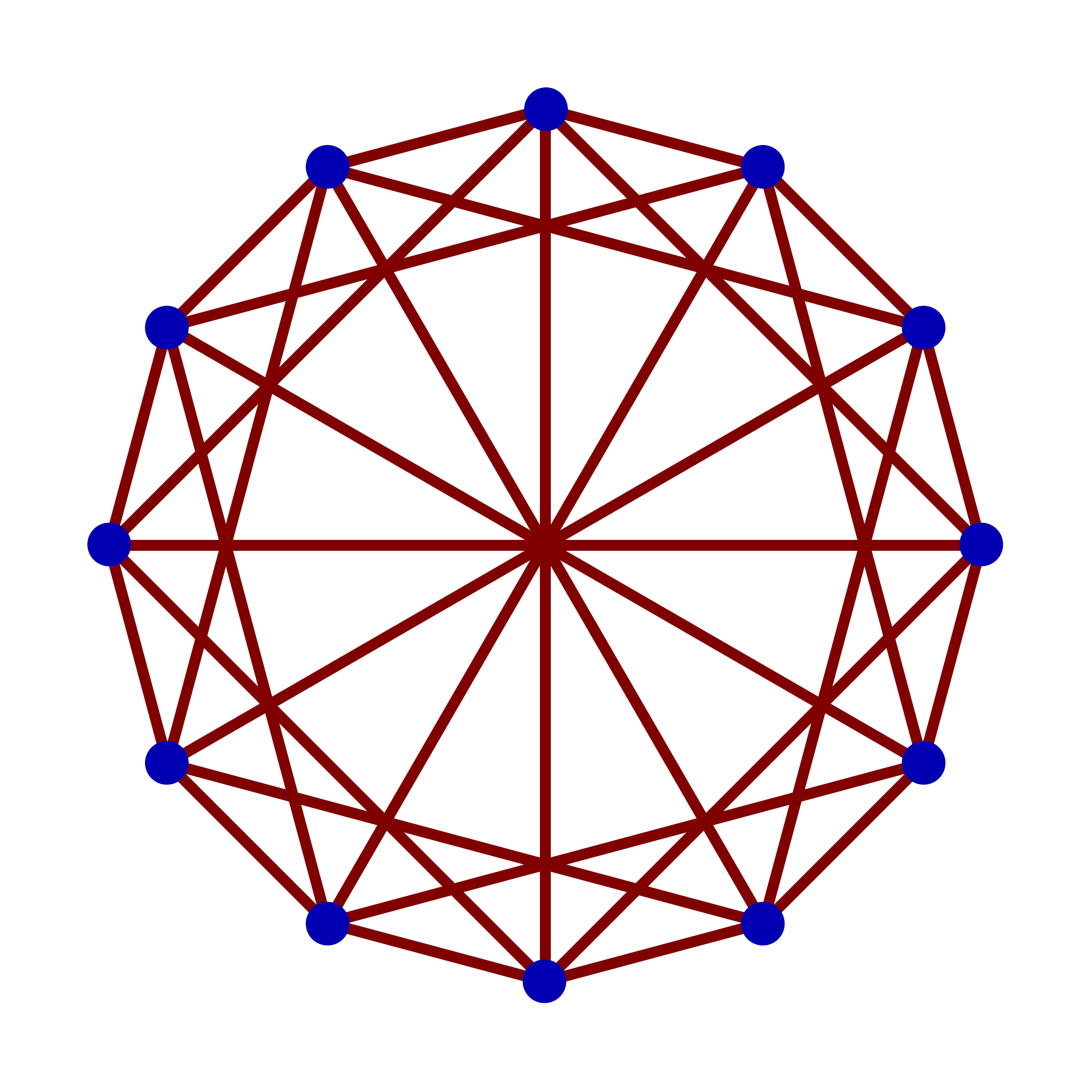}
\caption{Graph of \eqref{dodeca}, a representative from one of the T-equivalence classes associated with the dodecacode. The corresponding  refined enumerator polynomial  is given by \eqref{W3-L}. As a binary isodual code it is isospectral with $h_{24}^+$. Via the twist construction it gives rise to a non-integer isodual lattice isospectral with the odd Leech lattice.}
\label{fig:dodecacode}
\end{figure}

There is a unique equivalence class of codes with the largest possible Hamming  distance $d=6$, the so-called the dodecacode \cite{glynn2004geometry}. Real codes belonging to this class split into three\footnote{\label{3} Strictly speaking we should say at least three, as potentially there could be isospectral  classes of  T-equivalent codes, which are C-equivalent but not T-equivalent with each other.} classes of T-equivalent codes, with the refined enumerator polynomials 
\bea
W_{\rm I}&=&W_{\rm III}-2 W_1^2 {\cal R}^3, \\
W_{\rm II}&=&W_{\rm III}+4 W_1^2 {\cal R}^3, \\ \nonumber
W_{\rm III}&=&x^{12}+2 x^6 y^6+60 x^6 y^4 z^2+270 x^6 y^2 z^4+64 x^6 z^6+60 x^4 y^6 z^2+480 x^4 y^4 z^4+\\ && \nonumber 840 x^4 y^2 z^6+105 x^4 z^8+270 x^2 y^6 z^4+840 x^2 y^4 z^6+810 x^2 y^2 z^8+60 x^2 z^{10}+y^{12}+\\ && 64 y^6 z^6+105 y^4 z^8+60 y^2 z^{10}+4 z^{12}. \qquad
\label{W3-L}
\eea
Of course all three REPs correspond to the same enumerator polynomial 
\bea
W=x^{12}+396 x^6 y^6+1485 x^4 y^8+1980 x^2 y^{10}+234 y^{12}.
\eea
All three classes have $d=d_{\rm b}=6$. They can be brought to the canonical, or B-form, with many possible matrices $\rm B$. Here we give a representative from each  class of T-equivalent codes, in the notation of \eqref{Bk},
\bea
k_1&=&12020990775258723326,\\
k_2&=&8432846454558968306,\\
k_3&=&47473099643714589357. 
\label{dodeca}
\eea
Understood as binary codes, via the Gray map, these codes are isodual but not self-dual. For B-form codes this means the matrices $\rm B$ are symmetric but do not satisfy ${\rm B}\,{\rm B}^T={\rm I}$. It is therefore remarkable that enumerator polynomial of the third class, with the graph shown in Fig.~\ref{fig:dodecacode}, understood as classical binary code, is the same as the enumerator polynomial of the odd Golay code $h_{24}^+$,
\bea
\nonumber
W_{\rm III}(x^2,y^2,xy)=x^{24}+64 x^{18} y^6+375 x^{16} y^8+960 x^{14} y^{10}+1296 x^{12} y^{12}+\\ 960 x^{10} y^{14}+375 x^8 y^{16}+64 x^6 y^{18}+y^{24}.
\label{oddgolay}
\eea
The odd Golay code is a self-dual $[24,12,6]$ binary code. It is even but not doubly-even, which means corresponding Construction A lattice is self-dual and odd. Applying twist with $\vec{\delta}=\vec{1}/2/\sqrt{2}$, one obtains the odd Leech lattice, the unique self-dual odd lattice in $24$ dimensions, with shortest vector of length-squared $\ell^2=3$. Its theta-function, given by \eqref{oddgolay} and \eqref{twistedbinary}, is
\bea
\Theta_{\rm Odd\,  Leech}=1+4096\, q^{3/2}+98256\, q^2+1130496\, q^{5/2}+18384512\, q^3+\dots \label{OLT}
\eea
The odd Leech lattice can be understood as an odd self-dual Lorentzian lattice, which means that the generator of the $h^+_{24}$ code can be brought to the canonical form \eqref{canonical} with a symmetric $\rm B$, in which not all ${\rm B}_{ii}$ are zero.\footnote{If one defines $h_{24}^+$ using the generator matrix given in Fig.~12.1 of \cite{nebe2006self}, the permutation $\{1, 2, 3, 4, 5, 6, 7, 9, 10, 11, 20, 24, 15, 14, 13, 16, 19, 18, 17, 23, 22, 21, 8, 12\}$ brings it to a form that is self-dual with respect to \eqref{Mink}.}

Going back to the stabilizer codes from the third class \eqref{dodeca}, see Fig.~\ref{fig:dodecacode}, they correspond to an even self-dual Lorentzian lattice, which, understood as  a Euclidean lattice, is isodual and isospectral with the Construction A lattice of $h_{24}^+$. If we apply a twist  with $\vec{\delta}=\frac{\vec{1}}{2\sqrt{2}}$, we obtain an even self-dual Lorentzian lattice, which, as a Euclidean lattice, is isodual and isospectral with the odd Leech lattice. Its Siegel theta-function is given by \eqref{twisted} and reduces to \eqref{OLT} along the diagonal $\tau'=-\tau$.
The Narain CFT defined with this lattice has spectral gap $\Delta=d_{\rm b}/4=3/2$. It should also be noted that codes from two other classes, via the same twist procedure,  also lead to Narain theories with $\Delta=d_{\rm b}/4=3/2$ .

Finally, returning to the Golay code, its matrix $\rm B$ in the canonical form \eqref{BGolay} is symmetric but ${\rm B}_{ii}\neq 0$. This means that the Golay code can be interpreted as a self-dual stabilizer code, which is not real. We can use code equivalence to find C-equivalent real self-dual codes. There are three classes of T-dual codes (strictly speaking, at least three, see footnote \ref{3}): one has $d=4$ and $d_{\rm b}=6$ (and can be used to construct a Narian CFT with $\Delta=3/2$ via twist construction), and other two have a more modest $d=d_{\rm b}=4$. A curious observation here is that the matrix 
$\rm B$ \eqref{BGolay} with all diagonal elements set to zero gives rise to the stabilizer code, which shares the same REP with one of the  $d=d_{\rm b}=4$ classes mentioned above. This seems to indicate that for this matrix $\rm B$, C-equivalence  can  simply remove all non-zero diagonal matrix elements while leaving  everything else intact. This is an unusual situation, and it would be interesting to describe the class of matrices $\rm B$ for which this is possible.

\section{Conclusions}
\label{sec:con}
In this paper we have discussed a relation between quantum stabilizer codes, a particular class of quantum error-correcting codes, and a class of 2d conformal field theories. The key ingredient in our construction is the relation between stabilizer codes and Lorentzian lattices, which is the subject of Section \ref{sec:newA}.  Self-dual quantum stabilizer codes correspond to self-dual lattices, and real codes to even lattices. In this way, real self-dual codes define CFTs based on even self-dual lattices, which we call code theories. Basic properties of code CFTs are captured by the corresponding codes; in particular, the CFT partition function is given by the code's refined enumerator polynomial \eqref{ZcodeCFT}. 
Qualitatively, classical codes are  related to Euclidean lattices and chiral CFTs. In this paper we have shown that quantum codes correspond to Lorentzian lattices and non-chiral CFTs.

Our main focus has been on self-dual codes and lattices. The space of stabilizer codes is discrete, but in our construction it is embedded in the continuous space of Lorentzian self-dual lattices. This is an essential  difference from the case of classical codes, for which the space of Euclidean self-dual lattices is discrete. Within the Narain moduli space of all self-dual even Lorentzian lattices, we describe the set of real self-dual stabilizer codes as a group coset \eqref{CN}. There are other spin-off results  which can be formulated without references to CFTs. We have derived the Gilbert-Varshamov bound by averaging over all codes in canonical form, and calculated linear programming bounds on the largest binary Hamming distance, see Section \ref{sec:GV}. At the level of graphs, informed by the CFT interpretation, we outlined the importance of edge local complementation (ELC) equivalence classes, and classified all graphs on $n\leq 8$ nodes, see Appendix~\ref{sec:graphs}. Finally, we constructed an isodual non-integral lattice isospectral to odd Leech lattice in Section~\ref{sec:n=12}. 

Code theories form a subsector of Narain CFTs. T-duality transformations can map a code theory into another code theory, in which case the corresponding codes are necessarily equivalent in the code sense, as proved in Appendix~\ref{sec:TD}. Using T-duality transformations one can always bring any code CFT into the form of a compactification on an $n$-dimensional cube of ``unit'' size and  quantized $B$-flux, such that it is fully specified by a binary symmetric matrix ${\rm B}=B\, \,{\rm mod}\, \, 2$. The matrix $\rm B$ can be interpreted as the graph adjacency matrix. Thus code theories can be labeled by graphs, with  graphs of T-dual theories being related by edge local complementation. By classifying all classes of ELC equivalent graphs on $n\leq 8$ nodes we have found all physically distinct code CFTs with central charge $c=n\leq 8$.

Schematically, one can think of code theories as a particular ``ansatz'' which reduces modular invariance of the CFT partition function to a simple algebraic condition satisfied by a multivariate polynomial. In this way code theories provide a playground to probe several questions central to the conformal modular bootstrap program. As in the case of classical binary codes which give rise to optimal sphere packings in certain  dimensions, a particular code theory which we dub non-chiral $E_8$, and which is $\widehat{\rm SO(8)}_1$ WZW theory in disguise, attains the maximal value of the spectral gap among the Narain theories with central charge $c=n=4$. This theory is based on the root lattice of $E_8$, understood as a Lorentzian even self-dual lattice, see Section \ref{sec:n=4}. Other special lattices,  in particular the odd Leech lattice, also make an appearance, see Section \ref{sec:n=12}.  A drastic reduction of the modular invariance constraints at the level of code theories gives us a multitude of examples of ``fake'' CFT partition functions, modular invariant non-chiral functions $Z(\tau,\bar \tau)$, which admit expansions in terms of $U(1)^n\times U(1)^n$ characters with positive  integer coefficients (the first being $1$), yet do not correspond to any known theory. The number of fake $Z(\tau,\bar\tau)$'s quickly grows with $c=n$, which suggests one of two possibilities. It could be that these are not partition functions of any actual CFT, which means persistent allowed regions in  modular bootstrap exclusion plots  in fact might be empty. Another possibility is that these $Z(\tau,\bar\tau)$'s might correspond to actual CFTs from some new sector, most likely related to a family of (non-additive) codes. This would mean that the notion of a code CFT could be extended to include these  and perhaps other sets of theories.  (We also mention that a completely analogous construction exists for classical binary codes leading to examples of ``fake'' chiral  CFT partition functions for $c\geq 24$ divisible by $8$, see Section~\ref{sec:binary}.)

Finally, our analysis of stabilizer codes with small $n\leq 12$ reveals a growing number of isospectral but physically inequivalent Narain CFTs. From the mathematical point of view these are examples of isospectral but non-isomorphic Narain lattices. The first such example appears for $n=7$; it corresponds to a pair of isospectral even self-dual Lorentzian lattices in $\R^{7,7}$, see Fig.~\ref{fig:Fig7fish}.  For chiral CFTs based on Euclidean lattices, the lowest-dimensional pair of isospectral CFTs are the $E_8 \times E_8$ and ${\rm Spin}(32)/\Z_2$ lattice CFTs corresponding to Milnor's example of isospectral even self-dual lattices in 16 dimensions. 
In contrast to the Euclidean case, where next example occurs in $24$ dimensions,
there are many dozens of examples of isospectral $c=n=8$ theories, with the number presumably growing rapidly for larger $c=n$. 


Code CFTs may provide a useful framework for addressing the following two questions. The first is to understand the asymptotic behavior of the maximal spectral gap for Narain theories with $c=n\gg 1$ \cite{afkhami2020high,afkhami2020free}. At the level of code theories, the analog of the spectral gap is the binary Hamming distance $d_{\rm b}$, which can be effectively studied using linear programming methods. It is an open question though to relate quantum codes with large $d_{\rm b}$ to Lorentzian lattices with large shortest vector. To that end one needs to go beyond Construction A lattices, discussed in this paper, and introduce some analogs of constructions B, C etc.~developed for classical codes \cite{conway2013sphere}.  Another question is the recently proposed holographic duality between averaged Narain theories and certain Chern-Simons theories in the bulk \cite{afkhami2020free,maloney2020averaging}. We have argued in Section \ref{sec:GV} that the ensemble average over all code theories exhibits the same basic features as the average over  full Narain moduli space, suggesting a holographic interpretation. Thus, code theories may provide an additional testbed to verify and study this duality. 

There are several different ways in which classical codes may be associated with various chiral CFTs, both supersymmetric and not \cite{dolan1996conformal,gaiotto2018holomorphic}. We expect the construction outlined in this paper to be perhaps the simplest but not the only scheme relating quantum codes to non-chiral CFTs.  We already mentioned a possible connection between self-dual albeit non-real stabilizer codes, associated with self-dual odd Lorentzian lattices, and fermionic CFTs. But we expect that many other constructions are possible. Perhaps the most important aspect of the relation between  Euclidean lattices and chiral CFTs is that the former can be used to define consistent Vertex Operator Algebras (VOA). Thus, the VOA associated with the Leech lattice, and its Monster orbifold, exhibits  symmetries which go beyond pure geometric symmetries of the lattice \cite{frenkel1989vertex,borcherds1992monstrous}.
In light of our work, one of the immediate questions would be to study symmetries of the non-chiral VOAs associated with code CFTs, possibly leading to a non-chiral moonshine theory. 

Let us conclude with one more fundamental question: to what extent does the physical Hilbert space of a code theory exhibit quantum error-correcting properties related, or inherited, from the associated codes? Here we have in mind various properties, including ``quantum error
correction'' necessitated by the emergence of locality in the bulk
\cite{almheiri2015bulk} or related to the large N limit
\cite{Milekhin:2020zpg}, quantum information properties of CFT
ground states \cite{white2020conformal}, and probably many others.

\acknowledgments
We would like to thank Noam Elkies, Anton Gerasimov,  Nikita Nekrasov, Vasily Pestun,  and Eric Rains for discussions. AD is supported by the National Science Foundation under Grant No.~PHY-1720374. He would like to thank IHES for hospitality during the visit, which was supported  by funds from the European Research Council (ERC) under the European Union's Horizon 2020 research and innovation program (QUASIFT grant agreement 677368).



\appendix
\section{$E_7$ and $E_8$ lattices and codes}\label{appdx:e7e8}
In this section we show that root lattices of Lie algebras $E_7$ and $E_8$ are isomorphic to the Construction A lattices of Hamming $[7,4,3]$ code and the extended  $[8,4,4]$ code $e_8$. We start with the case of $E_8$ as it is more symmetric and simpler. For $E_8$ we also discuss equivalence of different Lorentz-signature metrics and the relation of non-chiral $E_8$ theory to the theory of eight free Majorana fermions. 

\subsection{$E_8$}\label{sec:aE8}
Root lattice of $D_n$ series is the ``checkerboard'' lattice of integer vectors $(x_1,\dots,x_n)\in \Z^n$ with the sum of all coordinates being even, $\sum_i x_i\, \, {\rm mod}\, \, 2=0$. We denote it as $D_n$. Vector $\vec{\delta}=\vec{1}/2$ does not belong to the lattice, but when $n$ is even, $2\delta$ does. In a procedure similar to the twist described in Section \ref{sec:binary}, we can define a new lattice 
\bea
D_n^+=D_n \cup (D_n+\delta),
\eea 
where $D_n+\delta$ is defined as  in \eqref{sumdelta}. For $n=8$ this lattice is the root lattice of algebra $E_8$. It includes vectors of the form $(x_1,\dots,x_n)$ where all $x_i$ are simultaneously either integer or half integer, and their sum is integer and even. One can choose 
\bea
\Uplambda_{E_8}=
\left(
\begin{array}{cccccccc}
 1 & 0 & 0 & 0 & 0 & 0 & -\frac{1}{2} & 0 \\
 -1 & 1 & 0 & 0 & 0 & 0 & -\frac{1}{2} & 0 \\
 0 & -1 & -1 & 0 & 0 & 0 & -\frac{1}{2} & 0 \\
 0 & 0 & 1 & -1 & 0 & 0 & -\frac{1}{2} & 0 \\
 0 & 0 & 0 & 1 & -1 & 0 & -\frac{1}{2} & 0 \\
 0 & 0 & 0 & 0 & 1 & 1 & -\frac{1}{2} & 1 \\
 0 & 0 & 0 & 0 & 0 & 1 & -\frac{1}{2} & -1 \\
 0 & 0 & 0 & 0 & 0 & 0 & -\frac{1}{2} & 0 \\
\end{array}
\right)
\eea
as a generator matrix, in which case gram matrix is the Cartan matrix of $E_8$,
\bea
\Uplambda_{E_8}^T\, \Uplambda_{E_8}=
\left(
\begin{array}{cccccccc}
 2 & -1 & 0 & 0 & 0 & 0 & 0 & 0 \\
 -1 & 2 & 1 & 0 & 0 & 0 & 0 & 0 \\
 0 & 1 & 2 & -1 & 0 & 0 & 0 & 0 \\
 0 & 0 & -1 & 2 & -1 & 0 & 0 & 0 \\
 0 & 0 & 0 & -1 & 2 & 1 & 0 & 1 \\
 0 & 0 & 0 & 0 & 1 & 2 & -1 & 0 \\
 0 & 0 & 0 & 0 & 0 & -1 & 2 & 0 \\
 0 & 0 & 0 & 0 & 1 & 0 & 0 & 2 \\
\end{array}
\right).
\eea

The generator matrix $\Uplambda_{E_8}$ is of course very different from the generator of the Construction A lattice $\Lambda(e_8)$ associated with the Hamming $[8,4,4]$ code. The latter is given by \eqref{Lambdabinary} with the matrix $\rm B$ given by \eqref{H8}. We will denote that matrix by $\Uplambda_{e_8}$. The lattices generated by 
$\Uplambda_{E_8}$ and $\Uplambda_{e_8}$ are not identical but isomorphic, which means there is a rotation ${\cal O}\in {\rm O}(8)$ and a matrix $Z\in {\rm GL}(8,\Z)$ such that 
\bea
\Uplambda_{E_8}Z={\cal O}\, \Uplambda_{e_8}. \label{difficultequation}
\eea
Finding ${\cal O}$ and $Z$ directly from \eqref{difficultequation} is difficult, and therefore the  \href{https://en.wikipedia.org/wiki/E8_lattice}{Wikipedia} calls the task of finding the explicit isomorphism ``not entirely trivial.''

The following trick saves the day. There are $240$ roots, vectors of length $\ell^2=2$, which can be written explicitly in both representations, in particular all columns of $ \Uplambda_{e_8}$ are roots. We  consider the Gram matrix 
\bea
\Uplambda_{e_8}^T\, \Uplambda_{e_8}=
\left(
\begin{array}{cccccccc}
 2 & 0 & 0 & 0 & 1 & 1 & 1 & 0 \\
 0 & 2 & 0 & 0 & 1 & 1 & 0 & 1 \\
 0 & 0 & 2 & 0 & 1 & 0 & 1 & 1 \\
 0 & 0 & 0 & 2 & 0 & 1 & 1 & 1 \\
 1 & 1 & 1 & 0 & 2 & 1 & 1 & 1 \\
 1 & 1 & 0 & 1 & 1 & 2 & 1 & 1 \\
 1 & 0 & 1 & 1 & 1 & 1 & 2 & 1 \\
 0 & 1 & 1 & 1 & 1 & 1 & 1 & 2 \\
\end{array}
\right). \label{Grm}
\eea
Our goal now is to choose $8$ roots from the list of $240$ roots of  $\Uplambda_{E_8}$ such that their scalar product is given by \eqref{Grm}. The procedure is iterative. Using computer algebra we calculate the $240\times 240$ matrix of scalar products.  First root is chosen at will. Then we choose second root at will from the set of those which have the desired scalar product with the first one. Third is chosen at will from the list of those which have desired scalar product with the first two, and so on. The procedure does not guarantee to succeed (we may not have a vector with the desired properties at a certain step), but since the lattice has many symmetries it works well in practice. 

Once the roots with the scalar product \eqref{Grm} are found, one can choose them to generate the lattice, which will be related to $\Uplambda_{E_8}$ by an appropriate  ${\rm GL}(8,\Z)$ transformation. That is the desired matrix $Z$. Once $Z$ is known, $\cal O$ follows from \eqref{difficultequation},
\bea
{\cal O}=\left(
\begin{array}{cccccccc}
 0 & 0 & 1 & -\frac{1}{2} & \,  \frac{1}{2} &\,  \frac{1}{2} & -\frac{1}{2} & 0 \\
 0 & 0 & -1 & -\frac{1}{2} &\,  \frac{1}{2} &\,  \frac{1}{2} & -\frac{1}{2} & 0 \\
 0 & 1 & 0 & \, \frac{1}{2} & -\frac{1}{2} & \, \frac{1}{2} & -\frac{1}{2} & 0 \\
 0 & -1 & 0 &\,  \frac{1}{2} & -\frac{1}{2} & \, \frac{1}{2} & -\frac{1}{2} & 0 \\
1 & 0 & 0 & \, \frac{1}{2} &\,  \frac{1}{2} & -\frac{1}{2} & -\frac{1}{2} & 0 \\
 -1 & 0 & 0 &\,  \frac{1}{2} &\,  \frac{1}{2} & -\frac{1}{2} & -\frac{1}{2} & 0 \\
 0 & 0 & 0 &\,  \frac{1}{2} &\,  \frac{1}{2} & \, \frac{1}{2} & \, \frac{1}{2} & -1 \\
 0 & 0 & 0 &\,  \frac{1}{2} &\,  \frac{1}{2} & \, \frac{1}{2} & \, \frac{1}{2} & 1 \\
\end{array}
\right)/\sqrt{2}. \label{Om}
\eea

One can take another route and ``guess'' \eqref{Om}.
Once $\cal O$ is known explicitly, it can be checked straightforwardly that $\cal O$ is orthogonal and solves \eqref{difficultequation} with some appropriate $Z$.

Lattice $E_8$ is even and self-dual, which follows from all diagonal matrix elements of \eqref{Grm} being even, while the matrix is integer and has determinant $1$. Curiously $E_8$ is also even and self-dual with respect to Lorentz signature metric \eqref{Mink}. Indeed, 
\bea
\Uplambda_{e_8}^T\, g\, \Uplambda_{e_8} =\left(
\begin{array}{cccccccc}
 0\, & 0\, & 0\, & 0\, & 0\, & 0\, & 0\, & 1 \\
 0\, & 0\, & 0\, & 0\, & 0\, & 0\, & 1\, & 0 \\
 0\, & 0\, & 0\, & 0\, & 0\, & 1\, & 0\, & 0 \\
 0\, & 0\, & 0\, & 0\, & 1 \,& 0\, & 0\, & 0 \\
 0\, & 0\, & 0\, & 1 \,& 0 \,& 1\, & 1 \,& 1 \\
 0\, & 0\, & 1\, & 0 \,& 1 \,& 0\, & 1 \,& 1 \\
 0\, & 1\,& 0\, & 0 \,& 1 \,& 1\, & 0 \,& 1 \\
 1\, & 0\, & 0\, & 0\, & 1\, & 1\, & 1\, & 0 \\
\end{array}
\right) \in {\rm GL}(8,\Z),
\eea
from where follows that it is self-dual. It is also even because all diagonal elements of the gram matrix are even. (Alternatively one can flip signs in $\rm B$ to make it antisymmetric. The lattice would remain the same, but  now $\Uplambda_{e_8}$ would be orthogonal matrix from ${\rm O}(4,4,\R)$, which guarantees that the lattice is even and self-dual.) In Section \eqref{sec:n=4} we used $E_8$ understood as a Lorentzian  lattice to define ``non-chiral $E_8$''  Narain CFT.

An immediate check reveals that the lattice generated by $\Uplambda_{E_8}$ is also even self-dual with respect to the same metric $g$. We leave the exercise of calculating $\Uplambda_{E_8}^T\, g\, \Uplambda_{E_8}$ to the reader. This is curios now, because it means lattice generated by $\Uplambda_{e_8}$ is even self-dual with respect to both metrics, $g$  and 
\bea
\eta={\cal  O}^T g\,  {\cal O}.
\eea
One can immediately ask, what is the Narain CFT defined with help of $\eta$? It turns out, this is  the same theory because of the lattice symmetry. We consider an orthogonal transformation of the form 
\bea
\label{THOH}
T={\cal H}({\cal O}_L \times {\cal O}_R){\cal H},
\eea
where ${\cal O}_{L,R}\in {\rm O}(4,\R)$, and $8\times 8$ block-matrix 
\bea
{\cal H}=\left(\begin{array}{cc}
{\rm I}\, &\,\, {\rm I}\\ {\rm I}\, & -{\rm I}
 \end{array}\right)/\sqrt{2},
\eea
performs the transformation \eqref{ab}. Then $T$ is a symmetry of $g$, $T^T g\, T=g$. (In physics terms, the transformation ${\cal O}_L \times {\cal O}_R$ is a part of T-duality group
which rotates $p_L$ and $p_R$.) Accordingly, the orthogonal matrix $S=T\, {\cal O}$ satisfies,
\bea
\eta=S^T g\, S.
\eea
It turns out that for the particular choice of 
\bea
{\cal O}_L=\left(
\begin{array}{cccc}
 0 & 0 & -1 & -1 \\
 -1 & -1 & 0 & 0 \\
 0 & 0 & -1 & 1 \\
 1 & -1 & 0 & 0 \\
\end{array}
\right),\qquad {\cal O}_R=\left(
\begin{array}{cccc}
 0 & 0 & 1 & 1 \\
 -1 & -1 & 0 & 0 \\
 0 & 0 & -1 & 1 \\
 -1 & 1 & 0 & 0 \\
\end{array}
\right),
\eea
$S$ is a symmetry of the lattice, 
\bea
S\, \Uplambda_{e_8} =\Uplambda_{e_8} Z_S,\qquad Z_S\in {\rm GL}(8,\Z).
\eea
Therefore Narain CFTs defined with the lattice $\Lambda(e_8)$ understood as the Lorentzian lattice with  metrics $g$ and $\eta$ are T-dual to each other. 

More generally, the $E_8$ lattice has a rich group of symmetries, most of which do not respect the Lorentzian metric, ``rotating'' it into a new one. Narain CFTs defined with any choice of the Lorentzian  metric are physically equivalent to each other.

Finally we discuss the equivalence of the non-chiral $E_8$ theory with the theory of eight free fermions with the diagonal GSO projection. The fermions  can be bosonised, leading to toroidal compactification  $D_4$ --  root lattice of SO(8) \cite{witten1983d} with the generator matrix 
\bea
\gamma_{D_4}=\left(
\begin{array}{cccc}
 1 & 0 & \, 0\, & 0 \\
 -1 & 1 & \, 0\, & 0 \\
 0 & -1 & \,1\, & 1 \\
 0 & 0 & -1 & 1 \\
\end{array}
\right).
\eea 
The B-field
\bea
B=\left(
\begin{array}{cccc}
 0 & -1 & -1 & 0 \\
 1 & 0 & -1 & 0 \\
 1 & 1 & 0 & 0 \\
 0 & 0 & 0 & 0 \\
\end{array}
\right)
\eea is chosen such that upper triangular parts of $\gamma_{D_4}^T \gamma_{D_4}$ and 
$\gamma_{D_4}^T B \gamma_{D_4}$ coincide, leading to ${\rm SO}(8)\times {\rm SO}(8)$ global symmetry \cite{englert1985non,elitzur1987aspects,lerchie1989lattices}. Resulting Lorentzian lattice with the generator 
\bea
\Uplambda_{\widehat{\rm SO(8)}_1}=\left(\begin{array}{c|c}
2(\gamma_{D_4}^T)^{-1} & B \gamma_{D_4} \\ \hline
0 & \gamma_{D_4}
\end{array}\right)/\sqrt{2}
\eea
describes $\widehat{\rm SO(8)}_1$ WZW theory as  a Narain CFT. It is related to the lattice generated by $\Uplambda_{e_8}$ by a T-duality transformation \eqref{THOH} with either ${\cal O}_L$ or ${\cal O}_R$ flipping sign of one arbitrary coordinate.


\subsection{$E_7$}\label{sec:aE7}
Root lattice $E_7$ can be defined via generator matrix
\bea
\Uplambda_{E_7}=
\left(
\begin{array}{ccccccc}
 1 & 0 & 0 & 0 & 0 & -\frac{1}{2} & 0 \\
 -1 & 1 & 0 & 0 & 0 & -\frac{1}{2} & 0 \\
 0 & -1 & 1 & 0 & 0 & -\frac{1}{2} & 0 \\
 0 & 0 & -1 & 1 & 0 & -\frac{1}{2} & 0 \\
 0 & 0 & 0 & -1 & 1 & -\frac{1}{2} & 1 \\
 0 & 0 & 0 & 0 & 1 & -\frac{1}{2} & -1 \\
 0 & 0 & 0 & 0 & 0 & \frac{1}{\sqrt{2}} & 0 \\
\end{array}
\right),
\eea
such that gram matrix is the Cartan matrix of $E_7$ Lie algebra,
\bea
\Uplambda_{E_7}^T \Uplambda_{E_7}=\left(
\begin{array}{ccccccc}
 2 & -1 & 0 & 0 & 0 & 0 & 0 \\
 -1 & 2 & -1 & 0 & 0 & 0 & 0 \\
 0 & -1 & 2 & -1 & 0 & 0 & 0 \\
 0 & 0 & -1 & 2 & -1 & 0 & -1 \\
 0 & 0 & 0 & -1 & 2 & -1 & 0 \\
 0 & 0 & 0 & 0 & -1 & 2 & 0 \\
 0 & 0 & 0 & -1 & 0 & 0 & 2 \\
\end{array}
\right). \label{E7cartan}
\eea

The generating matrix of the Hamming $[7,3,4]$ code is the transpose of \eqref{H7}. 
The generator matrix of the Construction A lattice of this code can be chosen as
\bea
\Uplambda_{e7}=
\left(
\begin{array}{ccccccc}
 2 & 0 & 0 & 0 & 1 & 0 & 0 \\
 0 & 2 & 0 & 0 & 0 & 1 & 0 \\
 0 & 0 & 2 & 0 & 1 & 1 & 0 \\
 0 & 0 & 0 & 2 & 0 & 0 & 1 \\
 0 & 0 & 0 & 0 & 1 & 0 & 1 \\
 0 & 0 & 0 & 0 & 0 & 1 & 1 \\
 0 & 0 & 0 & 0 & 1 & 1 & 1 \\
\end{array}
\right)/\sqrt{2}.
\eea
To match the the lattice generated  by $\Uplambda_{e_7}$ with the one generated by 
$\Uplambda_{E_7}$, we will employ the procedure analogous to the one used in the previous section. We construct $126$ roots of the code lattice, which include $14$ vectors of the form $(\pm 2,0^6)/\sqrt{2}$ (and permutations), and $2^4 \times 7$ vectors obtained from the $7$ codewords of Hamming weight $4$. Then we calculate the $126\times 126$ scalar product matrix, and start choosing roots one by one such that their scalar product is equal to \eqref{E7cartan}. The process does not need to succeed and in practice we had to experiment with a few different candidates for the fifth vector, before the process could be completed. 
Once those roots are identified, we can solve a system of linear equations to find a matrix $Z^{-1}\in {\rm GL}(7,\Z)$ which expresses those roots in terms of $\Uplambda_{e_7}$. After that an orthogonal matrix ${\cal O}$ satisfying 
\bea
\Uplambda_{E_8}={\cal O} \Uplambda_{e_8} Z^{-1},
\eea
can be easily found
\bea
\left(
\begin{array}{ccccccc}
 1 & 0 & 0 & 0 & -1 & 0 & 0 \\
 -1 & 0 & 0 & 0 & -1 & 0 & 0 \\
 0 & 0 & 1 & 0 & 0 & 0 & 1 \\
 0 & 0 & -1 & 0 & 0 & 0 & 1 \\
 0 & 1 & 0 & 0 & 0 & 1 & 0 \\
 0 & -1 & 0 & 0 & 0 & 1 & 0 \\
 0 & 0 & 0 & -\sqrt{2} & 0 & 0 & 0 \\
\end{array}
\right).
\eea

\section{Golay code and Leech lattice}\label{appdx:g24}
Binary extended $[24,12,8]$ Golay code $g_{24}$ can be defined using generator matrix in the canonical form \eqref{canonical} with 
\bea
{\rm B}=\left(
\begin{array}{cccccccccccc}
 1\, & 0\, & 0\, & 1\, & 1\, & 1\, & 1\, & 1\, & 0\, & 0\, & 0\, & 1\, \\
 0 & 1 & 0 & 0 & 1 & 1 & 1 & 1 & 1 & 0 & 1 & 0 \\
 0 & 0 & 1 & 0 & 0 & 1 & 1 & 1 & 1 & 1 & 0 & 1 \\
 1 & 0 & 0 & 1 & 0 & 0 & 1 & 1 & 1 & 1 & 1 & 0 \\
 1 & 1 & 0 & 0 & 1 & 0 & 0 & 1 & 1 & 1 & 0 & 1 \\
 1 & 1 & 1 & 0 & 0 & 1 & 0 & 0 & 1 & 1 & 1 & 0 \\
 1 & 1 & 1 & 1 & 0 & 0 & 1 & 0 & 0 & 1 & 0 & 1 \\
 1 & 1 & 1 & 1 & 1 & 0 & 0 & 1 & 0 & 0 & 1 & 0 \\
 0 & 1 & 1 & 1 & 1 & 1 & 0 & 0 & 1 & 0 & 0 & 1 \\
 0 & 0 & 1 & 1 & 1 & 1 & 1 & 0 & 0 & 1 & 1 & 0 \\
 0 & 1 & 0 & 1 & 0 & 1 & 0 & 1 & 0 & 1 & 1 & 1 \\
 1 & 0 & 1 & 0 & 1 & 0 & 1 & 0 & 1 & 0 & 1 & 1 \\
\end{array}
\right).
\label{BGolay}
\eea
This is a self-dual code as follows from ${\rm B}\, {\rm B}^T={\rm I}$, understood over ${\rm GF}(2)$.
Alternatively, one can define the generator matrix of the Construction A lattice $\Lambda(g_{24})$
\bea
\Uplambda_{g_{24}}=\left(\begin{array}{c|c}
2\,{\rm I}\,  &\,\, {\rm B}^T \\ \hline
\, 0\, \, &\, \,  {\rm I}\end{array}\right)/\sqrt{2},
\eea
and check that $\Uplambda^T \Uplambda$ is integer, unimodular, and with even numbers of the diagonal. 

Leech lattice can be obtained from $\Lambda(g_{24})$ by applying twist \eqref{shift} with the vector $\vec{\delta}=\vec{1}/2/\sqrt{2}$.

We have seen in Section \ref{sec:aE8} that $E_8$ lattice can be understood as a Lorentzian even self-dual lattice. It can be used to define non-chiral CFT with the largest spectral gap $\Delta=1$ for the given value of central charge (and $U(1)^4\times U(1)^4$ symmetry). This extremal property  can be traced to the lattice $E_8$ being the optimal sphere packing in $8$ Euclidean dimensions, with the spectral gap being specified by the maximal possible length of the shortest lattice vector, $2\Delta=\ell^2=2$. Given that Leech lattice yields the optimal sphere packing in $24$ dimensions with the shortest vector of length $\ell^2=4$, provided it can be interpreted as the Lorentzian lattice, it would lead to a non-chiral theory with the spectral gap $\Delta=2$. It has been recently shown using numerical modular bootstrap that the spectral gap for all theories with $n=12$ (and $U(1)^{12}\times U(1)^{12}$ symmetry)  is strictly smaller than $2$ \cite{afkhami2020free}. This indirectly proves Leech lattice is not an even self-dual lattice for any Lorentzian metric with the $\R^{n,n}$ signature. Here we provide an independent and more explicit consideration, underscoring the difference between  Leech lattice and $E_8$.

Our starting point is the Golay code $g_{24}$. If we could interpret it, via Gray map, as the self-dual stabilizer code, that would immediately show that $\Lambda(g_{24})$ is even self-dual Lorentzian lattice. Then applying twist with the same $\vec{\delta}=\vec{1}/2/\sqrt{2}$ would immediately yield Leech lattice, now as the Lorentzian even and self-dual. In other words we would like to interpret the generator matrix $G^T=(\, {\rm I}\, |\, {\rm B})$ of the binary Golay code as the generator matrix of the real stabilizer code. For that we need ${\rm B}={\rm B}^T$, which is satisfied, but also ${\rm B}_{ii}=0$, which is not. In other words, understood as the stabilizer code,  
Golay code is self-dual but not real. Therefore corresponding lattice $\Lambda(g_{24})$, understood as a Lorentzian lattice, is self-dual but odd (one can check that $\Uplambda_{g_{24}}^T\, g\, \Uplambda_{g_{24}}$ is integer unimodular matrix with odd numbers on the diagonal). Proceeding to define Leech lattice via $\delta$-twist would yield an odd self-dual lattice.  

One may wonder if one can use code equivalences to define a new code with ${\rm B}$ being symmetric and $B_{ii}=0$. The transformations of $\rm B$ include permutations ${\rm B}\rightarrow {\rm B}\,{\cal O}_p$, as well as (compare with \eqref{elc})
\bea
{\rm B}=\left(
\begin{array}{c|c}
b_{11} &\, b_{12} \\ \hline
b_{21}\, &\, \, b_{22}
\end{array}
\right)\quad \rightarrow 
\quad {\rm B}'=\left(
\begin{array}{c|c}
b_{11}^{-1}\, & b_{11}^{-1} b_{12} \\ \hline
b_{21} b_{11}^{-1}\, &\, b_{22}+b_{21}\, b_{11}^{-1}\, b_{12}
\end{array}
\right),   \label{binaryBdual}
\eea
where all algebra is over ${\rm GF}(2)$. It is assumed in \eqref{binaryBdual} that sub-matrix $b_{11}$ is not degenerate. 

Matrix ${\rm B}$ is not necessarily symmetric and may have non-zero diagonal elements. But if ${\rm B}={\rm B}^T$ and ${\rm B}_{ii}=0$, \eqref{binaryBdual} respects this property. Therefore if we hope to bring \eqref{BGolay} to the form ${\rm B}={\rm B}^T, {\rm B}_{ii}=0$, we must do it solely using permutations ${\rm B}\rightarrow {\rm B}\,{\cal O}_p$. It can be easily seen, this is not possible. 

To summarize, Leech lattice, as a Lorentzian lattice, is self-dual and odd.

\section{Any Narain CFT is a toroidal compactification}
\label{sec:ToroidalC}
We want to show that using symmetries of the physical theory, namely ${\rm O}(d)\times {\rm O}(d)$ transformations, any even self-dual Lorentzian lattice (the so called Narain lattice), can be brought to the form \eqref{TCN}.

Our starting point is the equation  \eqref{gm}, which states that any Narain lattice can be obtained from the cubic lattice  with the generator matrix ${\rm I}$ by an appropriate transformation from
${\rm O}(d,d)$. 
Let us denote  first $d$ vectors (columns) of the generator matrix   ${\rm I}$ by $u_i$ and last $d$ vectors (columns) by $\tilde{u}_i$. They satisfy 
\bea
u_i\cdot u_j=0,\quad \tilde{u}_i\cdot \tilde{u}_j=0,\quad u_i\cdot \tilde{u}_j=\delta_{ij}. \label{mainp}
\eea
Since the transformation from ${\rm O}(d,d)$ leaves metric invariant, we can say that an arbitrary lattice $\Lambda$ is generated by $2d$ vectors $u_i,\tilde{u}_j$ satisfying \eqref{mainp}. Let's start with $u_1$. It is a null-vector, $|u_1|^2=0$, and therefore if we represent it in the $\vec{u}_1=(\vec{k}^1_L,\vec{k}^1_R)$ coordinates, vectors $\vec{k}^1_L$ and $\vec{k}^1_R$ will have the same length. Using a transformation  from ${\rm O}(d)$ we can bring $\vec{k}^1_R$ to be equal to $\vec{k}^1_L$ (and will be denoted simply as $\vec{k}_1$). Next we consider vector $\vec{u}_2=(\vec{k}^2_L,\vec{k}^2_R)$. For the same reason $|\vec{k}^2_L|=|\vec{k}^2_R|$ and moreover  
\bea
\vec{k}_1\cdot \vec{k}_L^2=\vec{k}_1\cdot \vec{k}_R^2.
\eea
By an orthogonal transformation in the directions orthogonal to $\vec{k}_1$ we can make $\vec{k}_L^2=\vec{k}_R^2=\vec{k}_2$. Continuing this logic, we find
\bea
\vec{u}_i=(\vec{k}_i,\vec{k}_i).
\eea
We can repeat the same procedure for the vectors $\tilde{u}_i$, but in this case orthogonal transformation acting on $\vec{u}_i$ will bring them to the form 
\bea
\vec{u}_i=(\vec{k}_i,{\mathcal  O}\vec{k}_i),\qquad 
\vec{\tilde{u}}_i=(\vec{\tilde{k}}_i,\vec{\tilde{k}}_i).
\eea
where ${\cal O}\in {\rm O}(d)$. We can find an orthogonal matrix ${\cal Q}$ satisfying ${\cal Q}^2{\cal O}=-{\rm I}$, and after a diagonal transformation ${\cal Q}\times {\cal Q}\in {\rm O}(d)\times {\rm O}(d)$ and a trivial redefinition of $\vec{k}_i,\vec{\tilde{k}}_i$ obtain 
\bea
\vec{u}_i=({\mathcal  Q}\vec{k}_i,-{\mathcal  Q}^{-1}\vec{k}_i),\qquad 
\vec{\tilde{u}}_i=(\vec{\tilde{k}}_i,\vec{\tilde{k}}_i).
\eea
Last step is to impose $u_i\cdot \tilde{u}_j=\delta_{ij}$.
Vectors $\vec{\tilde{k}}_i$ define a lattice, which we can take to be $\Gamma^*$. Vectors $\vec{k}_i$ satisfy
\bea
\vec{\tilde{k}}_i\cdot ({\mathcal  Q}+{\mathcal  Q}^T) \vec{k}_j=\delta_{ij}.
\eea
Therefore vectors $\vec{e}_i=({\mathcal  Q}+{\mathcal  Q}^{T}) \vec{k}_i$ form  lattice $\Gamma$, which is dual to $\Gamma^*$, and antisymmetric matrix $B$ from \eqref{tc} is given by 
\bea
B=({\mathcal  Q}-{\mathcal  Q}^T)({\mathcal  Q}+{\mathcal  Q}^T)^{-1}.
\eea

\section{T-duality as code equivalence}\label{sec:TD}
Starting from a particular code CFT with the code generator matrix \eqref{Gm}, 
\bea
G^T=\left(
\begin{array}{c|c}
\alpha_1\, &\, \beta_1\\
\dots  & \dots \\
\alpha_{n}\, &\, \beta_{n}
\end{array}
\right), \label{Gmsd}
\eea
we would like identify all possible transformations from ${\rm O}(n)\times {\rm O}(n)$ which would  map the code lattice $\Lambda(\C)$ into another code lattice $\Lambda(\C')$ for some other code $\C'$. An element ${\cal O}_L \times {\cal O}_R \in {\rm O}(n)\times {\rm O}(n)$ would act on $G$ as follows
\bea
G \rightarrow G' ={\cal Q}\, G,\qquad {\cal Q}=
{1\over 2}\left(\begin{array}{c|c}
{\cal O}_L +{\cal O}_R\, &\, {\cal O}_L -{\cal O}_R\\
\hline 
{\cal O}_L -{\cal O}_R\, &\, {\cal O}_L +{\cal O}_R 
\end{array}\right) \in {\rm O}(n,n,\R). \label{transform}
\eea
We also remind the reader that $\alpha_i,\beta_i$ are equivalent (define the same code and the same lattice) upon shifting components by even number,
\bea
\label{tild}
G \sim G+2 \tilde{G},\qquad \tilde{G}\in {\rm Mat}(2n,n,\Z).
\eea 
Another way to represent \eqref{transform} is
\bea
\label{L}
{\vec p}_L={\alpha+\beta \over 2}\rightarrow {\cal O}_L{\alpha+\beta \over 2},\\
{\vec p}_R={\alpha-\beta \over 2}\rightarrow {\cal O}_R{\alpha-\beta \over 2}.
\label{R}
\eea

We already saw in Section \ref{sec:codeCFTs} that simultaneous permutations ${\cal O}_p\times {\cal O}_p \in {\rm O}(n)\times {\rm O}(n)$, as well as sign flips ${\rm I}\times {\cal O}_i \in {\rm O}(n)\times {\rm O}(n)$, $({\cal O}_i)_{kl}=\delta_{kl}-2\delta_{ik}\delta_{il}$, are code equivalences (map a code to an equivalent code). 

Next we consider the sign flips of the form ${\cal O}_i \times {\rm I}$. It is easy to see that the simultaneous sign flip  ${\cal O}_i \times {\cal O}_i$ of a particular component of $\alpha_i$ and $\beta_i$ (applied to all $1\leq i\leq n$) is a symmetry of the lattice. Therefore flipping the sign with ${\cal O}_L$ or with  ${\cal O}_R$ is equivalent.  

A pair of arbitrary  permutations ${\cal O}^1_p \times  {\cal O}^2_p$ can be represented as 
$({\cal O}_p \times { \rm I})({\cal O}^2_p \times {\cal O}^2_p)$, where ${\cal O}_p={\cal O}^1_p ({\cal O}^2_p)^{-1}$. The diagonal part has been already discussed, and we only need to analyze 
$({\cal O}_p \times { \rm I})$. For the vectors ${p}_L,{p}_R$ to correspond to a code lattice, $i$-th component of $p_L$ and $p_R$ must be sententiously integer or half-integer. Since the transformation $({\cal O}_p \times { \rm I})$ leaves $p_R$ invariant, permutation ${\cal O}_p$ must only reshuffle integer or half-integer components of $p_L$ with each other. In other words, provided there is a subset $w\subseteq \{1,\dots,n\}$ such that for all $n$ codewords $(\alpha_i,\beta_i)$, all components of $p_{L,i}^k$, $k\in w$ are simultaneously integer or half-integer, 
\bea
2p_{L,i}^k=2p_{L,i}^l\, \, {\rm mod}\, \, 2\quad {\rm for}\, \, k,l\in w,\quad  {p}^k_{L,i}={\alpha^k_i+\beta^k_i\over 2},\quad 1\leq i\leq n, 
\eea
then ${\cal O}_p$ is an arbitrary permutation of indexes within $w$. For simplicity we can assume $w$ includes first $k$ indexes, in which case all generators $\g$ are of the form 
\bea
\label{gengi}
\g_i =\epsilon_i\, \sigma_{\nu^i_1} \otimes \dots \otimes \sigma_{\nu^i_k} \otimes \dots
\eea
where all $\nu^i_l$ for $1\leq l \leq k$ are even or odd. If, for the given $i$, all $\nu_l^i$ are odd, vector $p_{l,i}=(\underbrace{1/2,\dots,1/2}_k,\dots)$ and ${\cal O}_p$ acts  on it trivially. 
Otherwise, when all $\nu_l^i$ are even, first $k$ components of $\vec{p}_{L,i}$ are either zeros or ones, which are reshuffled by  ${\cal O}_p$. Going back to the generator \eqref{gengi}, in the first case the generator remains invariant, in the second case first $k$ matrices are either $\rm I$ or $\sigma_y$ which are reshuffled by ${\cal O}_p$. 

If we now take a particular $\g_i$ such that first $k$  matrices are either $\rm I$ or $\sigma_y$ and reshuffle them, new vector will trivially commute with all  $\g_j$, provided $\g_i$ was. Therefor the new reshuffled $\g_i$ would belong to the code, since the code is self-dual. We therefore conclude that any transformation of the form ${\cal O}_p\times {\rm I}$ which transforms a code (lattice) into another code (lattice), is in fact a symmetry of that code (lattice). 

To summarize, we have shown that any transformation of the form ${\cal O}\times {\cal O}$ for ${\cal O}\in{\rm O}(n,\Z)$ acts on all codes by transforming them  into equivalent codes.
Furthermore, if a transformation
\bea
{\cal O}_L\times {\cal O}_R \in {\rm O}(n,\Z)\times {\rm O}(n,\Z) \label{OZ}
\eea
transforms a given code into another code, the codes are equivalent in the code equivalence sense. 

So far we have only considered the transformations of the form \eqref{OZ}, which is too restrictive. Going back to (\ref{L},\ref{R}) and taking into account that $\alpha,\beta$ can be shifted by arbitrary even-valued vectors, $\alpha\rightarrow \alpha +2a$, $a\in \Z^n$, while $p_L,p_R$ must always be integer or half-integer, we immediately conclude that all matrix elements of ${\cal O}_L$ and ${\cal O}_R$ are integer or half-integer. Since ${\cal O}_{L,R}$ are orthogonal,  $({\cal O}_{L,R})_{kl}$ is integer, it must be equal to $\pm 1$, and all other components of $k$-th row and $l$-th column must be zero. 
Because of the symmetry \eqref{tild}, all components of $2{\cal Q}$ must be integer. Therefore, if ${\cal O}_{L}$ is integer, so must be ${\cal O}_{R}$, and if ${\cal O}_{L}$ is half-integer, so must be ${\cal O}_{R}$. Finally, if $({\cal O}_{L,R})_{kl}$ is half-integer, it is equal $\pm 1/2$ and there are three other components in $k$-th row and $l$-th column of $({\cal O}_{L,R})_{kl}$ which also must be equal to $\pm 1/2$. Combining all this together and using diagonal transformation ${\cal O}\times {\cal O}$, ${\cal O}\in{\rm O}(n,\Z)$, which maps codes into equivalent codes, we can always take ${\cal O}_{L,R}$ to be block-diagonal matrices where each block being either: i) a $4\times 4$ matrix with all elements being $\pm 1/2$, ii) orthogonal matrix from ${\rm O}(k,\Z)$, $k\leq n$.  Both  ${\cal O}_{L,R}$ must have the same block structure.  

If ${\cal O}_{L,R}$ has no half-integer blocks, this is the case of \eqref{OZ} considered above. In what follows we assume ${\cal O}_{L,R}$ has at least one half-integer $4\times 4$  block, which, without loss of generality we can assume to be located in the upper-left corner. 
Since the diagonal permutations ${\cal O}_p \times {\cal O}_p$, combined with the reshuffling of columns of $G$ would not change the canonical form of $G$, without loss of generality we can assume $(\alpha_i)={\rm I}$ and $(\beta_i)={\rm B}$ is a symmetric matrix and focus on the $4\times 4$ left-upper corner block,
\bea
(\alpha_i)={\rm I}\quad  \rightarrow\quad  (\alpha'_i)=\left({H_L +H_R\over 2}\, {\rm I} + {H_L -H_R\over 2}\, {\rm B}\right)\, \,{\rm mod}\,\, 2,
\eea
and similarly for $\beta_i$. Here $H_L,H_R$ are orthogonal $4\times 4$ and $|(H_{L,R})_{kl}|=1/2$.
Provided new $\alpha'_i,\beta'_i$ define a new code, it can be brought to the canonical form using permutations ${\cal O}_p \times {\cal O}_p$, sign-flips and row operations. Sign flips can be absorbed into $H_L,H_R$ and permutations won't change the canonical form, we therefor can assume matrix $(\alpha'_i)$ is not degenerate. Similar logic with sign flips and permutations can be used to bring $H_L$ to be the Hadamard matrix, while $H_R$ would be one of $768$ possible combinations of signs. For each choice of $H_R$ we can scan through all $2^6$ possible choices of ${\rm B}$\footnote{Since all vectors of the form $2(a,b)$ for $a,b\in \Z^4$ belong to the code lattice, it can be shown code $(\,{\rm I}\,|\,{\rm B})$ must be even. There are only $8$ such codes.}, to conclude that whenever all $\alpha'_i$ happen to be integer-valued (which is necessary for new $G'$ to define code lattice), as a matrix it is degenerate. This concludes our proof.

8
\section{Classification of graphs on $n\leq 8$ nodes}
\label{sec:graphs}
We parametrize graphs with help of their adjacency matrix $\rm B$, an $\n\times n$ symmetric matrix with ${\rm B}_{ii}=0$ and ${\rm B}_{ij}=1$ if the vertices $i$ and $j$ are connected. All matrices $\rm B$ can be parametrized by integer numbers $0\leq k\leq 2^{n(n-1)/2}-1$ using the following non-degenerate map
\bea
{\rm B}\quad  \leftrightarrow\quad k=\sum_{i=1}^n \sum_{j=i+1}^n {\rm B}_{ij}\, 2^{(n+1)(n-2)/2-j+i+1-(i-1)(2n-i)/2}. \label{Bk}
\eea

Our results are summarized as a Wolfram Mathematica lists in the file {\tt graphs8} available \href{https://github.com/dymarsky/QEC}{\tt here}. It contains one variable $\tt ELiELCiI$ which is a nested list of lists.
It has $8$ components, which contain information about LC equivalence classes split into ELC classes, which in turn split into graph isomorphism equivalence classes for the graphs on $1\leq n\leq 8$ nodes. $\tt ELiELCiI[[n]]$ is a list of $t_n^{\rm LC}$ elements, first $i_n^{\rm LC}$ correspond to decomposable graphs, last $t_n^{\rm LC}-i_n^{\rm LC}$ to indecomposable graphs, see Table~\ref{table:LC}. For each $1\leq i\leq t_n^{\rm LC}$, $\tt ELiELCiI[[n,i]]$ with each entry corresponding to a particular ELC equivalence class within given LC equivalence class. Each element of $\tt ELiELCiI[[n,i,j]]$ is a list with each entry corresponding to the graph isomorphism class, within given ELC equivalence class. Each graph isomorphism equivalence class is labeled by the {\it maximal} number $k$ \eqref{Bk} among all numbers associated with  graphs  within this class. A simple consistency check confirms correct number of ELC classes $t_n^{\rm ELC}$ and $i_n^{\rm ELC}$, see Table~\ref{table:ELC}, and the correct number of graph isomorphism classes, see Table~\ref{table:I}.

\begin{table}[t]
\begin{center}
\begin{tabular}{c|cccccccccccc}
$n$ & 1 &  2 & 3 & 4 &5 & 6 &7 &8 & 9 & 10 & 11 &12\\  \hline  \\[-13pt]
$t_n^{\rm I}$ & 1 & 2&  4& 11& 34& 156& 1044& 12346& 274668& 12005168& 1018997864& 165091172592\\  \hline \\[-13pt]
$i_n^{\rm I}$  &  1 & 1 & 2&  6&  21& 112& 853& 11117& 261080& 11716571& 1006700565& 164059830476
\end{tabular}
\end{center}
\caption{Number of inequivalent graphs on $n$ nodes $t_n^{\rm I}$ (number of graph isomorphism equivalence  classes),  for $n\leq 12$. Number of inequivalent indecomposable graphs $i_n^{\rm I}$. Integer sequences \href{https://oeis.org/A000088}{\tt A000088} and
\href{https://oeis.org/A001349}{\tt A001349} correspondingly. }
\label{table:I}
\end{table}

\bibliographystyle{JHEP}
\bibliography{QLC}

%
%
%
%
%
%
%
%
\end{document}